\documentclass[12pt,english]{article}
\usepackage[T1]{fontenc}
\usepackage[latin9]{inputenc}
\usepackage[margin=1.25in]{geometry}
\usepackage{color}
\usepackage{lipsum}
\usepackage{babel}
\usepackage{booktabs}
\usepackage{mathtools}
\usepackage{amsmath}
\usepackage{xcolor}
\usepackage{amssymb}
\usepackage[nodisplayskipstretch]{setspace}
\usepackage{floatrow}
\usepackage{natbib}
\usepackage[unicode=true,pdfusetitle,
 bookmarks=true,bookmarksnumbered=false,bookmarksopen=false,
 breaklinks=false,pdfborder={0 0 0},pdfborderstyle={},backref=false,colorlinks=true]{hyperref}
\usepackage{color}
\usepackage{babel}
\usepackage{multirow}
\usepackage{mathrsfs}
\usepackage{wasysym}
\usepackage{amsfonts}
\usepackage{setspace}
\usepackage{mathrsfs}
\usepackage{wasysym}
\usepackage{amsmath, amssymb}
\usepackage{aecompl}
\usepackage{algorithm2e}
\usepackage{float}
\usepackage{bm}
\usepackage{amsthm}
\usepackage{threeparttable}
\usepackage{lscape}
\usepackage{appendix}
\usepackage{bm}
\usepackage{babel}
\usepackage{dsfont}

\hypersetup{
 pdfborderstyle={},pdfborderstyle={},pdfborderstyle={},pdfborderstyle={},pdfborderstyle={},pdfborderstyle={},pdfborderstyle={},pdfborderstyle={},pdfborderstyle={},pdfborderstyle={},pdfborderstyle={},pdfborderstyle={},pdfborderstyle={},pdfborderstyle={},pdfborderstyle={},pdfborderstyle={},pdfborderstyle={},pdfborderstyle={},pdfborderstyle={},pdfborderstyle={},pdfborderstyle={},pdfborderstyle={},pdfborderstyle={},pdfborderstyle={},pdfborderstyle={},pdfborderstyle={},pdfborderstyle={},pdfborderstyle={},pdfborderstyle={},pdfborderstyle={},pdfborderstyle={},pdfborderstyle={},pdfborderstyle={},pdfborderstyle={},pdfborderstyle={},pdfborderstyle={},pdfborderstyle={},pdfborderstyle={},pdfborderstyle={},pdfborderstyle={},pdfborderstyle={},pdfborderstyle={},pdfborderstyle={},pdfborderstyle={},pdfborderstyle={},pdfborderstyle={},pdfborderstyle={},pdfborderstyle={},pdfborderstyle={},linkcolor=blue,citecolor=blue,urlcolor=blue,pdfstartview=XYZ,plainpages=false,pdfpagelabels}
\makeatletter

\theoremstyle{plain}
\newtheorem{lemma}{Lemma}[section]
\newtheorem{assumption}{Assumption}[section]
\newtheorem{proposition}{Proposition}[section]
\newtheorem{theorem}{Theorem}[section]
\newtheorem{corollary}{Corollary}[section]

\theoremstyle{definition}

\newtheorem{example}{Example}[section]
\newtheorem{remark}{Remark}[section]

\makeatother
\begin{document}
\title{Partial Identification in Moment Models with Incomplete Data---A
Conditional Optimal Transport Approach\thanks{This is a substantially revised version of the previously circulated paper titled ``Partial Identification in Moment Models with Incomplete Data via Optimal
Transport'' available at: https://arxiv.org/abs/2503.16098v1, March 21, 2025. }}
\author{Yanqin Fan,\thanks{Department of Economics, University of Washington}
\ Hyeonseok Park,\thanks{Center for Industrial and Business Organization and Institute for Advanced Economic Research, Dongbei University of Finance
and Economics} \ Brendan Pass,\thanks{Department of Mathematical and Statistical Sciences, University of
Alberta} \ and Xuetao Shi\thanks{School of Economics, University of Sydney} }
\date{\today}
\maketitle
\begin{abstract}
In this paper, we develop a unified approach to study partial identification
of a finite-dimensional parameter defined by a general moment
model with incomplete data. We establish a novel characterization
of the identified set for the true parameter in terms of a continuum
of inequalities defined by conditional optimal transport. For the
special case of an affine moment model, we show that the identified
set is convex and that its support function can be easily computed by
solving a conditional optimal transport problem. For parameters that may not satisfy the moment model, we propose a
two-step procedure to construct its identified set. Finally, we demonstrate the generality and effectiveness of our
approach through several running examples.

\end{abstract}
\newpage{}
\section{Introduction}

\subsection{Model and Motivation}

Many parameters of interest in economic models satisfy a finite number of moment conditions. The seminal paper of \citet{hansen1982large} developed a systematic approach to identification, estimation, and inference when all economic variables in the moment condition model are observable in one data set. Since then, the generalized method of moments has become an influential tool in empirical research in economics. However, researchers often do not have access to a single data set that contains observations on all variables and must rely on multiple data sources that cannot be matched; see the discussion in Section \ref{subsec::related-work}. This paper develops the first systematic method to study the identification of parameters in moment condition models using multiple data sets. 

Specifically, consider the following moment model,
\begin{equation}
\mathbb{E}_{o}\left[m\left(Y_{1},Y_{0},X;\theta^{\ast}\right)\right]=\boldsymbol{0}\text{,}\label{UnconditionalM}
\end{equation}
where $Y_{1}\in\mathcal{Y}_{1}\subseteq\mathbb{R}^{d_{1}},Y_{0}\in\mathcal{Y}_{0}\subseteq\mathbb{R}^{d_{0}}$,
and $X\in\mathcal{X}\subseteq\mathbb{R}^{d_{x}}$ denote three distinct
random variables; $\theta^{\ast}\in\Theta\subseteq\mathbb{R}^{d_{\theta}}$
is the true unknown parameter; $m$ is a known $k$ dimensional moment
function; and $\mathbb{E}_{o}$ denotes the expectation with respect
to the true but unknown joint distribution $\mu_{o}$ of $\left(Y_{1},Y_{0},X\right)$. 

This paper studies identification of
$\theta^{\ast}$ (and known functions of $\theta^{\ast}$)\footnote{When $\delta^{\ast}=G\left(\theta^{\ast}\right)$ is the parameter
of interest for some function $G\left(\cdot\right)$, we say $G\left(\cdot\right)$ is
known if either its functional form is known or (point) identifiable.} in a prevalent
case of \textit{incomplete data} in empirical research, where the
relevant information on $\left(Y_{1},Y_{0},X\right)$ is contained
in two datasets, and the units of observations in different datasets
cannot be matched: the first dataset identifies the distribution of
$\left(Y_{1},X\right)$ and the second dataset identifies the distribution
of $\left(Y_{0},X\right)$.
We develop the first general unified methodology for studying
the sharp identified set (identified set, hereafter) of $\theta^{\ast}$
(and known functions of $\theta^{\ast}$) in the general moment model
(\ref{UnconditionalM}) with incomplete data. Our method allows for (non-additively
separable) moment function of \textit{any finite dimension}, parameter
$\theta^{*}$ of \textit{any finite dimension}, and two arbitrary
random variables $Y_{1}$ and $Y_{0}$, each of \textit{any finite dimension}.

\subsection{Main Contributions}

\paragraph*{General Contributions}

The available sample information in the incomplete data scenario that we study in this paper 
allows identification of the distribution functions of $\left(Y_{1},X\right)$
and $\left(Y_{0},X\right)$, but does not identify the distribution
of $\left(Y_{1},Y_{0},X\right)$ without additional assumptions on
the joint distribution of $\left(Y_{1},Y_{0},X\right)$. 
For the general moment model (\ref{UnconditionalM}), we establish
a characterization of the identified set of $\theta^{\ast}$ in terms
of a continuum of inequalities involving conditional
optimal transport (COT hereafter).\footnote{Conditional optimal transport is an optimal transport between two
conditional measures; see \cite{carlier2016vector} and \cite{hosseini2025conditional} for rigorous accounts of COT. For expositional simplicity, we often use optimal transport even though it is a conditional optimal transport in this paper.}  

For an important class of \textit{affine} moment models, where the
moment function $m\left(\cdot;\theta\right)$ can be written as an
affine transformation of $\theta$ with the expectation of the linear
transformation matrix being identifiable, but not the translation vector,
we show that by applying our general characterization, the identified
set for $\theta^{\ast}$ in this class of models is characterized
by a continuum of linear inequalities and is thus convex. Furthermore,
we derive the expression of the support function of the convex identified
set and show that the value of the support function in each direction
can be obtained by solving a COT problem. As we demonstrate using
several running examples, our novel characterization through COT allows
us to make use of theoretical and computational tools for optimal
transport (OT hereafter; e.g., \citet{rachev2006mass},
\citet{villani2008optimal}, \citet{santambrogio2015optimal}, and
\citet{peyre2019computational}) to further simplify the characterization and develop computationally more tractable identified sets for specific models
than the existing ones in the literature.\footnote{ Other works that have employed optimal transport in partial identification
analysis in different set-ups from ours include \citet{galichon2011set}, \citet{galichon2016optimal}, and
\citet{d2025partially}. We refer
interested readers to \citet{molinari2020microeconometrics} for a
recent survey.} The well-known Fr\'echet problem reviewed in Section \ref{subsec::related-work} is a simple example of an affine moment model to which our characterization of the identified set applies.
We show in Section \ref{subsec:Convexity-and-Support Function} that bounding causal effect in \citet{meango2025combining} is an example of the Fr\'echet problem and the COTs in our characterization provide closed-form expressions for the dual optimizations in Corollary 1 in \citet{meango2025combining}.\footnote{Mathematically, the bounds problem in \citet{meango2025combining} is similar to that in \citet{fan2014identifying, fan2016estimation}. All are variations of the short and long regression in \cite{cross2002regressions}; see also \cite{molinari2006generalization}.
}

To take full advantage of our results for affine moment models, we
propose a two-step procedure to obtain the identified set of the parameter
of interest. In the first step of this procedure, we construct an
auxiliary parameter that satisfies an affine moment model and establish
its identified set. The auxiliary parameter is chosen so that the
parameter of interest can be written as a known transformation of
the auxiliary parameter. In the second step, we compute the identified
set of the parameter of interest from the known transformation of
the convex identified set obtained in the first step. The
two-step procedure substantially broadens the applicability of our
results, since it applies to parameters of interest that may not even
satisfy the moment model (\ref{UnconditionalM}) as we illustrate
via the algorithmic measures in Examples \ref{Example 3::Demographic-Disparity} and \ref{Example 4::True-Positive-Rate-Disparity}.

\paragraph*{Results for Running Examples}

To demonstrate the versatility and advantages of our general method
and the proposed two-step procedure, we study three examples of partially
identified parameters with incomplete data and establish their identified sets using our approach. We show that our method
leads to novel and significantly improved results in all examples compared with those in the literature.

Example \ref{Example2:: LP2} is a linear projection model. We consider two different types of sample information, where the first extends that in \citet{d2024linear} and \citet{hwang2023bounding}; the second data structure extends that in \cite{kitagawa2023linear}.
The moment function is, in general, not affine, and the identified set of the parameter of interest may not be convex. We employ the
proposed two-step procedure, where like \citet{hwang2023bounding}, we express the parameter of interest as a known function of an unknown
parameter $\theta^{\ast}$ satisfying an affine moment model. In the first step, we show that the identified set of $\theta^{\ast}$ is
convex and its support function is given by an integral of COT with quadratic costs, the most studied OT problem; see \citet{brenier1991polar}, \citet{santambrogio2015optimal}, and
\citet{peyre2019computational}. 
In the second step, we obtain the identified set of the
parameter of interest from that of $\theta^{*}$. Using Proposition 2.17 in \citet{santambrogio2015optimal}, we show that when specialized to the model in \citet{d2024linear}, our identified set reduces to that in \citet{d2024linear}. When specialized to
\citet{hwang2023bounding}, our result leads to the identified set that is a proper subset of the superset proposed in \citet{hwang2023bounding}. 
Based on our characterization
of the identified set, we also propose a computationally simpler outer
set that is also a subset of the superset in \citet{hwang2023bounding}.

Examples \ref{Example 3::Demographic-Disparity} and \ref{Example 4::True-Positive-Rate-Disparity} are two algorithmic measures first studied in \citet{kallus2022assessing}
(KMZ, hereafter) in the data combination setup. The measures themselves
do not satisfy the moment model (\ref{UnconditionalM}). To apply
our results, we first express them as known transformations of some
auxiliary parameter $\theta^{*}$, such that $\theta^{*}$ satisfies
model (\ref{UnconditionalM}) with an affine moment function. Consequently,
the two-step procedure applies, and the results we establish for the
identified set of affine models apply to $\theta^{*}$ in both examples.
Specifically, for Example \ref{Example 3::Demographic-Disparity} on the demographic disparity (DD) measure,
we provide a \textit{\emph{closed-form expression}}\emph{ }for the
support function of the identified set for \textit{\emph{any}}\emph{
}collection of DD measures and\emph{ }\textit{\emph{any}} number of
protected classes. Furthermore, we show that the identified set is a polytope with
a closed-form expression for its vertices. For the specific collection
of DD measures studied in KMZ, our closed-form expression solves the
infinite dimensional linear programming formulation in KMZ. We analyze
the theoretical time complexity of both our support function and vertex-based
procedures and show that they are significantly faster than KMZ's
method. Using the empirical example in KMZ, we demonstrate that the
improvement can be over 15,000-fold.

For the true-positive rate disparity (TPRD) measure in Example \ref{Example 4::True-Positive-Rate-Disparity},
KMZ propose to construct the convex hull of the identified set. They
characterize the support function of the convex hull through a computationally
costly, NP-hard non-convex optimization problem. Choosing the auxiliary
parameter vector $\theta^{\ast}$ cleverly, we express the identified
set for any finite number of TPRD measures as a continuous map of
the identified set for $\theta^{\ast}$, where the identified set
for $\theta^{\ast}$ is convex. Consequently, the identified set for
any finite number of TPRD measures is connected, and any single
TPRD measure is a closed interval regardless of the
number of protected classes. For a single TPRD measure, we provide
closed-form expressions for the lower and upper bounds of the closed
interval extending the result in KMZ developed for two
protected classes. For multiple TPRD measures, we demonstrate that the
support function of the identified set for the auxiliary parameter
$\theta^{\ast}$ can be computed through a conditional \textit{partial
optimal transport} problem. Technically, we show that the conditional partial OT problem always
admits a solution with monotone support. Exploring the monotone support of the solution, we develop a novel algorithm, called
\textit{Dual Rank Equilibration AlgorithM (DREAM)}, for computing
the identified set for $\theta^{\ast}$ and for any collection of
TPRD measures. We establish its theoretical time complexity and demonstrate
that DREAM is faster than linear programming using the empirical example
in KMZ. 

\subsection{Related Work} 
\label{subsec::related-work}

Data combination problems are pervasive in applied work. Examples
include studies of the effect of age at school entry on the years of schooling
by combining data from the US censuses in 1960 and 1980 (\citet{angrist1992effect});
differentiated product demand using both micro and macro survey data
(\citet{berry2004differentiated}); studies of long-run returns to
college attendance using PSID/NLSY and Addhealth data (\citet{fan2014identifying,fan2016estimation}); repeated cross-sectional data in difference-in-differences designs
(\citet{fanManzanares2017partial}); evaluations of long-term treatment
effects combining experimental and observational data (\citet{athey2019surrogate});
algorithmic fairness analysis where decision outcomes and protected
attributes are recorded in separate datasets (\citet{kallus2022assessing});
study of neighborhood effects, where longitudinal
residence information and information on individual heterogeneity
are contained in separate datasets (\citet{hwang2023bounding});
inter-generational income mobility, where linked income data across
generations are often unavailable (\citet{santavirta2024name}); consumption
research, where income and consumption are often measured in two different
datasets (\citet{crossley2022regression}); and counterfactual analyses
of actual choice using stated and revealed preference data (\citet{meango2025combining}).

Existing works in the incomplete data case fall into two broad categories. The first category considers the case where the moment function is additively
separable in $Y_{1}$ and $Y_{0}$ and establishes point identification
of $\theta^{\ast}$ under weak conditions.  See, e.g., \citet{angrist1992effect},
\citet{hahn1998role}, \citet{chen2008semiparametric}, \citet{graham2016efficient},
and \citet{athey2019surrogate}.

The second category develops partial identification results for specific models.
Except for the running examples we describe below, most existing work
studying partial identification under data combination can be reformulated
as examples of a special case of model (\ref{UnconditionalM}), where
the moment function is 
\begin{equation}
m\left(y_{1},y_{0},x;\theta\right)=\theta-h\left(y_{1},y_{0}\right)\label{Eq: Univariate}
\end{equation}
for a known function $h$ of dimension $k=1$. Equivalently, $\theta^{\ast}=\mathbb{E}_{o}\left[h\left(Y_{1},Y_{0}\right)\right]$.
For univariate $Y_{1}$ and $Y_{0}$ with given marginals, computing lower
and upper bounds for $\theta^{*}$ for a known function $h$ has a
long history in probability literature, and is referred to as the (general)
Fr\'{e}chet problem; see \cite{ridder2007econometrics} and Section 3 in \citet{fan2014copulas}
for a detailed discussion of this problem, including early references and its relation to copulas. These bounds have been used to establish identified sets for $\theta^{*}$
in a broad range of applications, including distributional treatment
effects in program evaluation and bivariate option pricing in finance;
see e.g., \citet{fan2010sharp,fan2012confidence},
\citet{fan2010partial}, \citet{fan2017partial}, and \citet{firpo2019partial}.
More recently, \citet{fan2023partial}
study identification in model (\ref{UnconditionalM}) with general
moment function $m$ and univariate $Y_{1}$ and $Y_{0}$ via the copula approach and propose a valid inference procedure.\footnote{\citet{fan2023partial} include a point-identified nuisance parameter
$\gamma^{\ast}$ in their moment function to separate the parameter
of interest $\theta^{\ast}$ from $\gamma^{\ast}$. Model (\ref{UnconditionalM})
can be accommodated to having the nuisance parameter by appending
$\gamma^{\ast}$ to $\theta^{\ast}$.} However, the copula approach is limited to univariate $Y_{1}$ and $Y_{0}$. For
multivariate $Y_{1}$ and $Y_{0}$, the lower and upper bounds on $\theta^{*}$
are defined by COTs with ground cost function $h$
or $-h$. \citet{lin2025estimation} propose consistent estimation
using the primal formulation of COT, while \citet{ji2023model}
study estimation and inference using their dual formulation.

\paragraph*{Organization}
 Section \ref{sec:Model-and-Motivating} introduces our sample information and three motivating examples. In
Section \ref{sec:The-Identified-Set}, we first characterize
the identified set for $\theta^{\ast}$ in our model (\ref{UnconditionalM})
in terms of a continuum of inequalities, and then study the
properties of the identified set for the special class of affine moment models. Finally, we apply our characterization to the Fr\'{e}chet problem in (\ref{Eq: Univariate}) and revisit  \citet{meango2025combining}. 
Sections \ref{sec:Linear-Projection-Models} to \ref{sec:Identified set -- TPRD}
apply our general results to each of the motivating examples. The
last section offers some concluding remarks. Technical proofs are
relegated to Appendix \ref{sec:Proofs}. Appendix \ref{sec::additional}
contains additional results on each motivating example.

\paragraph*{Notation}
We let $\boldsymbol{0}$ denote the zero vector and $I_{d}$ denote
the identity matrix of dimension $d\times d$. We use $\mathds{1}\left\{ \cdot\right\} $ to denote the indicator
function. For any vector $v\in\mathbb{R}^{d}$,
we use $\left\Vert v\right\Vert $ to denote its Euclidean norm. We
denote $\mathbb{S}^{d}$ as the unit sphere of dimension $d$. For any function $h$ and measure $\mu$, we denote $\int \vert h\vert d\mu < \infty$ by $h \in L^{1}(\mu)$. For
any cumulative distribution function $F$ defined on $\mathbb{R}$,
we let $F^{-1}\left(t\right)\equiv\inf\left\{ x:F\left(x\right)\geq t\right\} $
denote the quantile function. For any two random variables $W$ and
$V$, we use $F_{W\mid v}^{-1}\left(\cdot\right)$ to denote the quantile
function derived from the distribution of $W$ conditioning on $V=v$.

\section{Model and Motivating Examples }

\label{sec:Model-and-Motivating}

Our model is defined by (\ref{UnconditionalM}) and Assumption \ref{Assu:: DGP}
below, where $\mu_{1X}$ and $\mu_{0X}$ denote probability distributions
of $\left(Y_{1},X\right)$ and $\left(Y_{0},X\right)$, respectively.

\begin{assumption} \label{Assu:: DGP} (i) $d_{1}\geq1$, $d_{0}\geq1$,
and $d_{x}\geq0$. (ii) The distributions $\mu_{1X}$ and $\mu_{0X}$
are identifiable from the sample information contained in two separate
datasets, but the joint distribution $\mu_{o}$ is not identifiable.
(iii) The projections of $\mu_{o}$ on $\left(Y_{1},X\right)$ and
$\left(Y_{0},X\right)$ are $\mu_{1X}$ and $\mu_{0X}$, respectively.
\end{assumption}

We maintain Assumption \ref{Assu:: DGP} throughout the paper. To
avoid the trivial case, we assume that $d_{1}\geq1$ and $d_{0}\geq1$,
but $d_{x}\geq0$. If $d_{x}=0$, then there is no common $X$ in
both datasets. Assumption \ref{Assu:: DGP} (iii) can be interpreted
as a comparability assumption. Consider, for example, the first dataset
that contains only observations on $\left(Y_{1},X\right)$. Since
$Y_{0}$ is missing, the distribution of $\left(Y_{0},X\right)$ is
not identifiable. Assumption \ref{Assu:: DGP} (iii) requires the
underlying probability distribution of $\left(Y_{0},X\right)$ in
the first dataset to be the same as the one of $\left(Y_{0},X\right)$
in the second dataset. The latter probability distribution $\mu_{0X}$
is identifiable because the second dataset contains observations on
$\left(Y_{0},X\right)$. Furthermore, it implies that the distribution
of $X$ in both datasets is the same.

In the rest of this section, we present three motivating examples and use them to illustrate our main results throughout
the paper.
 
\begin{example}[Linear Projection Model]
\label{Example2:: LP2}
Let $Y_{1}^{\top}\equiv\left(Y_{1s},Y_{1r}^{\top}\right)$
and $X^{\top}=\left(X_{p}^{\top},X_{np}^{\top}\right)$, where $Y_{1s}\in\mathbb{R}$,
$Y_{1r}\in\mathbb{R}^{d_{1}-1}$, $X_{p}\in\mathbb{R}^{d_{x_{p}}}$,
and $X_{np}\in\mathbb{R}^{d_{x_{np}}}$. Consider the following linear
projection model: 
\[
Y_{1s}=\left(Y_{0}^{^{\top}},X_{p}^{\top},Y_{1r}^{\top}\right)\delta^{\ast}+\epsilon\text{ and }\mathbb{E}\left[\epsilon\left(Y_{0}^{^{\top}},X_{p}^{\top},Y_{1r}^{\top}\right)\right]=\boldsymbol{0},
\]
where $\delta^{\ast}\in\mathbb{R}^{d_{0}+d_{x_{p}}+d_{1}-1}$ is the
parameter of interest.

We study the identification of $\delta^*$ under two different types of sample information. The first data type is
motivated by the short and long regression in \cite{cross2002regressions}
and adopted in  \cite{pacini2019two},
\cite{d2024linear}, and \cite{hwang2023bounding} for linear projection
models, where researchers observe $\left(Y_{1},X\right)$ and $\left(Y_{0},X\right)$
in two separate data sets.\footnote{We note that the variable $X_{np}$ is
observed in both datasets, but does not enter the linear projection
model. In the textbook complete data case, $X_{np}$ is an
irrelevant variable. But as
we discuss in Remark \ref{Re:Irrelevant}, in the incomplete data case,  $X_{np}$ may shrink the identified set and
becomes relevant.} When $X=X_{p}$, i.e., there is no $X_{np}$, the
model reduces to that in \cite{hwang2023bounding}; When $d_{1}=1$,
i.e., there is no $Y_{1r}$, the model becomes that in \cite{d2024linear} which includes \cite{pacini2019two} as a special case.
\cite{d2024linear} establish the identified set in their model, and note that their approach does not apply to the case in \cite{hwang2023bounding}.
On the other hand, \cite{hwang2023bounding} construct supersets in
her model.

The second type of sample information is motivated by \cite{kitagawa2023linear}, where $\left(Y_{1s},Y_{0},X_{p}\right)$ is available
in one dataset and $\left(Y_{1s},Y_{1r},X_{p}\right)$ is in another.
This fits into our model by relabeling the random variables. Assuming
that only limited information is available, such as the projection
coefficients of $Y_{1s}$ on $\left(Y_{0},X_{p}\right)$ and $Y_{1s}$
on $\left(Y_{1r},X_{p}\right)$ and variance-covariance matrices of
the random variables within each dataset, \cite{kitagawa2023linear}
construct the identified set of $\delta^*$ (given their sample information) through
a non-convex optimization. Instead, we establish the identified set of $\delta^*$ by making full use of the sample information, which 
identifies distributions of $\left(Y_{1s},Y_{0},X_{p}\right)$ and $\left(Y_{1s},Y_{1r},X_{p}\right)$
separately. 

Our approach is applicable to both data types. Same to \cite{hwang2023bounding}, we characterize
the identified set of $\delta^{*}$ in two steps. First we characterize
the identified set of a newly defined parameter $\theta^{\ast}$ such
that $\delta^{\ast}=G\left(\theta^{\ast}\right)$ for some continuous
function $G\left(\cdot\right)$ and then deduce the identified set
of $\delta^{*}$ from that of $\theta^{*}$. Specifically, we let $\theta^{*}=\left(\theta_{s}^{*\top},\theta_{r,1}^{*\top},\ldots,\theta_{r,\left(d_{1}-1\right)}^{*\top}\right)^{\top}\in\mathbb{R}^{d_{0}d_{1}}$,
where $\theta_{s}^{*}=\mathbb{E}_{o}\left[Y_{0}Y_{1s}\right]$ and
$\theta_{r,j}^{*}=\mathbb{E}_{o}\left[Y_{0}Y_{1r,j}\right]$ in which
$Y_{1r,j}$ for $j=1,\ldots,d_{1}-1$ are the elements of $Y_{1r}$. It is easy to see that $\theta^{*}$ satisfies the moment condition:
$\mathbb{E}_{o}\left[m\left(Y_{1},Y_{0},X;\theta^{\ast}\right)\right]=\boldsymbol{0}$,
where 
\[
m\left(y_{1},y_{0},x;\theta\right)=\theta-\left(y_{0}^{\top}y_{1s},y_{0}^{\top}y_{1r,1},\ldots,y_{0}^{\top}y_{1r,\left(d_{1}-1\right)}\right)^{\top}\text{.}
\]

Let $\theta_{r}^{\ast}\equiv\left[\theta_{r,1}^{\ast},\ldots,\theta_{r,\left(d_{1}-1\right)}^{\ast}\right]\in\mathbb{R}^{d_{0} \times \left(d_{1}-1\right)}$.
When $d_{1}>1$, $\delta^{\ast}$ can be expressed as 
\begin{equation}
\delta^{\ast}
=
\begin{pmatrix}
\mathbb{E}\left[Y_{0} Y_{0}^{^{\top}}\right] & \mathbb{E}\left[Y_{0}X_{p}^{\top}\right] & \theta_{r}^{\ast}\\
\mathbb{E}\left[X_{p}Y_{0}^{^{\top}}\right] & \mathbb{E}\left[X_{p}X_{p}^{\top}\right] & \mathbb{E}\left[X_{p}Y_{1r}^{\top}\right]\\
\theta_{r}^{\ast\top} & \mathbb{E}\left[Y_{1r}X_{p}^{\top}\right] & \mathbb{E}\left[Y_{1r}Y_{1r}^{\top}\right]
\end{pmatrix}^{-1}
\begin{pmatrix}
\theta_{s}^{\ast}\\
\mathbb{E}\left[X_{p}Y_{1s}\right]\\
\mathbb{E}\left[Y_{1r}Y_{1s}\right]
\end{pmatrix}
\equiv G\left(\theta^{\ast}\right),\label{Eq:GHuang}
\end{equation}
where the expectations in the definition of $G\left(\cdot\right)$
are identifiable from both types of data. 
When $d_{1}=1$, we have that $Y_{1}=Y_{1s}$, $\theta^{*}=\theta_{s}^{*}$, and
\begin{equation}
\delta^{\ast}
=
\begin{pmatrix}
\mathbb{E}\left[Y_{0}Y_{0}^{^{\top}}\right] & \mathbb{E}\left[Y_{0}X_{p}^{\top}\right]\\
\mathbb{E}\left[X_{p}Y_{0}^{^{\top}}\right] & \mathbb{E}\left[X_{p}X_{p}^{\top}\right]
\end{pmatrix}^{-1}
\begin{pmatrix}
\theta^{\ast}\\
\mathbb{E}\left[X_{p}Y_{1}\right]
\end{pmatrix}
.\label{Eq:DHau}
\end{equation}

Furthermore, 
for the data structure in \cite{kitagawa2023linear},
$\theta_s^*=\mathbb{E}\left[Y_{0}Y_{1s}\right]$ is point identified
from the first dataset $\left(Y_{1s},Y_{0},X_{p}\right)$. We can redefine $\theta_{r}^{\ast}$ as $\theta^{\ast}$ and express $\delta^{\ast}$ as
\begin{equation}
\delta^{\ast}
=
\begin{pmatrix}
\mathbb{E}\left[Y_{0} Y_{0}^{^{\top}}\right] & \mathbb{E}\left[Y_{0} X_{p}^{\top}\right] & \theta^{\ast}\\
\mathbb{E}\left[X_{p}Y_{0}^{^{\top}}\right] & \mathbb{E}\left[X_{p}X_{p}^{\top}\right] & \mathbb{E}\left[X_{p}Y_{1r}^{\top}\right]\\
\theta^{\ast\top} & \mathbb{E}\left[Y_{1r}X_{p}^{\top}\right] & \mathbb{E}\left[Y_{1r}Y_{1r}^{\top}\right]
\end{pmatrix}^{-1}
\begin{pmatrix}
\mathbb{E}\left[Y_{0}Y_{1s}\right]\\
\mathbb{E}\left[X_{p}Y_{1s}\right]\\
\mathbb{E}\left[Y_{1r}Y_{1s}\right]
\end{pmatrix} . \label{eq:Kitagawa}
\end{equation}

\end{example}


\begin{example}[Demographic Disparity in KMZ]
\label{Example 3::Demographic-Disparity}

We illustrate the applicability of our methodology in assessing algorithmic
fairness through data combination studied in KMZ. We focus on the
demographic disparity (DD) measure in this example and the true-positive
rate disparity (TPRD) measure in the next example. Other measures, such as the true-negative rate disparity, can be studied in the same way.
The assumptions on the data imposed in both examples align with KMZ.

Let $Y_{1}\in\left\{ 0,1\right\} $ denote the binary decision outcome
obtained from human decision making or machine learning algorithms.
For instance, $Y_{1}=1$ represents approval of a loan application.
Let $Y_{0}$ be the protected attribute, such as race or gender, that
takes values in $\left\{ a_{1},...,a_{J}\right\} $. Researchers might
be interested in knowing the disparity in within-class average loan
approval rates. This measure is called demographic disparity and
is defined as 
\[
\delta_{DD}^{\ast}\left(j,j^{\dagger}\right)=\Pr\left(Y_{1}=1\mid Y_{0}=a_{j}\right)-\Pr\left(Y_{1}=1\mid Y_{0}=a_{j^{\dagger}}\right)
\]
between classes $a_{j}$ and $a_{j^{\dagger}}$.

Denote $X$ as the set of additional observed covariates. Assume that
we observe $\left(Y_{1},X\right)$ and $\left(Y_{0},X\right)$ separately.
KMZ study  $\left[\delta_{DD}^{\ast}\left(1,J\right),\ldots,\delta_{DD}^{\ast}\left(J-1,J\right)\right]$
directly, where $a_{J}$ is treated as an advantage/reference group.
For $J=2$, they provide a closed-form for the identified set for
$\delta_{DD}^{\ast}\left(1,2\right)$. For $J>2$, they state that the identified
set is convex and characterize its support function evaluated at each
direction as an infinite dimensional linear programming.

Noting that the DD measure itself does not satisfy the moment model (\ref{UnconditionalM}),
we apply the two-step procedure by introducing an auxiliary parameter
$\theta^{\ast}\equiv\left(\theta_{1}^{\ast},...,\theta_{J}^{\ast}\right)^{\top}$.
Specifically, we define $\theta_{j}^{\ast}\equiv\Pr\left(Y_{1}=1\mid Y_{0}=a_{j}\right)$
for $j=1,\ldots,J$. The DD measure $\delta_{DD}^{\ast}\left(j,j^{\dagger}\right)$
for any $j\neq j^{\dagger}$ can be expressed as $\delta_{DD}^{\ast}\left(j,j^{\dagger}\right)\equiv\theta_{j}^{\ast}-\theta_{j^{\dagger}}^{\ast}$.
We characterize $\theta^{\ast}$ via the moment model (\ref{UnconditionalM})
with the following moment function: 
\[
m\left(y_{1},y_{0},x;\theta\right)
=
\begin{pmatrix}
\theta_{1}\mathds{1}\left\{ y_{0}=a_{1}\right\} -\mathds{1}\left\{ y_{1}=1,y_{0}=a_{1}\right\} \\
\vdots\\
\theta_{J}\mathds{1}\left\{ y_{0}=a_{J}\right\} -\mathds{1}\left\{ y_{1}=1,y_{0}=a_{J}\right\} 
\end{pmatrix}
,
\]
where $\theta\equiv\left(\theta_{1},...,\theta_{J}\right)^{\top}$.
It is easy to see that that $\mathbb{E}_{o}\left[m\left(Y_{1},Y_{0},X;\theta^{\ast}\right)\right]=\boldsymbol{0}$.

Let $e_{+}\left(j\right)\in\mathbb{R}^{J}$ ($e_{-}\left(j\right)\in\mathbb{R}^{J}$)
be a row vector such that the $j$-th element of $e_{+}\left(j\right)$
($e_{-}\left(j\right)$) is $1$ ($-1$) and all the remaining elements
are zero. Any DD measure $\delta_{DD}^{\ast}\left(j,j^{\dagger}\right)$
can be expressed as $\left[e_{+}\left(j\right)+e_{-}\left(j^{\dagger}\right)\right]\theta^{\ast}$.
Suppose we are interested in $K$ different DD measures, where $K$
can be smaller than, greater than, or equal to $J-1$. The vector
of DD measures can be written as $E\theta^{\ast}$, where $E\in\mathbb{R}^{K\times J}$
is a matrix such that each row of $E$ is of the form $e_{+}\left(j\right)+e_{-}\left(j^{\dagger}\right)$.
The identified set for $E\theta^{\ast}$ follows from that of $\theta^{\ast}$.
\end{example}

\begin{example}[True-Positive Rate Disparity in KMZ]
\label{Example 4::True-Positive-Rate-Disparity}

Let $Y_{1s}\in\left\{ 0,1\right\} $ be the decision outcome and $Y_{1r}\in\left\{ 0,1\right\} $
be the true outcome. The true outcome is the target that justifies
an optimal decision. $Y_{0}$ and $X$ denote the protected attribute
and the proxy variable introduced in the previous section. True-positive
rate disparity measures the disparity in the proportions of people
who correctly get approved in loan applications between two classes,
given their true non-default outcome. The TPRD measure between any
two classes $a_{j}$ and $a_{j^{\dagger}}$ is defined as 
\[
\delta_{TPRD}^{\ast}\left(j,j^{\dagger}\right)\equiv\Pr\left(Y_{1s}=1\mid Y_{1r}=1,Y_{0}=a_{j}\right)-\Pr\left(Y_{1s}=1\mid Y_{1r}=1,Y_{0}=a_{j^{\dagger}}\right).
\]
Let $Y_{1}\equiv\left(Y_{1s},Y_{1r}\right)$. We assume that $\left(Y_{1},X\right)$
and $\left(Y_{0},X\right)$ can be observed separately.

KMZ study the identified set for $\left[\delta_{TPRD}^{\ast}\left(1,J\right),\ldots,\delta_{TPRD}^{\ast}\left(J-1,J\right)\right]$.
When $J=2$, they establish sharp bounds on $\delta_{TPRD}^{\ast}\left(1,2\right)$.
For cases where $J>2$, KMZ state that the identified set is non-convex
and provide the support function of its convex hull through rather
complicated non-convex optimizations. As a result, it is difficult
to directly analyze the properties of the identified set, such as its
connectedness. Moreover, solving the optimizations can be computationally
intense.

Instead of directly analyzing the identified set for the TPRD measures
which may be non-convex, we apply the two-step procedure by expressing
the TPRD measures as a continuous nonlinear function of some auxiliary
parameter $\theta^{\ast}$ such that $\theta^{*}$ satisfies model
(\ref{UnconditionalM}) and its identified set is convex.
Specifically, let $\theta^{\ast}\equiv\left(\theta_{1}^{\ast},...,\theta_{2J}^{\ast}\right)$,
where for $j=1,...,J,$ we define 
\[
\theta_{j}^{\ast}\equiv\Pr\left(Y_{1s}=1,Y_{1r}=1,Y_{0}=a_{j}\right)\textrm{ and }\theta_{J+j}^{\ast}\equiv\Pr\left(Y_{1s}=0,Y_{1r}=1,Y_{0}=a_{j}\right).
\]
For any $j\neq j^{\dagger}$, define a nonlinear map $g_{j,j^{\dagger}}:\left[0,1\right]^{2J}\rightarrow\left[-1,1\right]$
as $
g_{j,j^{\dagger}}\left(\theta^{\ast}\right)=\frac{\theta_{j}^{\ast}}{\theta_{j}^{\ast}+\theta_{J+j}^{\ast}}-\frac{\theta_{j^{\dagger}}^{\ast}}{\theta_{j^{\dagger}}^{\ast}+\theta_{J+j^{\dagger}}^{\ast}}$. The TPRD measure $\delta_{TPRD}^{\ast}\left(j,j^{\dagger}\right)$ between
classes $a_{j}$ and $a_{j^{\dagger}}$ can be expressed as $\delta_{TPRD}^{\ast}\left(j,j^{\dagger}\right)=g_{j,j^{\dagger}}\left(\theta^{\ast}\right)$.
And we can represent any $K$ different TPRD measures as $G\left(\theta^{\ast}\right)$,
where $G:\left[0,1\right]^{2J}\rightarrow\left[-1,1\right]^{K}$ is
a multidimensional nonlinear map such that each row of $G\left(\cdot\right)$
takes the form of $g_{j,j^{\dagger}}\left(\cdot\right)$.

Let $y_{1}\equiv\left(y_{1s},y_{1r}\right)$. For $\theta=\left(\theta_{1},...,\theta_{2J}\right)$,
it holds that $\mathbb{E}_{o}\left[m\left(Y_{1},Y_{0},X;\theta^{\ast}\right)\right]=\boldsymbol{0}$,
where \begin{align*}
m\left(y_{1},y_{0},x;\theta\right)= & \left(\theta_{1}-\mathds{1}\left\{ y_{1}=\left(1,1\right),y_{0}=a_{1}\right\} ,\ldots,\theta_{J}-\mathds{1}\left\{ y_{1}=\left(1,1\right),y_{0}=a_{J}\right\} ,\right.\\
 & \left.\theta_{J+1}-\mathds{1}\left\{ y_{1}=\left(0,1\right),y_{0}=a_{1}\right\} ,\ldots,\theta_{2J}-\mathds{1}\left\{ y_{1}=\left(0,1\right),y_{0}=a_{J}\right\} \right)^{\top}.
\end{align*}

\end{example}

\section{Identified Set for $\theta^{\ast}$ and its COT Characterization }

\label{sec:The-Identified-Set}

Let $\Theta_{I}\subseteq\Theta$ denote the identified set for $\theta^{\ast}$.
It is defined as 
\begin{equation}
\Theta_{I}\equiv\left\{ \theta\in\Theta:\mathbb{E}_{\mu}\left[m\left(Y_{1},Y_{0},X;\theta\right)\right]=\boldsymbol{0}\text{ for some }\mu\in\mathcal{M}\left(\mu_{1X},\mu_{0X}\right)\right\} ,\label{Def:: Theta I}
\end{equation}
where $\mathbb{E}_{\mu}$ denotes the expectation taken with respect
to some probability measure $\mu\in\mathcal{M}\left(\mu_{1X},\mu_{0X}\right)$,
and $\mathcal{M}\left(\mu_{1X},\mu_{0X}\right)$ is the class of probability
measures whose projections on $\left(Y_{1},X\right)$ and $\left(Y_{0},X\right)$
are $\mu_{1X}$ and $\mu_{0X}$.

Equivalently, the identified set $\Theta_{I}$ can be defined via
conditional distributions. Let $\mu_{1\mid x}$ and $\mu_{0\mid x}$
be the conditional distributions of $Y_{1}$ on $X=x$ and $Y_{0}$
on $X=x$, respectively. Denote $\mu_{X}$ as the probability distribution
of $X$. Recall that $\mathcal{Y}_{1}$, $\mathcal{Y}_{0}$, and $\mathcal{X}$
are the supports of random variables $Y_{1}$, $Y_{0}$, and $X$.
For any $x\in\mathcal{X}$, let $\mathcal{M}\left(\mu_{1\mid x},\mu_{0\mid x}\right)$
denote the class of probability distributions whose projections on
$Y_{1}$ and $Y_{0}$ are $\mu_{1\mid x}$ and $\mu_{0\mid x}$.\footnote{All the statement with ``for any $x\in\mathcal{X}$'' can be relaxed
to ``for almost any $x\in\mathcal{X}$ with respect to $\mu_{X}$
measure''. We ignore such mathematical subtlety to simplify the discussion.} The identified set $\Theta_{I}$ defined in (\ref{Def:: Theta I})
can be equivalently expressed as 
\[\Theta_{I}=\left\{ \theta\in\Theta:\begin{array}{c}
\int\left[\int\int m\left(y_{1},y_{0},x;\theta\right)d\mu_{10\mid x}\left(y_{1},y_{0}\right)\right]d\mu_{X}\left(x\right)=\boldsymbol{0}\text{ }\\
\text{ for some }\mu_{10\mid x}\in\mathcal{M}\left(\mu_{1\mid x},\mu_{0\mid x}\right)\textrm{ for any }x\in\mathcal{X}
\end{array}\right\} .\label{Def:: Theta_I Conditioning}
\]
To simplify the notation, from now on, we omit the integration variables when there is no confusion.

\subsection{A General Characterization of $\Theta_{I}$ via COT }

\label{subsec: Identification - general}

This section provides a COT approach to characterize
the identified set for $\theta^{*}$ in model (\ref{UnconditionalM}). Consider the following set defined through a continuum of inequalities:
\begin{align}
 & \Theta_{o}=\left\{ \theta\in\Theta:\int\mathcal{KT}_{t^{\top}m}\left(\mu_{1\mid x},\mu_{0\mid x};x,\theta\right)d\mu_{X}\leq0\text{ for all }t\in\mathbb{S}^{k}\right\} \textrm{, where }\label{Def::Theta_o Conditioning}\\
 & \mathcal{KT}_{t^{\top}m}\left(\mu_{1\mid x},\mu_{0\mid x};x,\theta\right)\equiv\inf_{\mu_{10\mid x}\in\mathcal{M}\left(\mu_{1\mid x},\mu_{0\mid x}\right)}\int\int t^{\top}m\left(y_{1},y_{0},x;\theta\right)d\mu_{10\mid x}.\nonumber 
\end{align}
For any given $x\in\mathcal{X}$ and $\theta\in\Theta$, $\mathcal{KT}_{t^{\top}m}\left(\mu_{1\mid x},\mu_{0\mid x};x,\theta\right)$
is the COT cost between conditional measures $\mu_{1\mid x}$ and
$\mu_{0\mid x}$ with (ground) cost function $t^{\top}m\left(y_{1},y_{0},x;\theta\right)$.
Here we state the explicit dependence of $\mathcal{KT}_{t^{\top}m}\left(\mu_{1\mid x},\mu_{0\mid x};x,\theta\right)$
on $x$ and $\theta$ because the cost function $t^{\top}m\left(y_{1},y_{0},x;\theta\right)$
is allowed to depend on $x$ and $\theta$. With the following assumption, we show that $\Theta_{I}=\Theta_{o}$.

\begin{assumption} \label{assump::Smoothness-conditioning} For every
$\theta\in\Theta$, the following two conditions
hold. (i) For almost every $x\in\mathcal{X}$ with respect to $\mu_{X}$ measure, $m\left(y_{1},y_{0},x;\theta\right)$ is continuous in $\left(y_{1},y_{0}\right)$ on
$\mathcal{Y}_{1}\times\mathcal{Y}_{0}$. (ii) There exist non-negative continuous
functions $h_{0}\left(y_{0},x;\theta\right)\in L^{1}\left(\mu_{0X}\right)$
and $h_{1}\left(y_{1},x;\theta\right)\in L^{1}\left(\mu_{1X}\right)$
such that 
$
\left\Vert m\left(y_{1},y_{0},x;\theta\right)\right\Vert \leq h_{0}\left(y_{0},x;\theta\right)+h_{1}\left(y_{1},x;\theta\right)$.
\end{assumption}

Assumption \ref{assump::Smoothness-conditioning} (i) holds automatically
when the supports of $\mu_{0\mid x}$ and $\mu_{1\mid x}$ are finite
for every $x\in\mathcal{X}$, i.e., $\mu_{0\mid x}$ and $\mu_{1\mid x}$ are discrete measures, because any function on a finite set is continuous.
 Assumption \ref{assump::Smoothness-conditioning} (ii)
is satisfied if $m$ is bounded. It also holds if $m$ does not depend on $x$, both $\mathcal{Y}_{1}$
and $\mathcal{Y}_{0}$ are bounded, and Assumption \ref{assump::Smoothness-conditioning}
(i) holds. As a result,  Examples \ref{Example 3::Demographic-Disparity} and \ref{Example 4::True-Positive-Rate-Disparity}  satisfy Assumption
\ref{assump::Smoothness-conditioning}. For Example \ref{Example2:: LP2}, the moment function satisfies Assumption
\ref{assump::Smoothness-conditioning} (ii) if $\mathbb{E}\left[\left\Vert Y_{1}\right\Vert ^{2}\right]<\infty$
and $\mathbb{E}\left[\left\Vert Y_{0}\right\Vert ^{2}\right]<\infty$.

\begin{theorem}\label{Thm:: Equivalence of Theta I and Theta o}
It holds that $\Theta_{I}\subseteq\Theta_{o}$. If Assumption \ref{assump::Smoothness-conditioning}
holds, then $\Theta_{I}=\Theta_{o}$. \end{theorem}

Proofs of theorems and propositions in this paper are provided in the appendix. The proof of Theorem \ref{Thm:: Equivalence of Theta I and Theta o} relies on
the minimax theorem derived in \citet{vianney2015minmax} and the
result on the weak compactness of $\mathcal{M}\left(\mu_{1\mid x},\mu_{0\mid x}\right)$
in the proof of Proposition 2.1 in \citet{villani2003topics}.

Theorem \ref{Thm:: Equivalence of Theta I and Theta o} characterizes
the identified set for $\theta^{\ast}$ via a continuum of inequality
constraints on the integral of the COT cost $\mathcal{KT}_{t^{\top}m}\left(\mu_{1\mid x},\mu_{0\mid x};x,\theta\right)$
with respect to $\mu_{X}$. This novel characterization allows us
to explore both theoretical and computational tools, including duality theory developed in the OT literature, to study the
identified set $\Theta_{I}$; see e.g., \citet{rachev2006mass}, \citet{villani2008optimal},
\citet{santambrogio2015optimal}, and \citet{peyre2019computational}.

\paragraph*{Covariate-Assisted Identified Set}

The conditioning variable $X$ provides information on the joint distribution
of $\left(Y_{1},Y_{0}\right)$ by restricting the conditional distributions
of $Y_{1}$ on $X=x$ and $Y_{0}$ on $X=x$ to be $\mu_{1\mid x}$
and $\mu_{0\mid x}$, which are identifiable from the datasets. The identified set obtained using only subvectors of $X$ can be potentially
larger than the covariate-assisted identified set, which explores the information of the full vector $X$.

Partition $X\equiv\left(X_{p},X_{np}\right)$, where $X_{p}\in\mathbb{R}^{d_{x_{p}}}$
and $X_{np}\in\mathbb{R}^{d_{x_{np}}}$. Suppose that the moment function
only depends on $X_{p}=x_{p}$ and denote it as $m\left(y_{1},y_{0},x_{p};\theta\right)$.
This includes Fr\'{e}chet problem (\ref{Eq: Univariate}) and Examples
\ref{Example2:: LP2}-\ref{Example 4::True-Positive-Rate-Disparity}
for $d_{x_{p}}=0$. Denote by $\mu_{1\mid x_{p}}$ and $\mu_{0\mid x_{p}}$
the conditional distributions of $Y_{1}$ and $Y_{0}$ on $X_{p}=x_{p}$, respectively. Let $\mu_{X_{p}}$ be the probability distribution of
$X_{p}$. Define the following set constructed using only $X_{p}$:
\begin{align*}
 & \Theta^{O}=\left\{ \theta\in\Theta:\int\mathcal{KT}_{t^{\top}m}\left(\mu_{1\mid x_{p}},\mu_{0\mid x_{p}};x_{p},\theta\right)d\mu_{X_{p}}\leq0\text{ for all }t\in\mathbb{S}^{k}\right\} \textrm{, where }\\
 & \mathcal{KT}_{t^{\top}m}\left(\mu_{1\mid x_{p}},\mu_{0\mid x_{p}};x_{p},\theta\right)\equiv\inf_{\mu_{10\mid x_{p}}\in\mathcal{M}\left(\mu_{1\mid x_{p}},\mu_{0\mid x_{p}}\right)}\int\int t^{\top}m\left(y_{1},y_{0},x_{p};\theta\right)d\mu_{10\mid x_{p}}.
\end{align*}
The following proposition shows that $\Theta_{I}$ is at most as large
as $\Theta^{O}$.

\begin{proposition} \label{Prop:: Theta O} Under Assumption \ref{assump::Smoothness-conditioning},
it holds that $\Theta_{I}\subseteq\Theta^{O}$. Additionally, if $\mu_{0\mid x}$ is Dirac at $g_{0}(x)$, then
$
\Theta_{I}=\left\{ \theta\in\Theta:\mathbb{E}\left[m\left(Y_{1},g_{0}\left(X\right),X;\theta\right)\right]=\boldsymbol{0}\right\} .
$
\end{proposition}

Proposition \ref{Prop:: Theta O} extends Theorem 3.3 in \cite{fan2017partial}
established for Fr\'{e}chet problem (\ref{Eq: Univariate}) with univariate $Y_{1}$ and $Y_{0}$.
As long as $X_{np}$ is not independent of both $Y_{1}$ and $Y_{0}$,
we expect the covariate-assisted identified set $\Theta_{I}$ to be
smaller than $\Theta^{O}$. 
In general, the stronger $X$ and $Y_{1}$ (or $Y_{0}$) are dependent
on each other, the smaller is the covariate-assisted identified set
for $\theta^{\ast}$. The second part of the proposition shows that if the
conditional measure $\mu_{0\mid x}$ is Dirac at $g_{0}(x)$, then
$\Theta_{I}$ reduces to the identified set for the moment model with
complete data $(Y_{1},X)$. Therefore, when one of $Y_{1}$ and $Y_{0}$ is perfectly
dependent on $X$, $\theta^{\ast}$ is point identified under the
standard rank condition. Without the common $X$ in both datasets,
this would only be possible in the extreme case that one of $Y_{1}$
and $Y_{0}$ is constant almost surely.

\subsection{Affine Moment Model}

\label{subsec:Convexity-and-Support Function}

In this section, we show that if the moment function $m$ is affine in $\theta$, then $\Theta_{I}$ is convex with a simple expression for its support function.
Throughout the discussion, we assume that Assumption \ref{assump::Smoothness-conditioning}
holds so that Theorem \ref{Thm:: Equivalence of Theta I and Theta o}
applies.

\begin{assumption} \label{assmp:: m linear in theta} Let $m_{a}\left(y_{0},x\right)$
and $m_{a}\left(y_{1},x\right)$ be matrix-valued functions of dimension
$k\times d_{\theta}$ and $m_{b}\left(y_{1},y_{0},x\right)$ be a
vector-valued function of dimension $k$. One of the following decompositions
of $m$ holds: (i) $m\left(y_{1},y_{0},x;\theta\right)=m_{a}\left(y_{0},x\right)\theta+m_{b}\left(y_{1},y_{0},x\right)$; (ii) $m\left(y_{1},y_{0},x;\theta\right)=m_{a}\left(y_{1},x\right)\theta+m_{b}\left(y_{1},y_{0},x\right)$.
\end{assumption}

For each running example, we formulate the moment functions and $\theta$ in a way that ensures Assumption \ref{assmp:: m linear in theta} holds. Without loss of generality, we focus on
the first decomposition in the following discussion. Define 
\begin{equation}
\mathcal{KT}_{t^{\top}m_{b}}\left(\mu_{1\mid x},\mu_{0\mid x};x\right)\equiv\inf_{\mu_{10\mid x}\in\mathcal{M}\left(\mu_{1\mid x},\mu_{0\mid x}\right)}\int\int t^{\top}m_{b}\left(y_{1},y_{0},x\right)d\mu_{10\mid x}.\label{eq:AInner}
\end{equation}
Note that $\mathcal{KT}_{t^{\top}m_{b}}\left(\mu_{1\mid x},\mu_{0\mid x};x\right)$
does not depend on $\theta$, because the cost function $t^{\top}m_{b}\left(y_{1},y_{0},x\right)$
does not depend on $\theta$.

\begin{assumption} \label{assump::Parameter space} $\Theta$ is
compact and convex with a nonempty interior. \end{assumption}

\begin{theorem} \label{Thm:: Theta_o equivalent expression} 
Suppose Assumptions \ref{assump::Smoothness-conditioning} and \ref{assmp:: m linear in theta}
hold.  (i) $\Theta_{I}$ can be rewritten as 
\[
\Theta_{I}=\left\{
\theta\in\Theta:t^{\top}\mathbb{E}\left[m_{a}\left(Y_{0},X\right)\right]\theta\leq-\int\mathcal{KT}_{t^{\top}m_{b}}\left(\mu_{1\mid x},\mu_{0\mid x};x\right)d\mu_{X}\text{ for all }t\in\mathbb{S}^{k}\right\} .
\]
(ii) If Assumption \ref{assump::Parameter space} also
holds, then $\Theta_{I}$ is closed and convex. \end{theorem}

Under Assumption \ref{assump::Parameter space},
the convexity of $\Theta_{I}$ follows immediately from Theorem \ref{Thm:: Theta_o equivalent expression}
(i), because the constraints in the expression of $\Theta_{I}$ in
Theorem \ref{Thm:: Theta_o equivalent expression} (i) are affine
in $\theta$.

\begin{example}[General Fr\'{e}chet Problem and Bounds on Causal Effect in \citet{meango2025combining}]
\label{subsec:Monotone-Rearrangement-Inequalit}

As a straightforward application of Theorem \ref{Thm:: Theta_o equivalent expression}, consider the Fr\'{e}chet problem in (\ref{Eq: Univariate}). For this example, $k=1$ and $t\in\left\{ -1,1\right\} $. Under Assumptions \ref{assump::Smoothness-conditioning} and \ref{assump::Parameter space}, $\Theta_{I}$ reduces to the closed interval
in \citet{ji2023model} and \citet{lin2025estimation}. In addition, the COT costs for a given $x\in\mathcal{X}$ are lower and upper bounds on the conditional parameter defined as  $\theta^{\ast}\left(x\right)=\mathbb{E}_{o}\left[h\left(Y_{1},Y_{0}\right)\mid X=x\right]$
for a given $x\in\mathcal{X}$ and have closed-from expressions when  $Y_{1}$ and $Y_{0}$ are univariate
and $h\left(y_{1},y_{0}\right)$ is supermodular (see e.g., \cite{cambanis1976inequalities}). Specifically, for each $x\in\mathcal{X}$, $\theta^{\ast}\left(x\right)\in\left[\theta_{L}\left(x\right),\theta_{U}\left(x\right)\right]$,
where 
\begin{align}
\theta_{L}\left(x\right) & =\int_{0}^{1}h\left(F_{1\mid x}^{-1}\left(u\mid x\right),F_{0\mid x}^{-1}\left(1-u\mid x\right)\right)du,\label{Eq:CI}\nonumber \\
\theta_{U}\left(x\right) & =\int_{0}^{1}h\left(F_{1\mid x}^{-1}\left(u\mid x\right),F_{0\mid x}^{-1}\left(u\mid x\right)\right)du
\end{align}
in which $F_{1\mid x}^{-1}\left(\cdot\mid x\right)$ and $F_{0\mid x}^{-1}\left(\cdot\mid x\right)$
are conditional quantile functions of $Y_{1}$ given $X=x$ and $Y_{0}$
given $X=x$, respectively.

Recently, \citet{meango2025combining} study a conditional parameter that can be shown to be of this form, where $h$ is a product function. Let $D$ denote the agent's binary decision based on an endogenous
decision relevant attribute $X$. Using our notation, the parameter of interest in \citet{meango2025combining}
is $\theta^{\ast}\left(x\right)=\mathbb{E}\left[D\left(x\right)\right]$,
where $D\left(x\right)$ is the potential decision when the decision
relevant attribute is exogenously set to $x$. Let $P$ denote the
vector of stated preference reports. Proposition 1 in \citet{meango2025combining}
shows that under their Assumptions 1 and 2, for each conditional distribution
$F_{D,P\mid X}$ for $\left(D,P\right)$ given $X$, the average structural
function $\theta^*\left(x\right)$ is identified through 
\[
\theta^*\left(x\right)=\int\mathbb{E}\left[D\mid X=x,P=p\right]dF_{P}\left(p\right).
\]

Since the sample information can only identify $F_{D\mid x}$ and
$F_{P\mid x}$, which are the conditional distributions of $D$ and
of $P$ given $X=x$, respectively, $\theta^*\left(x\right)$ is generally
not point identified. Corollary 1 in \citet{meango2025combining}
expresses sharp lower and upper bounds on $\theta^*\left(x\right)$ for
each fixed $x\in\mathcal{X}$ in terms of infinite dimensional optimization
problems. We show below that the sharp lower and upper bounds can be obtained
from (\ref{Eq:CI}) similarly to \citet{fan2014identifying,fan2016estimation}.

Applying the law of iterated expectations to the first displayed equation
in Section 1.4.2 in \citet{meango2025combining}, we obtain an equivalent
expression for $\theta^*\left(x\right)$ as 
\[
\theta^*\left(x\right)=\mathbb{E}_{o}\left[\left(\frac{f_{P}\left(P\right)}{f_{P\mid x}\left(P\mid x\right)}\right)D\mid X=x\right].
\]
We observe separate samples on $\left(D,X\right)$ and $\left(P,X\right)$.
Consequently, (\ref{Eq:CI}) applies to $\theta^*\left(x\right)$ above with $h$ being a product function and the marginal distribution functions
being respectively the conditional distribution functions of $\frac{f_{P}\left(P\right)}{f_{P\mid x}\left(P\mid x\right)}$
given $X=x$ and $D$ given $X=x$. This establishes
explicit expressions for the value functions of the dual optimizations in Corollary 1 in \citet{meango2025combining}.
\end{example}

\subsubsection{The Support Function of $\Theta_{I}$}

For a convex set $\Psi\subseteq\mathbb{R}^{d_{\theta}}$, its support
function $h_{\Psi}\left(\cdot\right):\mathbb{S}^{d_{\theta}}\rightarrow\mathbb{R}$
is defined pointwise by $h_{\Psi}\left(q\right)=\sup_{\psi\in\Psi}q^{\top}\psi$. Our novel characterization of the identified set in Theorem \ref{Thm:: Theta_o equivalent expression}
provides a straightforward way to obtain its support function. Define 
\[
\Theta_{R}\equiv\left\{
\theta\in\mathbb{R}^{d_{\theta}}:t^{\top}\mathbb{E}\left[m_{a}\left(Y_{0},X\right)\right]\theta\leq-\int\mathcal{KT}_{t^{\top}m_{b}}\left(\mu_{1\mid x},\mu_{0\mid x};x\right)d\mu_{X}\text{ for all }t\in\mathbb{S}^{k}\right\} ,
\]
where $\mathcal{KT}_{t^{\top}m_{b}}\left(\mu_{1\mid x},\mu_{0\mid x};x\right)$
is defined in (\ref{eq:AInner}). We
make the following two additional assumptions. The first assumption is on the parameter
space $\Theta$.

\begin{assumption} \label{assump:: Theta_o interior} It holds that
$\Theta_{I}=\Theta_{R}$.\end{assumption}
By definition, we have that $\Theta_{R}\cap\Theta=\Theta_{I}$. Assumption
\ref{assump:: Theta_o interior} requires that the parameter space
$\Theta$ does not provide additional information on the identified
set $\Theta_{I}$ given $\Theta_{R}$. It implies that the boundary
of $\Theta_{I}$ is completely determined by the continuum of inequalities
in Theorem \ref{Thm:: Theta_o equivalent expression} (i). For all
the running examples, we can easily choose $\Theta$ to make Assumption \ref{assump:: Theta_o interior}
hold.  The second assumption is on the moment function.
\begin{assumption} \label{assump::Full rank} Suppose $k=d_{\theta}$
and $\mathbb{E}\left[m_{a}\left(Y_{0},X\right)\right]$ has full rank.
\end{assumption}

Assumption \ref{assump::Full rank} requires that the number of moments
is the same as the number of parameters, and that the coefficient matrix
is of full rank. It implies that if the joint distribution $\mu_{o}$
of $\left(Y_{1},Y_{0},X\right)$ is identifiable from the data, then
$\theta^{\ast}$ is just identified but not overly-identified. For the examples in the paper, Assumption \ref{assump::Full rank} is
satisfied either automatically because $\mathbb{E}\left[m_{a}\left(Y_{0},X\right)\right]$
is a diagonal matrix with positive diagonal entries or under very
mild conditions.

Under Assumption \ref{assump::Full rank}, the expression of $\Theta_{I}$
in Theorem \ref{Thm:: Theta_o equivalent expression} (i) is equivalent
to 
\begin{align}
\Theta_{I}= & \left\{
\theta\in\Theta:q^{\top}\theta\leq s\left(q\right)\text{ for all }q\in\mathbb{S}^{d_{\theta}}\right\} \textrm{, where }\nonumber \\
s\left(q\right)\equiv & -\int\mathcal{KT}_{q^{\top}\mathbb{E}\left[m_{a}\left(Y_{0},X\right)\right]^{-1}m_{b}}\left(\mu_{1\mid x},\mu_{0\mid x};x\right)d\mu_{X}\label{eq:support function - general}\\
= & -\int\left[\inf_{\mu_{10\mid x}\in\mathcal{M}\left(\mu_{1\mid x},\mu_{0\mid x}\right)}\int\int q^{\top}\mathbb{E}\left[m_{a}\left(Y_{0},X\right)\right]^{-1}m_{b}\left(y_{1},y_{0},x\right)d\mu_{10\mid x}\right]d\mu_{X}.\nonumber 
\end{align}
As a result, we have that $q^{\top}\theta\leq s\left(q\right)$ for
all $\theta\in\Theta_{I}$. 
Moreover, applying the result in \cite{rockafellar1997ConvexAnalysis}, we establish that $s(q)$ is indeed the support function of $\Theta_{I}$, as it is a positively homogeneous, proper, closed, and convex function satisfying $s(0) = 0$.

\begin{proposition}\label{Prop:: Support function -  general form}
Under Assumptions  \ref{assump::Smoothness-conditioning}-\ref{assump::Full rank},
it holds that $h_{\Theta_{I}}\left(q\right)=s\left(q\right)$.\end{proposition}

Proposition \ref{Prop:: Support function -  general form} shows that we can obtain the value of the support function for any direction
$q$ by computing the OT cost $\mathcal{KT}_{q^{\top}\mathbb{E}\left[m_{a}\left(Y_{0},X\right)\right]^{-1}m_{b}}\left(\mu_{1\mid x},\mu_{0\mid x};x\right)$
for each $x$ and then integrating with respect to $\mu_{X}$. Moreover, if the parameter of interest is expressed as a linear map from $\theta^{\ast}$
through the linear operator $L$, then based on Proposition \ref{Prop:: Support function -  general form}, the support function of the identified set for the parameter of interest
can be simply calculated as $s\left(L^{\star}q\right)$, where $L^{\star}$
is the adjoint operator with respect to the inner product. 

\begin{remark}
The identification results in Section \ref{subsec: Identification - general}
apply to the most general model defined in (\ref{UnconditionalM}).
Even if any of Assumptions \ref{assmp:: m linear in theta}-\ref{assump::Full rank}
fails, one can still use the original definition of $\Theta_{o}$
to study the identified set $\Theta_{I}$. However, if Assumptions
\ref{assmp:: m linear in theta}-\ref{assump::Full rank} are satisfied,
then the result in Section \ref{subsec:Convexity-and-Support Function}
provides a mathematically and computationally attractive way to obtain
the support function of $\Theta_{I}$. In both cases, the computational
bottleneck lies in the evaluation of the OT cost: $\mathcal{KT}_{t^{\top}m}(\mu_{1\mid x},\mu_{0\mid x};x,\theta)$
in the general case and $\mathcal{KT}_{t^{\top}m_{b}}(\mu_{1\mid x},\mu_{0\mid x};x)$
in the affine case.
\end{remark}

\section{The Linear Projection Model}\label{sec:Linear-Projection-Models}

In this and the next two sections, we construct identified sets for the parameters
in Examples \ref{Example2:: LP2}-\ref{Example 4::True-Positive-Rate-Disparity}
by applying our general results, and compare them with the existing results for each example. For Example \ref{Example2:: LP2}, we maintain the assumption that $\mathbb{E}\left[\left\Vert Y_{1}\right\Vert ^{2}\right]<\infty$
and $\mathbb{E}\left[\left\Vert Y_{0}\right\Vert ^{2}\right]<\infty$, an assumption that is also imposed in both \cite{d2024linear} and \cite{hwang2023bounding}.

\subsection{Identified Sets for $\theta^{*}$ and $\delta^{*}$ }

\label{sec:Identified set: Example-2:-LP}

Given the identified set $\Theta_{I}$ for $\theta^{\ast}$, the identified
set for $\delta^{*}$ can be obtained by mapping each element in $\Theta_{I}$ through (\ref{Eq:GHuang}), (\ref{Eq:DHau}), or (\ref{eq:Kitagawa}), depending on the model.  We focus our discussion on constructing the identified
set $\Theta_{I}$ in this section. We provide a detailed analysis for the first data type and summarize the identified set for the second data type in Remark \ref{re: Second}.

The moment function for $\theta^{*}$ is affine by its functional
form: \begin{align*}
 & m\left(y_{1},y_{0},x;\theta\right)=\theta+m_{b}\left(y_{1},y_{0},x\right)\textrm{, where}\\
 & m_{b}\left(y_{1},y_{0},x\right)\equiv-\left(y_{0}^{\top}y_{1s},y_{0}^{\top}y_{1r,1},\ldots,y_{0}^{\top}y_{1r,\left(d_{1}-1\right)}\right)^{\top}.
\end{align*}
Theorem \ref{Thm:: Theta_o equivalent expression} implies that the
identified set for $\theta^{\ast}$ is 
\begin{equation}
\Theta_{I}=\left\{ \theta\in\Theta:t^{\top}\theta\leq-\int\mathcal{KT}_{t^{\top}m_{b}}\left(\mu_{1\mid x},\mu_{0\mid x};x\right)d\mu_{X}\text{ for all }t\in\mathbb{S}^{d_{0}d_{1}}\right\} .\label{eq: LP Theta I general}
\end{equation}
Partitioning $t$ as $\left(t_{s}^{\top},t_{r,1}^{\top},\ldots,t_{r,\left(d_{1}-1\right)}^{\top}\right)^{\top}$
with $t_{s}\in\mathbb{R}^{d_{0}}$ and $t_{r,j}\in\mathbb{R}^{d_{0}}$
for $j=1,\ldots,d_{1}-1$, we can express $\mathcal{KT}_{t^{\top}m_{b}}\left(\mu_{1\mid x},\mu_{0\mid x};x\right)$
as a COT with a product cost function: 
\begin{align}
 & \mathcal{KT}_{t^{\top}m_{b}}\left(\mu_{1\mid x},\mu_{0\mid x};x\right)\nonumber \\
= & \inf_{\mu_{10\mid x}\in\mathcal{M}\left(\mu_{1\mid x},\mu_{0\mid x}\right)}\int\int-\left(t_{s}^{\top}y_{0},t_{r,1}^{\top}y_{0},\cdots,t_{r,\left(d_{1}-1\right)}^{\top}y_{0}\right)y_{1}d\mu_{10\mid x}\label{eq:LP Full OT}\nonumber \\
= & \frac{1}{2}\inf_{\mu_{10\mid x}\in\mathcal{M}\left(\mu_{1\mid x},\mu_{0\mid x}\right)}\int\int\left\Vert \left(t_{s}^{\top}y_{0},t_{r,1}^{\top}y_{0},\cdots,t_{r,\left(d_{1}-1\right)}^{\top}y_{0}\right)-y_{1}\right\Vert ^{2}d\mu_{10\mid x}\nonumber \\
 & -\frac{1}{2}\int\left\Vert \left(t_{s}^{\top}y_{0},t_{r,1}^{\top}y_{0},\cdots,t_{r,\left(d_{1}-1\right)}^{\top}y_{0}\right)\right\Vert ^{2}d\mu_{0\mid x}-\frac{1}{2}\int\left\Vert y_{1}\right\Vert ^{2}d\mu_{1\mid x}.
\end{align}
The first term on the right-hand side of Equation (\ref{eq:LP Full OT})
is half of the Wasserstein distance between the conditional distribution
of $\left(t_{s}^{\top}Y_{0},t_{r,1}^{\top}Y_{0},\cdots,t_{r,\left(d_{1}-1\right)}^{\top}Y_{0}\right)$
and that of $Y_{1}$ given $X=x$. Wasserstein distance is the value
function of the most studied and best understood OT problem. Recent
breakthroughs in computational OT have contributed to the surge in
its applications in diverse disciplines; see \cite{peyre2019computational}
for a comprehensive account of computational methods and applications.
In addition, when both marginals are Gaussian, closed-form expressions
for Wasserstein distance are known; see the discussion in \cite{santambrogio2015optimal}
and \cite{galichon2016optimal}.

One important feature of the OT problem in (\ref{eq:LP Full OT}) is that the `effective' dimension of the marginal measures is equal
to the dimension of $Y_{1}$ regardless of the dimension of $Y_{0}$,
which can be high. Consequently, the computational burden only depends
on $d_{1}$ and is independent of $d_{0}$. Moreover, in Appendix \ref{subsec:LP - Alternative-Formulation}, we show that by reordering
elements of $\theta^{\ast}$, we can obtain a similar but alternative
characterization of $\Theta_{I}$ such that the COT consists of a
Wasserstein distance between the conditional distribution of $\left(t_{1}^{\top}Y_{1},\cdots,t_{d_{0}}^{\top}Y_{1}\right)$
and that of $Y_{0}$ given $X=x$ with $t_{j}\in\mathbb{R}^{d_{1}}$
for $j=1,\ldots,d_{0}$. For such a characterization, the computational
burden would only depend on $d_{0}$.

When $d_{1}>1$ as in \cite{hwang2023bounding}, the value function
of Equation (\ref{eq:LP Full OT}) does not have a closed-from solution.
However, a celebrated theorem of \cite{brenier1991polar} implies
that under mild regularity condition on $\mu_{1|x}$, the solution
to (\ref{eq:LP Full OT}) concentrates on the gradient of a convex
function, and the value function can be written as $\int-\nabla\phi_{t}\left(y_{1},x\right)y_{1}d\mu_{1\mid x},$
where $\nabla\phi_{t}\left(Y_{1},x\right)$ denotes the gradient of
$\phi_{t}\left(\cdot,x\right)$, a convex function for each $t$ and
$x$.

\begin{proposition} \label{Prop:: Linear Projection Model2 identified set}
Suppose $\mu_{1|x}$ is absolutely continuous with respect to the
Lebesgue measure for almost every $x\in\mathcal{X}$ with respect
to $\mu_{X}$ measure. Then the support function of $\Theta_{I}$
is $h_{\Theta_{I}}\left(q\right)=\mathbb{E}\left[\nabla\phi_{q}\left(Y_{1},X\right)Y_{1}\right]$
and the identified set $\Delta_{I}$ for $\delta^{\ast}$ is 
\[
\Delta_{I}=\left\{ \delta=G\left(\theta\right):q^{\top}\theta\leq\mathbb{E}\left[\nabla\phi_{q}\left(Y_{1},X\right)Y_{1}\right]\textrm{ for all }q\in\mathbb{S}^{d_{0}d_{1}}\right\} ,
\]
where $G$ is defined in Equation (\ref{Eq:GHuang}). \end{proposition}

Due to the functional form of $G\left(\cdot\right)$, the resulting
set $\Delta_{I}$ may not be convex. On the other hand, because $\Theta_{I}$
is convex and $G\left(\cdot\right)$ is continuous, the set $\Delta_{I}$
is connected and the projection of $\Delta_{I}$ onto any one-dimensional subspace is a connected interval.

\begin{remark} In \cite{hwang2023bounding}, $X=X_{p}$. By the definition
of $\Theta_{I}$ in (\ref{Def:: Theta I}), we
obtain that 
\begin{align*}
\Theta_{I}=\left\{ \theta\in\Theta:\theta-\mathbb{E}_{\mu}\left[m_{b}\left(Y_{1},Y_{0},X\right)\right]=\boldsymbol{0}\text{ for some }\mu\in\mathcal{M}\left(\mu_{1X},\mu_{0X}\right)\right\} .
\end{align*}
Its support function is defined as $h_{\Theta_{I}}\left(q\right)=\sup_{\mu\in\mathcal{M}\left(\mu_{1X},\mu_{0X}\right)}\mathbb{E}_{\mu}\left[q^{\top}m_{b}\left(Y_{1},Y_{0},X\right)\right]$.
This is the support function in Equation (7) in \cite{hwang2023bounding}.
\cite{hwang2023bounding} proposes to discretize elements in $\mathcal{M}\left(\mu_{1X},\mu_{0X}\right)$
in order to compute the support function via linear programming. When
the dimension of $Y_{1}$, $Y_{0}$, or $X$ is high, linear programming facilitated
by discretization may be computationally challenging, which motivates
\cite{hwang2023bounding} to propose supersets of $\Theta_{I}$. On
the other hand, depending on the formulation, the computational cost
of our characterization of the identified set or support function
can be independent of $d_{1}$, the dimension of $Y_{1}$; or of $d_{0}$,
the dimension of $Y_{0}$. When both $d_{1}$ and $d_{0}$ are large
and the computational burden is a concern, we can also construct outer
sets for $\Theta_{I}$ from our novel characterization in (\ref{eq: LP Theta I general})
by choosing specific values of $t$. In Appendix \ref{Appendix:: LP Hwang},
we provide one such outer set for $\Theta_{I}$, which is computationally
attractive and, importantly, is a subset of the superset proposed
in \cite{hwang2023bounding}. \end{remark} 

\begin{remark}\label{re: Second}
For the second data type, the identified
set $\Theta_{I}$ for $\theta^{\ast}$ can be established in the same
way as for the first data type with the COT $\mathcal{KT}_{t^{\top}m_{b}}\left(\mu_{1\mid x},\mu_{0\mid x};x\right)$
being $\inf_{\mu_{10\mid x}\in\mathcal{M}\left(\mu_{1\mid x},\mu_{0\mid x}\right)}\protect\int\protect\int-\left(t_{r,1}^{\top}y_{0},\cdots,t_{r,\left(d_{1}-1\right)}^{\top}y_{0}\right)y_{1}d\mu_{10\mid x}$,
where $t_{r,j}\in\mathbb{R}^{d_{0}}$ for $j=1,\ldots,d_{1}-1$. The
identified set for $\delta^{\ast}$ is then obtained by mapping $\Theta_{I}$
through (\ref{eq:Kitagawa}).
\end{remark}

\subsection{The Model in \cite{d2024linear}}

In \cite{d2024linear}, $d_{1}=1$.  Proposition 2.17 in \citet{santambrogio2015optimal} implies that Equation (\ref{eq:LP Full OT})
has a closed-form solution and 
\begin{align*}
\mathcal{KT}_{t^{\top}m_{b}}\left(\mu_{1\mid x},\mu_{0\mid x};x\right)= & \inf_{\mu_{10\mid x}\in\mathcal{M}\left(\mu_{1\mid x},\mu_{0\mid x}\right)}\int\int-\left(t^{\top}y_{0}\right)y_{1}d\mu_{10\mid x}\\
= & -\int_{0}^{1}F_{t^{\top}Y_{0}\mid x}^{-1}\left(u\right)F_{Y_{1}\mid x}^{-1}\left(u\right)du.
\end{align*}
Consequently, we derive the identified set $\Theta_{I}$ from (\ref{eq: LP Theta I general})
as 
\begin{equation}
\Theta_{I}=\left\{ \theta\in\Theta:t^{\top}\theta\leq\int\int_{0}^{1}F_{t^{\top}Y_{0}\mid x}^{-1}\left(u\right)F_{Y_{1}\mid x}^{-1}\left(u\right)dud\mu_{X}\text{ for all }t\in\mathbb{S}^{d_{0}}\right\} .\label{eq: DHau Theta I}
\end{equation}

Following \cite{d2024linear}, we let $\Theta=\mathbb{R}^{d_{0}}$.
Assumptions \ref{assump::Parameter space} and \ref{assump:: Theta_o interior}
are both satisfied. For any $q_{0}\in\mathbb{S}^{d_{0}}$, Proposition
\ref{Prop:: Support function -  general form} provides the support
function of $\Theta_{I}$: 
\begin{equation}
h_{\Theta_{I}}\left(q_{0}\right)=\int\int_{0}^{1}F_{q_{0}^{\top}Y_{0}\mid x}^{-1}\left(u\right)F_{Y_{1}\mid x}^{-1}\left(u\right)dud\mu_{X}.\label{eq:DHau Theta I suppor}
\end{equation}

Assume that the inverse of the second moment matrix $\mathbb{E}\left[\left(Y_{0}^{\top},X_{p}^{\top}\right)^{\top}\left(Y_{0}^{\top},X_{p}^{\top}\right)\right]$
exists, and denote its block partition by $\left(A_{ij}\right)_{i,j\in\left\{ 0,p\right\} }$.
The same assumption is imposed in \cite{d2024linear}. It follows
from Equation (\ref{Eq:DHau}) that $\delta^{\ast}=\left(A_{00}^{\top},A_{p0}^{\top}\right)^{\top}\theta^{\ast}+\left(A_{0p}^{\top},A_{pp}^{\top}\right)^{\top}\mathbb{E}\left[X_{p}Y_{1}\right]$,
which is an affine map. The standard result on the support function
provides the following proposition on the support function of $\Delta_{I}$,
which coincides with Theorems 1 and 2 of \cite{d2024linear}.

\begin{proposition} \label{Prop:: Linear Projection Model identified set}
It holds that $\Delta_{I}=\left\{ \delta\in\mathbb{R}^{d_{0}+d_{x_{p}}}:q^{\top}\delta\leq h_{\Delta_{I}}\left(q\right)\text{ for all }q\in\mathbb{S}^{d_{0}+d_{x_{p}}}\right\}$,
where for $q\equiv\left(q_{0}^{\top},q_{X_{p}}^{\top}\right)^{\top}$
such that $q_{0}\in\mathbb{R}^{d_{0}}$ and $q_{X_{p}}\in\mathbb{R}^{d_{x_{p}}}$,
\[
h_{\Delta_{I}}\left(q\right)=\int\int_{0}^{1}F_{\left(q_{0}^{\top}A_{00}+q_{X_{p}}^{\top}A_{p0}\right)Y_{0}\mid x}^{-1}\left(u\right)F_{Y_{1}\mid x}^{-1}\left(u\right)dud\mu_{X}+\left(q_{0}^{\top}A_{0p}+q_{X_{p}}^{\top}A_{pp}\right)\mathbb{E}\left[Y_{1}X_{p}\right].
\]
\end{proposition}

\begin{remark}
\label{Re:Irrelevant}
It follows from Proposition
\ref{Prop:: Theta O} that the identified set $\Theta_I$ in (\ref{eq: DHau Theta I}) is generally tighter than the outer set using the distributions of $t^\top Y_0$ and $Y_1$ conditional on $X_p$ only. As a result, the `irrelevant variable' $X_{np}$ in the complete data case may tighten the identified set of $\delta^*$ and hence becomes `relevant' in the incomplete data set-up, see \cite{d2024linear} for a similar discussion. 
\end{remark}

\begin{remark}
In the model studied in \citet{pacini2019two}, there is no $X_{np}$.
When $Y_{0}$ is a scalar, it can be shown that our $\Delta_{I}$ is equal
to the set obtained in \citet{pacini2019two}. On the
other hand, when $Y_{0}$ is multivariate, the bound in \citet{pacini2019two}
is not tight; see \citet{d2024linear} and Appendix
\ref{appendix:: LP Pacini} for numerical comparisons. In fact, the
set in \citet{pacini2019two} is equivalent to the set mapped from an outer set of $\Theta_I$ defined by
the same inequality constraint as in $\Theta_{I}$ in 
(\ref{eq: DHau Theta I}) but for $t\in\mathbb{U}^{d_{0}}$
rather than for all $t\in\mathbb{S}^{d_{0}}$, where $\mathbb{U}^{d_{0}} \equiv\left\{ t\in\mathbb{S}^{d_{0}}:\textrm{only one element in }t\textrm{ is nonzero}\right\}$.
\end{remark}

\section{Demographic Disparity Measures in KMZ}

\label{sec: Identified set - DD}

\subsection{Identified Sets for DD Measures and Its Vertex Representation}

\label{subsec:Identified-Set-Theta-DD}

 The moment function $m$ is affine with 
\begin{align*}
 & m_{a}\left(y_{0},x\right)=\mathrm{diag}\left(\mathds{1}\left\{ y_{0}=a_{1}\right\} ,\ldots,\mathds{1}\left\{ y_{0}=a_{J}\right\} \right)\textrm{ and }\\
 & m_{b}\left(y_{0},y_{1},x\right)=-\left(\mathds{1}\left\{ y_{1}=1,y_{0}=a_{1}\right\} ,\ldots,\mathds{1}\left\{ y_{1}=1,y_{0}=a_{J}\right\} \right)^{\top}.
\end{align*}
The identified set $\Theta_{I}$ of $\theta^{\ast}$ follows
from Theorem \ref{Thm:: Theta_o equivalent expression} (i), where
\[
\mathcal{KT}_{t^{\top}m_{b}}\left(\mu_{1\mid x},\mu_{0\mid x};x\right)=\inf_{\mu_{10\mid x}\in\mathcal{M}\left(\mu_{1\mid x},\mu_{0\mid x}\right)}\int\int\sum_{j=1}^{J}-t_{j}\mathds{1}\left\{ y_{1}=1,y_{0}=a_{j}\right\} d\mu_{10\mid x}.
\]
Let $\mathtt{\boldsymbol{d}}\left(y_{1}\right)=\mathds{1}\left\{ y_{1}=1\right\} $
and $\mathtt{\boldsymbol{d}}_{t}\left(y_{0}\right)=\sum_{j=1}^{J}t_{j}\mathds{1}\left\{ y_{0}=a_{j}\right\} $.
Then $\mathcal{KT}_{t^{\top}m_{b}}\left(\cdot\right)$ equals to 
\[
\inf_{\mu_{10\mid x}\in\mathcal{M}\left(\mu_{1\mid x},\mu_{0\mid x}\right)}\int\int-\mathtt{\mathtt{\boldsymbol{d}}}\left(y_{1}\right)\times\mathtt{\mathtt{\boldsymbol{d}}}_{t}\left(y_{0}\right)d\mu_{10\mid x}\left(y_{1},y_{0}\right)=-\int_{\Pr\left(Y_{1}=0\mid x\right)}^{1}F_{D_{t}\mid x}^{-1}\left(u\right)du
\]
with $D_{t}\equiv \mathtt{\boldsymbol{d}}_{t}\left(Y_{0}\right)$.
The last equality follows from the monotone rearrangement inequality.

\begin{proposition} \label{Prop:: DD identified set} The identified set of any $K$ different DD measures $E\theta^{\ast}$
is $\Delta_{DD}=\left\{ E\theta:\theta\in\Theta_{I}\right\} $, where
\[
\Theta_{I}=\left\{ \theta\in\Theta:\sum_{j=1}^{J}t_{j}\theta_{j}\Pr\left(Y_{0}=a_{j}\right)\leq\int\int_{\Pr\left(Y_{1}=0\mid x\right)}^{1}F_{D_{t}\mid x}^{-1}\left(u\right)dud\mu_{X}\text{ for all }t\in\mathbb{S}^{J}\right\} .
\]
Moreover, if $\Pr\left(Y_{0}=a_{j}\right)>0$ for $j=1,\ldots,J$,
then $\Delta_{DD}$ is convex. For any $p\in\mathbb{S}^{K}$, let
$D_{E^{\top}p}\equiv\sum_{j=1}^{J}\left(E^{\top}p\right)_{j}\mathds{1}\left\{ Y_{0}=a_{j}\right\} \Pr\left(Y_{0}=a_{j}\right)^{-1}$,
where $\left(E^{\top}p\right)_{j}$ denotes the $j$-th element of
$E^{\top}p$. The support function of $\Delta_{DD}$ is
\begin{equation}
h_{\Delta_{DD}}\left(p\right)=\int\int_{\Pr\left(Y_{1}=0\mid x\right)}^{1}F_{D_{E^{\top}p}\mid x}^{-1}\left(u\right)dud\mu_{X}.\label{eq:DD support function}
\end{equation}
\end{proposition}
Given $p$, $D_{E^{\top}p}$ is a discrete random variable. As a result, we can easily obtain its quantile function. The closed-form expression of the support function
of $\Delta_{DD}$ in (\ref{eq:DD support function}) provides an easy
way to obtain the identified set for any single DD measure $\delta_{DD}^{\ast}\left(j,j^{\dagger}\right)$
for $J\geq2$; see Appendix \ref{appendix:: Single DD}
for details.
Furthermore, the analytical expression of the support function in (\ref{eq:DD support function}) allows us to obtain the vertex representation of the identified set $\Delta_{DD}$, which substantially simplifies its computation.

In the following, we prove that both $\Theta_{I}$ and $\Delta_{DD}$
are polytopes and provide their vertex representations. 
We first derive
the vertices for $\Theta_{I}$ whose support function is
\begin{equation}
h_{\Theta_{I}}\left(q\right)=\int\int_{\Pr\left(Y_{1}=0\mid x\right)}^{1}F_{D_q\mid x}^{-1}\left(u\right)dud\mu_{X}.\label{eq:theta support function}
\end{equation}
The ones for $\Delta_{DD}$ can be
obtained by premultiplying $E$ to the vertices for $\Theta_{I}$.

We need to introduce some notation. For the set $M\equiv\left\{ 1,\ldots,J\right\} $, define $\mathfrak{S}_{M}$
as its symmetric group, which is the group of all permutations of
$M$. Let $\sigma$ denote a generic element of $\mathfrak{S}_{M}$,
i.e., a permutation of $M$. The order of $\mathfrak{S}_{M}$, denoted
as $\left|\mathfrak{S}_{M}\right|$, equals $J!$. We can enumerate
each element of $\mathfrak{S}_{M}$ as $\sigma_{s}$ for $s=1,\ldots,\left|\mathfrak{S}_{M}\right|$.
The values of $\Pr\left(Y_{0}=a_{j}\right)$ for $j\in M$ are identifiable
from the sample information. For any $\sigma_{s}\in\mathfrak{S}_{M}$,
we define the following polyhedral region of $q$: 
\[
R_{s}\equiv\left\{ q\in\mathbb{R}^{J}:q_{\sigma_{s}\left(1\right)}\Pr\left(Y_{0}=a_{\sigma_{s}\left(1\right)}\right)^{-1}\leq\ldots\leq q_{\sigma_{s}\left(J\right)}\Pr\left(Y_{0}=a_{\sigma_{s}\left(J\right)}\right)^{-1}\right\} .
\]
Given $q\in R_{s}$, for any $s=1,\ldots,\left|\mathfrak{S}_{M}\right|$,
the quantile function $F_{D_{q}\mid x}^{-1}\left(u\right)$ is a step
function taking the value $\sum_{j=1}^{J}\frac{q_{\sigma_{s}\left(j\right)}}{\Pr\left(Y_{0}=a_{\sigma_{s}\left(j\right)}\right)}U\left(u,j,x\right)$,
where 
\[
U\left(u,j,x\right)\equiv\mathds{1}\left\{ \sum_{j=1}^{j-1}\Pr\left(Y_{0}=a_{\sigma_{s}\left(j\right)}\mid x\right)<u<\sum_{j=1}^{j}\Pr\left(Y_{0}=a_{\sigma_{s}\left(j\right)}\mid x\right)\right\} 
\]
and $\Pr\left(Y_{0}=a_{j}\mid x\right)$ denotes the conditional probability
of $Y_{0}=a_{j}$ given $X=x$. This is a linear function of $q$.
Since the integrations with respect to $u$ and $\mu_{X}\left(x\right)$
in Equation (\ref{eq:theta support function}) do not affect linearity,
it holds that $h_{\Theta_{I}}\left(q\right)$ is linear in $q$ for
$q\in R_{s}$. Because each $R_{s}$ is a convex cone and $\cup_{s=1}^{\left|\mathfrak{S}_{M}\right|}R_{s}=\mathbb{R}^{J}$,
$h_{\Theta_{I}}\left(q\right)$ is a piecewise linear function in
$q\in\mathbb{R}^{J}$ on a finite number of convex cones. Theorem
2.3.4 in \cite{hug2010course} shows that $\Theta_{I}$ is a polytope.
Furthermore, for each $s=1,\ldots,\left|\mathfrak{S}_{M}\right|$,
because $h_{\Theta_{I}}\left(q\right)$ is linear in $q$ for $q\in R_{s}$,
there is a vector $v_{s}$ such that $h_{\Theta_{I}}\left(q\right)=q^{\top}v_{s}$
for $q\in R_{s}$. Let $v_{s}\equiv\left(v_{s,1},\ldots,v_{s,J}\right)$.
Then each element $v_{s,\sigma_{s}\left(j\right)}$ for $j=1,\ldots,J$
can be computed by 
\[
\int\int_{\Pr\left(Y_{1}=0\mid x\right)}^{1}\frac{1}{\Pr\left(Y_{0}=a_{\sigma_{s}\left(j\right)}\right)}U\left(u,j,x\right)dud\mu_{X}.
\]
There are $\left|\mathfrak{S}_{M}\right|$ number of vectors $v_{s}$
(not necessarily unique) such that $h_{\Theta_{I}}\left(q\right)=q^{\top}v_{s}$
for $q\in R_{s}$. The set $\left\{ v_{1},\ldots,v_{\left|\mathfrak{S}_{M}\right|}\right\} $
contains all the vertices of the identified set $\Theta_{I}$.\footnote{Not all $v_{s}$ for $s=1,\ldots,\left|\mathfrak{S}_{M}\right|$ are
vertices. Some may lie inside the convex hull of others.} This proves the following proposition.

\begin{proposition}\label{Prop:: DD vertexRepre} Let $\textrm{conv}$
denotes the convex hull. Then the vertex representations of $\Theta_{I}$
and $\Delta_{DD}$ are $\textrm{conv}\left\{ v_{1},\ldots,v_{\left|\mathfrak{S}_{M}\right|}\right\} $
and $\textrm{conv}\left\{ Ev_{1},\ldots,Ev_{\left|\mathfrak{S}_{M}\right|}\right\} $,
respectively.\footnote{Here we derive the vertex representation of $\Delta_{DD}$ from that
of $\Theta_{I}$. Given the analytic expression of the support function
of $\Delta_{DD}$, we can also obtain the vertex representation of
$\Delta_{DD}$ directly from its support function.} \end{proposition}

\subsection{Numerical Comparison with KMZ }

\label{subsec:Comparison-DD}

KMZ study a specific collection of DD measures: $E^{\star}\theta^{\ast}=\left[\delta_{DD}^{\ast}\left(1,J\right),\ldots,\delta_{DD}^{\ast}\left(J-1,J\right)\right]$.
Denote $\Delta_{DD}^{\star}$ as the identified set for $E^{\star}\theta^{\ast}$.
They propose to construct $\Delta_{DD}^{\star}$ using its support
function as 
\[
\Delta_{DD}^{\star}\equiv\left\{ \delta :p^{\top}\delta\leq\int\varPhi_{K}\left(p,x\right)d\mu_{X}\left(x\right)\textrm{ for all }p\in\mathbb{S}^{J-1}\right\} ,
\]
where $\int\varPhi_{K}\left(p,x\right)d\mu_{X}\left(x\right)$ is
the value of the support function evaluated at $p$.
Based on their Proposition 10, KMZ propose a linear programming approach
to computing $\varPhi_{K}\left(p,x\right)$. For any given $p$ and
$x$, there are $2J$ variables and about $5J$ (equality and inequality)
constraints in their linear program. In Appendix \ref{subsec:Time-Complexity-of-DD},
we discuss the time complexity of the approach in KMZ and our methods, and show that both our support function approach and vertex representation approach are significantly more computationally efficient than the
KMZ method. 

For the numerical comparison, we use the empirical application in
KMZ, which studies the identified set for demographic disparity in
mortgage credit decisions using HMDA (Home Mortgage Disclosure Act)
dataset. Following KMZ, we focus on three racial groups: White, Black,
and Asian and Pacific Islander (API). We consider three sets of proxy
variables for race: geolocation (county) only, annual income only,
and both geolocation and annual income. To construct $\Delta_{DD}$,
we apply the vertex representation $\textrm{conv}\left\{ Ev_{1},\ldots,Ev_{\left|\mathfrak{S}_{M}\right|}\right\} $
as discussed in Proposition \ref{Prop:: DD vertexRepre}. In practice,
this representation requires only the conditional probabilities $\Pr\left(Y_{1}=0\mid x_{i}\right)$
and $\Pr\left(Y_{0}=a_{j}\mid x_{i}\right)$ for $j=1,\ldots,J$,
where $x_{i}$ for $i=1,\ldots,n$ are observations of $X$. Since
these values are provided in the KMZ code, we use them directly instead
of recomputing them from the raw data.\footnote{The computed conditional probabilities in Sections \ref{subsec:Comparison-DD} and \ref{subsec: DREAM and Kallus} can be downloaded from https://github.com/CausalML/FairnessWithUnobservedProtectedClass.}
For the KMZ method, we rely
on their published code and set the number of direction vectors to
100. As both methods require the same conditional probabilities, the
comparison of computational time considers only the steps after these
probabilities are obtained. Additional details on this empirical analysis
are provided in Appendix \ref{subsec:Additional-Details-Empirical}.

\begin{figure}
\centering \includegraphics[width=1\linewidth]{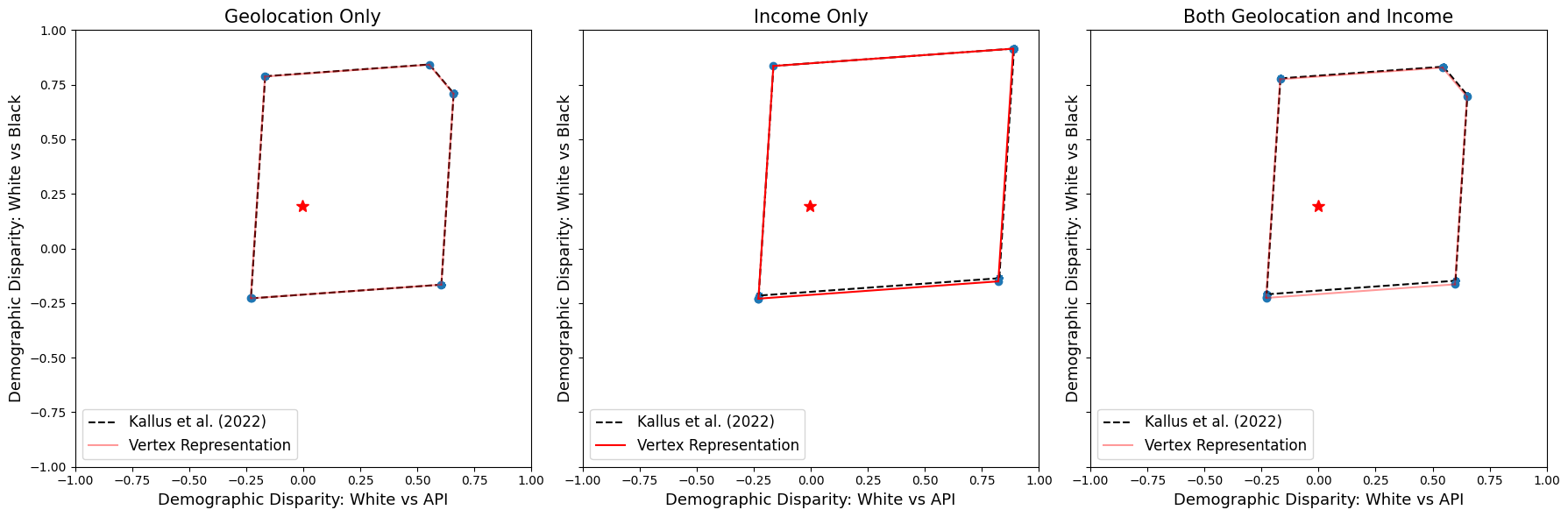}
\caption{Identified sets for demographic disparity in mortgage credit decision.
The red star represents the true demographic disparity calculated
in KMZ}
\label{fig:dd-comparison} 
\end{figure}

Figure \ref{fig:dd-comparison} shows that our method and the approach
in KMZ yield the same identified set. However, our method is substantially
faster. The vertex representation approach computes the identified
set for each proxy variable in under 70 milliseconds, whereas the KMZ method requires approximately 20 minutes. This corresponds to
a speed improvement of over 15,000-fold relative to the KMZ method.

\section{True-Positive Rate Disparity Measures in KMZ}
\label{sec:Identified set -- TPRD}
\subsection{Identified Sets for TPRD Measures---Partial Optimal Transport Characterization}

\label{subsec:Algorithm TPRD}

The moment function is affine with $m_{a}\left(y_{0},x\right)$ being
the $2J\times2J$ identity matrix and 
\begin{align*}
m_{b}\left(y_{1},y_{0},x;\theta\right)= & -\left(\mathds{1}\left\{ y_{1}=\left(1,1\right),y_{0}=a_{1}\right\} ,\ldots,\mathds{1}\left\{ y_{1}=\left(1,1\right),y_{0}=a_{J}\right\} ,\right.\\
 & \left.\mathds{1}\left\{ y_{1}=\left(0,1\right),y_{0}=a_{1}\right\} ,\ldots,\mathds{1}\left\{ y_{1}=\left(0,1\right),y_{0}=a_{J}\right\} \right)^{\top}.
\end{align*}
Consequently, the identified
set for $\theta^{*}$ is 
\[
\Theta_{I}=\left\{ \theta\in\Theta:\sum_{j=1}^{2J}t_{j}\theta_{j}\leq-\int\mathcal{KT}_{t^{\top}m_{b}}\left(\mu_{1\mid x},\mu_{0\mid x};x\right)d\mu_{X}\text{ for all }t\in\mathbb{S}^{2J}\right\} 
\]
with support function for any $q\in\mathbb{S}^{2J}$ given by 
\begin{equation}
h_{\Theta_{I}}\left(q\right)=-\int\mathcal{KT}_{q^{\top}m_{b}}\left(\mu_{1\mid x},\mu_{0\mid x};x\right)d\mu_{X}.\label{eq:TPRD support function}
\end{equation}

Let $\Delta_{TPRD}$ denote the identified set for $K$ different
TPRD measures represented by $G\left(\theta^{\ast}\right)$. We can
compute $\Delta_{TPRD}$ as a non-linear map from $\Theta_{I}$: 
\[
\Delta_{TPRD}=\left\{ \delta=G\left(\theta\right):\theta\in\Theta_{I}\right\} =\left\{ \delta=G\left(\theta\right):q^{\top}\theta\leq h_{\Theta_{I}}\left(q\right)\textrm{ for all }q\in\mathbb{S}^{2J}\right\} ,
\]
where $h_{\Theta_{I}}\left(\cdot\right)$ is defined in (\ref{eq:TPRD support function}).
Because $\Theta_{I}$ is convex and the map $G$ is continuous, we
know that $\Delta_{TPRD}$ is connected, which implies that the identified
set for any single TPRD measure is a closed interval. In Appendix \ref{Appendix::TPRD Single}, we derive the closed-form
expressions for the lower and upper endpoints of the interval, extending
the result in KMZ established for $J=2$.

For multiple TPRD measures, we need to compute $\mathcal{KT}_{q^{\top}m_{b}}\left(\cdot\right)$,
for which we propose a novel computationally efficient algorithm in
Section \ref{subsec: DREAM and Kallus}. Our algorithm solves the
reformulation of $\mathcal{KT}_{q^{\top}m_{b}}\left(\cdot\right)$
as the value function of a linear programming problem in (\ref{eq:Partial transport problem}).

Given $q\equiv\left(q_{1},\ldots,q_{2J}\right)$, define $c\left(i,j\right)\equiv-q_{\left(1-i\right)J+j}$
for $i=0,1$ and $j=1,\ldots,J$. Because $Y_{1}$ and $Y_{0}$ are discrete random variables, we define, with a slight abuse of notation, 
\begin{align*}
 & \mu_{0\mid x}\left(j\right)\equiv\int\mathds{1}\left\{ y_{0}=a_{j}\right\} d\mu_{0\mid x}\left(y_{0}\right)=\Pr\left(Y_{0}=a_{j}\mid X=x\right)\textrm{, }\\
 & \mu_{1\mid x}\left(i,1\right)\equiv\int\mathds{1}\left\{ y_{1s}=i,y_{1r}=1\right\} \mu_{1\mid x}\left(y_{1s},y_{1r}\right)=\Pr\left(Y_{1s}=i,Y_{1r}=1\mid X=x\right)\textrm{, and }\\
 & \mu_{10\mid x}\left(i,1,j\right)\equiv\int\int\mathds{1}\left\{ y_{1}=\left(i,1\right),y_{0}=a_{j}\right\} d\mu_{10\mid x}\left(y_{1},y_{0}\right).
\end{align*}
The COT cost $\mathcal{KT}_{t^{\top}m_{b}}\left(\cdot\right)$ can then be rewritten as the following optimization problem with the argument
$\mu_{10\mid x}\left(i,1,j\right)$: 
\begin{align}
 & \mathcal{KT}_{q^{\top}m_{b}}\left(\mu_{1\mid x},\mu_{0\mid x};x\right)=\min_{\mu_{10\mid x}}\sum_{j=1}^{J}\sum_{i=0}^{1}c\left(i,j\right)\mu_{10\mid x}\left(i,1,j\right)\label{eq:Partial transport problem}\\
\textrm{s.t. } & \textrm{\textrm{(i) }}\mu_{10\mid x}\left(i,1,j\right)\geq0\textrm{ for }i=0,1\textrm{ and }j=1,\ldots,J\textrm{;}\nonumber \\
 & \textrm{(ii) }\sum_{i=0}^{1}\mu_{10\mid x}\left(i,1,j\right)\leq\mu_{0|x}\left(j\right)\textrm{ for }j=1,\ldots,J\textrm{;}\nonumber \\
 & \textrm{(iii) }\sum_{j=1}^{J}\mu_{10\mid x}\left(i,1,j\right)=\mu_{1\mid x}\left(i,1\right)\textrm{ for }i=0,1.\nonumber 
\end{align}
Note that the second constraint is an inequality rather than an equality
because the COT cost $\mathcal{KT}_{q^{\top}m_{b}}\left(\cdot\right)$
does not involve $\mu_{10\mid x}\left(y_{1},y_{0}\right)$ for $y_{1}=\left(y_{1s},0\right)$.
Consequently, (\ref{eq:Partial transport problem}) is actually a
conditional \textit{partial optimal transport}, where $\mu_{0\mid x}\left(j\right)$
and $\mu_{1\mid x}\left(i,1\right)$ are identified from the sample
information. Furthermore, since $i$ in the optimization problem (\ref{eq:Partial transport problem})
takes two values only, we can always make $c\left(i,j\right)$ submodular
by relabeling the index $j$. Without loss of generality, we will assume from now on that the cost function $c\left(i,j\right)$ in (\ref{eq:Partial transport problem}) is submodular.

The following lemma shows that this reformulation of the partial optimal transport problem admits a solution whose support is monotone.

\begin{lemma} \label{lem:: Partial transport problem} 
There is a solution $\left\{ \mu_{10\mid x}^{\ast}\left(i,1,j\right):i=0,1\textrm{ and }j=1,\ldots,J\right\} $
to (\ref{eq:Partial transport problem}) with monotone support. That
is, for some $J^{\ast}\in\left\{ 1,\ldots,J\right\} $, $\mu_{10\mid x}^{\ast}\left(1,1,j\right)=0$
for all $j<J^{\ast}$ and $\mu_{10\mid x}^{\ast}\left(0,1,j\right)=0$
for all $j>J^{\ast}$. \end{lemma}


\subsection{Dual Rank Equilibration Algorithm}

\label{subsec: DREAM and Kallus}

Exploiting the structure of the solution, we propose Algorithm \ref{DREAM}
to compute $\mathcal{KT}_{q^{\top}m_{b}}\left(\cdot\right)$ and call it Dual Rank
Equilibration AlgorithM (DREAM). DREAM is built on an equivalent characterization
of (\ref{eq:Partial transport problem}) which allows it to use the
ranks of $c\left(i,j\right)$ to equilibrate between minimizing the
partial transport cost and respecting the constraints in (\ref{eq:Partial transport problem}).
The equilibration is done twice by first separately adjusting $\mu_{10\mid x}^{\ast}\left(0,1,j\right)$
and $\mu_{10\mid x}^{\ast}\left(1,1,j\right)$ across index $j$ and
then jointly adjusting $\mu_{10\mid x}^{\ast}\left(0,1,j\right)$
and $\mu_{10\mid x}^{\ast}\left(1,1,j\right)$ for a fixed $j$---Dual
Rank Equilibration AlgorithM. Appendix \ref{subsec:DREAM} provides
a detailed description of DREAM. DREAM involves only basic arithmetic, comparison, and logical operations, making it straightforward to implement
directly in code. In fact, it can be executed by hand. In Appendix \ref{subsec:Time-Complexity-TPRD}, we provide theoretical
time complexity analysis of DREAM and show that it is faster than
directly solving the linear programming (\ref{eq:Partial transport problem}),
particularly when $J$ is large.
\RestyleAlgo{ruled}
\SetKwComment{Comment}{/* }{ */}

\begin{algorithm}[hbt!]
\footnotesize
\caption{Dual Rank Equilibration AlgorithM (DREAM)}\label{DREAM}
\SetKwInOut{Input}{Input} 
\SetKwInOut{Output}{Output}
\Input{$c[i,j]$, $\mu_{1\mid x}[i,1]$, and $\mu_{0\mid x}[j]$ for $i=0,1$ and $j=1,\ldots,J$} 
\Output{$\mathcal{KT}_{q^\top m_{2}}(\mu_{1\mid x},\mu_{0\mid x};x)$}

\tcc{Initialization}
$d[j] \gets c[1,j]-c[0,j]$ for $j=1,\dots,J$\;
Sort the indices $j$ in ascending order of $d[j]$\;
Relabel the indices in $c[i,j]$ and $\mu_{0\mid x}[j]$ accordingly\;
$\mu_{10\mid x}[i,1,j] \gets 0$ for $i=0,1$ and $j=1,\dots,J$\;
$JL \gets 1$\;
$JU \gets J$\;

\tcc{Step 1: Determine $JL$ and $JU$}
\While{$\sum_{j=1}^{JL}\mu_{0\mid x}[j] < \mu_{1\mid x}[0,1]$}{
  $JL \gets JL+1$\;
}
\While{$\sum_{j=JU}^{J}\mu_{0\mid x}[j] < \mu_{1\mid x}[1,1]$}{
  $JU \gets JU-1$\;
}

\For{$jj = JL$ \KwTo $JU$}{
  \tcc{Step 2: First Rank Equilibration}

  $\mu_{10\mid x}[0,1,1:jj] \gets$ LPS$(c[0,1:jj],\mu_{0\mid x}[1:jj],\mu_{1\mid x}[0,1])$\;
   \tcc{Function LPS is defined in Algorithm \ref{LSP}}
  $\mu_{10\mid x}[1,1,jj:J] \gets$ LPS$(c[1,jj:J],\mu_{0\mid x}[jj:J],\mu_{1\mid x}[1,1])$\;
  $slack \gets \mu_{10\mid x}[0,1,jj]+\mu_{10\mid x}[1,1,jj]-\mu_{0\mid x}[jj]$\;

\If{$ slack > 0$}{

  \tcc{Step 3: Second Rank Equilibration}
  $d\_cost \gets [c[0,1:jj-1]-c[0,jj],c[1,jj+1:J]-c[1,jj]]$\;
  $d\_mass \gets [\mu_{0\mid x}[1:jj-1]-\mu_{10\mid x}[0,1,1:jj-1],\mu_{0\mid x}[jj+1:J]-\mu_{10\mid x}[1,1,jj+1:J]]$\;
  $a \gets$ LPS$(d\_cost, d\_mass,slack)$\;
  $\mu_{10\mid x}[0,1,1:jj-1] \gets \mu_{10\mid x}[0,1,1:jj-1]+a[1:jj-1]$\;
  $\mu_{10\mid x}[0,1,jj] \gets \mu_{10\mid x}[0,1,jj]-\sum_{j=1}^{jj-1}a[j]$\;
  $\mu_{10\mid x}[1,1,jj+1:J]  \gets \mu_{10\mid x}[1,1,jj+1:J]+a[jj:J-1]$\;
   $\mu_{10\mid x}[1,1,jj] \gets \mu_{10\mid x}[1,1,jj]-\sum_{j=jj}^{J-1}a[j]$\;
      }

$t\_cost\left[jj\right]\gets\sum_{j=1}^{J}\sum_{i=0}^{1}c\left[i,j\right]\mu_{10\mid x}\left[i,1,j\right]$\;
}
\textbf{return} $\mathcal{KT}_{q^{\top}m_{2}}(\mu_{1\mid x},\mu_{0\mid x};x)\gets\min\left\{ t\_cost\left[JL\right],\ldots,t\_cost\left[JU\right]\right\} $
\end{algorithm}

\RestyleAlgo{ruled}
\SetKwComment{Comment}{/* }{ */}

\begin{algorithm}[hbt!]
\footnotesize
\caption{Linear Programming Solver}\label{LSP}
\SetKwInOut{Function}{Function} 
\SetKwInOut{Output}{Output}
\Function{LPS $\left(cost,ineq,eq \right)$} 
$len \gets$ length($cost$)\;
$m \gets$ zeros($len$)\;
Compute ranking $rank[1],\dots,rank[len]$ for $cost[1],\dots,cost[len]$\;
  \For{$r \gets 1$ \KwTo $len$}{
    Let $w$ be the index where $rank[w]=r$\;
    $m[w] \gets \min\{eq,\, ineq[w]\}$\;
    $eq \gets eq-m[w]$\;
  }
\textbf{return} $m$
\end{algorithm}

Different from our approach, KMZ study the convex hull of the identified
set for a collection of TPRD measures instead of the identified set
itself. They focus on $\left[\delta_{TPRD}^{\ast}\left(1,J\right),\ldots,\delta_{TPRD}^{\ast}\left(J-1,J\right)\right]$,
which is one example of our TPRD measures with $G^{\star}\left(\theta^{\ast}\right)\equiv\left[g_{1,J}\left(\theta^{\ast}\right),\ldots,g_{J-1,J}\left(\theta^{\ast}\right)\right]$.
In their Proposition 11, KMZ propose to compute the support function
 of the convex hull at each direction from a complicated double
optimization problem, where the inner maximization problem is a linear
program and the outer maximization is a non-convex optimization problem,
which is known to be NP-hard.  

We applied DREAM and \cite{gurobi}, a general-purpose
linear programming solver, to the empirical application studied
in KMZ.\footnote{We tried and failed to replicate the figures in KMZ using their approach due to the computational challenge of the NP-hard non-convex optimization involved.} The application investigates TPRD in Warfarin
dosing with the ClinPGx/PharmGKB dataset used in \cite{IWPC_2009}.\footnote{The dataset can be downloaded from https://www.clinpgx.org/downloads.} 
The analysis focuses on three protected
attributes (White, Black, and Asian) and three specifications of proxy
variables: genetic factors alone, current medications alone, and both
genetic factors and current medications jointly. Because values of $\mu_{0\mid x}\left(\cdot\right)$ and $\mu_{1\mid x}\left(\cdot\right)$
in $(\ref{eq:Partial transport problem})$ for each observed $X$
have been computed in KMZ, we use them directly. During implementation, we first sample vectors $q_{1},\ldots,q_{N_{q}}$
uniformly from the $2J$-dimensional unit sphere, along with 
vectors $\theta_{1},\ldots,\theta_{N_{\theta}}$ from $\prod_{j=1}^{2J}\left[\theta_{j}^{L},\theta_{j}^{U}\right]$,
where $\theta_{j}^{L}$ and $\theta_{j}^{U}$ are lower and upper
bounds for each $\theta_{j}$ that are defined in Appendix Corollary \ref{Cor:: TPRD single TPRD}.
Then, we construct the set $\widehat{\Theta}_{I}=\left\{ \theta\in\left\{ \theta_{1},\ldots,\theta_{N_{\theta}}\right\} :q^{\top}\theta\leq h_{\Theta_{I}}\left(q\right)\textrm{ for all }q=q_{1},\ldots,q_{N_{q}}\right\}$. Finally, we apply the mapping to each $\theta\in\widehat{\Theta}_{I}$
to obtain an approximation of $\Delta_{TPRD}$: $\widehat{\Delta}_{TPRD}=\left\{ \delta=G\left(\theta\right):\theta\in\widehat{\Theta}_{I}\right\}$.\footnote{Based on DREAM, the solution to (\ref{eq:Partial transport problem})
depends only on the ranks of linear combinations of $c\left(i,j\right)$
for $i=0,1$ and $j=1,\ldots,J$ (or equivalently elements of the
direction vector $q$). As a result, one can show that $\Theta_{I}$
is a polytope, and obtain its vertex representation by following the
similar analysis in Section \ref{subsec:Identified-Set-Theta-DD}.
However, because the representation is complex and we still need to
sample $\theta$ to construct $\widehat{\Delta}_{TPRD}$, we recommend
using $\widehat{\Theta}_{I}$.} Appendix
\ref{subsec:Additional-Details-Empirical} provides more detail.

\begin{figure}[h]
\includegraphics[width=5cm,height=5cm]{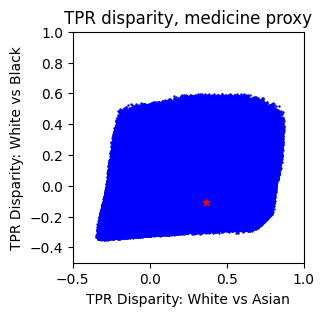}\includegraphics[width=5cm,height=5cm]{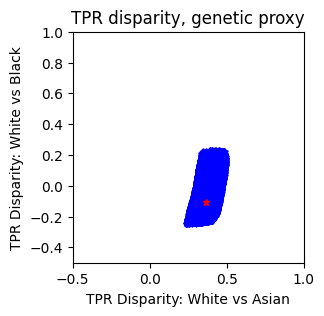}\includegraphics[width=5cm,height=5cm]{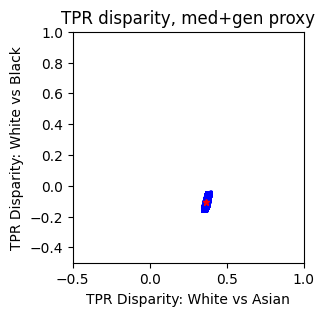}\caption{Identified sets from DREAM algorithm. The red star represents the
true TPRD calculated in \citet{kallus2022assessing}.}
\label{fig:tprd-dream-only} 
\end{figure}

Figure \ref{fig:tprd-dream-only} plots the identified
sets obtained from DREAM, which are identical to the sets produced by
the \cite{gurobi} solver up to a negligible numerical error.
The runtime comparison shows that our approach is more than
twice as fast as the commercial linear programming solver.

\section{Concluding Remarks }

\label{sec:Concluding-Remarks} In this paper, we have developed the
first unified approach to study the identified set for a finite-dimensional
parameter in a general moment equality model with incomplete data.\footnote{
The methods developed in this paper could, in principle, be extended
to the case of more than two datasets using results in \citet{pass2015multi}.
We leave extensions for future work.}
Through several examples, we have demonstrated the advantages and simplicity
of our approach. By exploring recent developments in optimal transport,
our method often leads to equivalent, yet computationally more tractable, identified
sets for specific models than the existing ones in the literature.

This paper provides the first step towards developing a complete set of econometric techniques for moment models under data combination.  Building on the identification analysis in this paper, the next step is to construct valid estimation and inference in the general moment model (\ref{UnconditionalM}) with incomplete data. 
For Example \ref{Example2:: LP2} without common covariate $X$, \citet{d2024linear}
develops estimation and inference for $\theta^{*}$. For the special case that $k=1$
and $\theta^{*}=\mathbb{E}\left[h\left(Y_{1},Y_{0},X)\right)\right]$, \citet{ji2023model}
and \citet{lin2025estimation} construct estimation and inference
using primal and dual formulation, respectively. We leave the general case for future work.

\section{Acknowledgment}

We thank St\'{e}phane Bonhomme, Xiaohong Chen, Marc Henry, Ruixuan Liu, and participants of the BIRS Workshop on Optimal Transport and Distributional
Robustness in March 2024, 2025 IAER Econometrics Workshop, and 2025
China Annual Conference of the Chinese Economists Association; seminar
participants in the statistics department at the University of Washington,
economics departments at Johns Hopkins University and Yale University,
as well as Amazon in October 2024 for useful feedback. Brendan Pass
is pleased to acknowledge the support of the Natural Sciences and Engineering
Research Council of Canada Discovery Grant numbers 04658-2018 and
04864-2024.

\newpage

 
\begin{appendices}

\section{Technical Proofs }

\label{sec:Proofs}

\begin{lemma} \label{lem::Equivalence result--non conditioning}
Consider $\Theta_{I}$ defined in (\ref{Def:: Theta I}). Define 
\begin{equation}
\Theta_{o}\equiv\left\{ \begin{array}{c}
\theta\in\Theta:\inf_{\mu\in\mathcal{M}\left(\mu_{1X},\mu_{0X}\right)}\mathbb{E}_{\mu}\left[t^{\top}m\left(Y_{1},Y_{0},X;\theta\right)\right]\leq0\textrm{ for all }t\in\mathbb{S}^{k}\end{array}\right\} .\label{Def ::Theta_o}
\end{equation}
(i) It holds that $\Theta_{I}\subseteq\Theta_{o}$. (ii) Conversely,
if $\theta_{o}\in\Theta_{o}$, then there exists a sequence of measures
$\mu^{\left(s\right)}\in\mathcal{M}\left(\mu_{1X},\mu_{0X}\right)$
such that $\mathbb{E}_{\mu^{\left(s\right)}}\left[m\left(Y_{1},Y_{0},X;\theta_{o}\right)\right]\rightarrow\mathbf{0}$
as $s\rightarrow\infty.$ \end{lemma}

\begin{proof}[Proof of Lemma \ref{lem::Equivalence result--non conditioning}]
For notational compactness, we introduce $W \equiv  (Y_1, Y_0, X)$, and $\mathbb{B}^k \equiv \{t \in \mathbb{R}^k: \Vert t\Vert \le 1\}$.  By the definition of $\Theta_{I}$ in (\ref{Def:: Theta I}),
for any $\theta_{I}\in\Theta_{I}$, it holds that $\mathbb{E}_{\mu}\left[t^{\top}m\left(W;\theta_{I}\right)\right]=\boldsymbol{0}$
for some $\mu\in\mathcal{M}\left(\mu_{1X},\mu_{0X}\right)$ and for
all $t\in\mathbb{S}^{k}$. Then it must be true that 
$
\inf_{\mu\in\mathcal{M}\left(\mu_{1X},\mu_{0X}\right)}\mathbb{E}_{\mu}\left[t^{\top}m\left(W;\theta_{I}\right)\right]\leq0
$
for all $t\in\mathbb{S}^{k}$. Hence we have $\theta_{I}\in\Theta_{o}$
and $\Theta_{I}\subseteq\Theta_{o}$. This proves the first part of the lemma.

We now prove the second part of the lemma. Because the left-hand side of the inequality in the definition
(\ref{Def ::Theta_o}) is positive homogeneous in $t$, $\Theta_{o}$
is equivalent to 
\[
\Theta_{o}=\left\{ 
\theta\in\Theta: 
\inf_{\mu\in\mathcal{M}\left(\mu_{1X},\mu_{0X}\right)}\mathbb{E}_{\mu}\left[t^{\top}m\left(W; \theta\right)\right]\leq0 
\text{ for all }t\in\mathbb{B}^{k}
\right\} .
\]
If $\theta_{o}\in\Theta_{o}$, then we have that for all $t\in\mathbb{B}^{k}$, $
\inf_{\mu\in\mathcal{M}\left(\mu_{1X},\mu_{0X}\right)}\mathbb{E}_{\mu}\left[t^{\top}m\left(W; \theta_{o}\right)\right]\leq0$.
Taking the supremum over all $t \in \mathbb{B}^k$,
we get 
\[
\sup_{t\in\mathbb{B}^{k}}\inf_{\mu\in\mathcal{M}\left(\mu_{1X},\mu_{0X}\right)}\mathbb{E}_{\mu}\left[t^{\top}m\left(W;\theta_{o}\right)\right]\leq0.
\]
The idea is now to apply a minimax theorem to the left-hand side of
the above inequality. The result in \citet{vianney2015minmax} is
sufficiently general. Because the function $\left(t,\mu\right)\mapsto\mathbb{E}_{\mu}\left[t^{\top}m\left(W;\theta_{o}\right)\right]$ is bilinear, it
is concave in $t$ and convex in $\mu$. The domains
of both $t$ and $\mu$ are convex, and $t$ is finite-dimensional
and bounded. For $t=\boldsymbol{0}$, the function is identically
$0$ and therefore bounded below. Therefore, Theorem 1 in \citet{vianney2015minmax}
applies and we obtain that 
\begin{align*}
 \sup_{t\in\mathbb{B}^{k}}\inf_{\mu\in\mathcal{M}\left(\mu_{1X},\mu_{0X}\right)}\mathbb{E}_{\mu}\left[t^{\top}m\left(W;\theta_{o}\right)\right]
= \inf_{\mu\in\mathcal{M}\left(\mu_{1X},\mu_{0X}\right)}\sup_{t\in\mathbb{B}^{k}}\mathbb{E}_{\mu}\left[t^{\top}m\left(W;\theta_{o}\right)\right]\leq0.
\end{align*}
For any $\mu\in\mathcal{M}\left(\mu_{1X},\mu_{0X}\right)$, we have $\sup_{t \in \mathbb{B}^k} \mathbb{E}_{\mu}[t^{\top} m(W_i; \theta_o)] \le 
\Vert\mathbb{E}_{\mu}[m(W_i, \theta_o)]\Vert$ by the Cauchy-Schwarz inequality, where the equality holds when $t$ is proportional to $E_{\mu}[m(W_i, \theta_o)]$. Thus, we have 
\begin{align*}
0 \geq\inf_{\mu\in\mathcal{M}\left(\mu_{1X},\mu_{0X}\right)}\sup_{t\in\mathbb{R}^{k},\left\Vert t\right\Vert \leq1}\mathbb{E}_{\mu}\left[t^{\top}m\left(W;\theta_{o}\right)\right]
=\inf_{\mu\in\mathcal{M}\left(\mu_{1X},\mu_{0X}\right)}\left\Vert \mathbb{E}_{\mu}\left[m\left(W;\theta_{o}\right)\right]\right\Vert \geq0,
\end{align*}
which implies that $\inf_{\mu\in\mathcal{M}\left(\mu_{1X},\mu_{0X}\right)}\left\Vert \mathbb{E}_{\mu}\left[m\left(W;\theta_{o}\right)\right]\right\Vert =0$.
This completes the proof for the second part of the lemma.
\end{proof}

\begin{lemma} \label{lem::Equivalence conditioning} Consider $\Theta_{o}$
defined in (\ref{Def::Theta_o Conditioning}). If $\theta_{o}\in\Theta_{o}$,
then for any $x\in\mathcal{X}$, there exists a sequence of measures
$\mu_{10\mid x}^{\left(s\right)}\in\mathcal{M}\left(\mu_{1\mid x},\mu_{0\mid x}\right)$
such that  as $s\rightarrow\infty$ it holds $\int\left[\int\int m\left(y_{1},y_{0},x;\theta_{o}\right)d\mu_{10\mid x}^{\left(s\right)}\right]d\mu_{X}\rightarrow\boldsymbol{0}
$.
\end{lemma}

\begin{proof}[Proof of Lemma \ref{lem::Equivalence conditioning}]

By Lemma \ref{lem::Equivalence result--non conditioning} (ii), there
exists a sequence of measures $\mu^{\left(s\right)}\in\mathcal{M}\left(\mu_{1X},\mu_{0X}\right)$
such that $\mathbb{E}_{\mu^{\left(s\right)}}\left[m\left(Y_{1},Y_{0},X;\theta_{o}\right)\right]\rightarrow\mathbf{0}$
as $s\rightarrow\infty.$ For any $s$, the measure $\mu^{\left(s\right)}$
represents a joint distribution of $\left(Y_{1},Y_{0},X\right)$ with
$\mu_{1X}$ and $\mu_{0X}$ being the distributions of $\left(Y_{1},X\right)$
and $\left(Y_{0},X\right)$ that do not depend on $s$. From such
a joint distribution of $\left(Y_{1},Y_{0},X\right)$, by the disintegration
theorem, we can obtain $\mu_{X}$ as the probability measure of $X$
and $\mu_{10\mid x}^{\left(s\right)}$ for each $x\in\mathcal{X}$
whose projections on $Y_{1}$ and $Y_{0}$ are $\mu_{1\mid x}$ and
$\mu_{0\mid x}$. None of the measures $\mu_{X}$, $\mu_{1\mid x}$
for all $x\in\mathcal{X}$, or $\mu_{0\mid x}$ for all $x\in\mathcal{X}$
change with $s$. Therefore, we obtain the sequence of measures $\mu_{10\mid x}^{\left(s\right)}\in\mathcal{M}\left(\mu_{1\mid x},\mu_{0\mid x}\right)$
for each $x\in\mathcal{X}$ that satisfies the condition in the lemma.
\end{proof}

\begin{lemma} \label{lem::Primitive for High-level} Assumption \ref{assump::Smoothness-conditioning}
implies that given every $\theta\in\Theta$
and almost every $x\in\mathcal{X}$ with respect to $\mu_{X}$ measure,
the functional 
$
\mu_{10\mid x}\mapsto\int\int m\left(y_{1},y_{0},x;\theta\right)d\mu_{10\mid x}\left(y_{1},y_{0}\right)
$
is continuous on $\mathcal{M}(\mu_{1|x}, \mu_{0|x})$
with respect to weak convergence of measures. Moreover, for every $\mu_{10\mid x}\in\mathcal{M}(\mu_{1|x}, \mu_{0|x})$, $\int\int m\left(y_{1},y_{0},x;\theta\right)d\mu_{10\mid x}\left(y_{1},y_{0}\right)$ is in  $L^{1}\left(\mu_{X}\right)$. \end{lemma}

\begin{proof}[Proof of Lemma \ref{lem::Primitive for High-level}]

Let $m\left(y_{1},y_{0},x;\theta\right)=\left(m_{1}\left(y_{1},y_{0},x;\theta\right),\dotsc,m_{k}\left(y_{1},y_{0},x;\theta\right)\right)^{\top}$.
For each $\theta\in\Theta$ and almost every $x\in\mathcal{X}$ with
respect to $\mu_{X}$ measure, $m_{i}\left(y_{1},y_{0},x;\theta\right)$
for $i=1,\dotsc,k$ is lower and upper bounded by $L^{1}$-integrable
continuous functions:
\[
-h_{0}\left(y_{0},x;\theta\right)-h_{1}\left(y_{1},x;\theta\right)\le m_{i}\left(y_{1},y_{0},x;\theta\right)\le h_{0}\left(y_{0},x;\theta\right)+h_{1}\left(y_{1},x;\theta\right).
\]
Additionally, we have that the following map
\begin{align*}
\mu_{10|x}\mapsto & \int\int\left(h_{0}\left(y_{0},x;\theta\right)+h_{1}\left(y_{1},x;\theta\right)\right)d\mu_{10|x}\\
 & =\int h_{0}\left(y_{0},x;\theta\right)d\mu_{0|x}+\int h_{1}\left(y_{1},x;\theta\right)d\mu_{1|x}
\end{align*}
is constant and therefore continuous on $\mathcal{M}\left(\mu_{1|x},\mu_{0|x}\right)$
with respect to weak convergence of measures. 
By applying Lemma 4.3 in \cite{villani2008optimal}, we can show that
$\mu_{10|x}\mapsto\int m_{i}\left(y_{1},y_{0},x;\theta\right)d\mu_{10|x}$
is both lower and upper semi-continuous (and therefore continuous)
on $\mathcal{M}\left(\mu_{1|x},\mu_{0|x}\right)$ with respect to
weak convergence of measures. 
This concludes the proof since $k$ is finite and fixed. 
\end{proof}

\begin{proof}[Proof of Theorem \ref{Thm:: Equivalence of Theta I and Theta o}]

By Lemma \ref{lem::Equivalence result--non conditioning} (i), it
remains to show that for any $\theta_{o}\in\Theta_{o}$, it holds
that $\theta_{o}\in\Theta_{I}$. Assumption \ref{assump::Smoothness-conditioning} and Lemma \ref{lem::Primitive for High-level} imply that
the functional 
$
\mu_{10\mid x}\mapsto\int\int m\left(y_{1},y_{0},x;\theta_{o}\right)d\mu_{10\mid x}\left(y_{1},y_{0}\right)
$
is continuous on $\mathcal{M}\left(\mu_{1\mid x},\mu_{0\mid x}\right)$ with respect to weak convergence of measures given every
$\theta_{o}\in\Theta$ and almost every $x\in\mathcal{X}$ with respect
to $\mu_{X}$ measure. By the proof of Proposition 2.1 in \citet{villani2003topics},
$\mathcal{M}\left(\mu_{1\mid x},\mu_{0\mid x}\right)$ is compact
with respect to the weak topology. In consequence, for almost every
$x\in\mathcal{X}$ with respect to $\mu_{X}$ measure, the sequence
of measures $\mu_{10\mid x}^{\left(k\right)}$ in Lemma \ref{lem::Equivalence conditioning}
has a subsequence converging to some $\mu_{10\mid x}^{\star}\in\mathcal{M}\left(\mu_{1\mid x},\mu_{0\mid x}\right)$.
Thus, there exists $\mu_{10\mid x}^{\star}\in\mathcal{M}\left(\mu_{1\mid x},\mu_{0\mid x}\right)$
for almost every $x\in\mathcal{X}$ such that 
$
\int\left[\int\int m\left(y_{1},y_{0},x;\theta_{o}\right)d\mu_{10\mid x}^{\star}\right]d\mu_{X}=\boldsymbol{0}.
$
Hence, $\theta_{o}\in\Theta_{I}$ and the theorem holds. 
\end{proof}

\begin{proof}[Proof of Proposition \ref{Prop:: Theta O}]

First, under Assumption \ref{assump::Smoothness-conditioning}, Theorem \ref{Thm:: Equivalence of Theta I and Theta o}
provides that $\Theta_{I}=\Theta_{o}$. It suffices to show that for
any $\theta\in\Theta_{o}$, it holds that $\theta\in\Theta^{O}$.
The weak compactness of $\mathcal{M}\left(\mu_{1\mid x},\mu_{0\mid x}\right)$
shown in the proof of Proposition 2.1 in \cite{villani2003topics}
provides that for any $t\in\mathbb{S}^{k}$ and $\theta\in\Theta$,
\begin{equation}
\int\mathcal{KT}_{t^{\top}m}\left(\mu_{1\mid x},\mu_{0\mid x};x,\theta\right)d\mu_{X}=\int\left[\int\int t^{\top}m\left(y_{1},y_{0},x_{p};\theta\right)d\mu_{10\mid x}^{\dagger}\right]d\mu_{X}\label{eq:Theta O 1}
\end{equation}
for some $\mu_{10\mid x}^{\dagger}\in\mathcal{M}\left(\mu_{1\mid x},\mu_{0\mid x}\right)$
for each $x\in\mathcal{X}$. Let $\mu_{X_{np}}$ denote the probability
distribution of $X_{np}\in\mathcal{X}_{np}$. By the definition of
conditional probability, we have that
\begin{equation}
\int\left[\int\int t^{\top}m\left(y_{1},y_{0},x_{p};\theta\right)d\mu_{10\mid x}^{\dagger}\right]d\mu_{X}=\int\left[\int\int t^{\top}m\left(y_{1},y_{0},x_{p};\theta\right)d\mu_{10\mid x_{p}}^{\dagger}\right]d\mu_{X_{p}},\label{eq:Theta O 2}
\end{equation}
where $\mu_{10\mid x_{p}}^{\dagger}\equiv\int_{x_{np}\in\mathcal{X}_{np}}\mu_{10\mid x}^{\dagger}d\mu_{X_{np}}\left(x_{np}\right)$.
Again, the definition of $\mu_{1\mid x}$, $\mu_{0\mid x}$, and
$\mu_{X_{np}}$ implies
that $\mu_{10\mid x_{p}}^{\dagger}\in\mathcal{M}\left(\mu_{1\mid x_{p}},\mu_{0\mid x_{p}}\right)$.
Thus, we have that for any $t\in\mathbb{S}^{k}$ and $\theta\in\Theta$,
\begin{align}
\int\left[\int\int t^{\top}m\left(y_{1},y_{0},x_{p};\theta\right)d\mu_{10\mid x_{p}}^{\dagger}\right]d\mu_{X_{p}} & \geq\int\mathcal{KT}_{t^{\top}m}\left(\mu_{1\mid x_{p}},\mu_{0\mid x_{p}};x_{p},\theta\right)d\mu_{X_{p}}.\label{eq:Theta O 3}
\end{align}
Combining (\ref{eq:Theta O 1}), (\ref{eq:Theta O 2}), and (\ref{eq:Theta O 3}),
it holds that if $\theta$ satisfies the inequality conditions in
the definition of $\Theta_{o}$ for each $t\in\mathbb{S}^{k}$, then
it would also satisfy the inequality conditions in $\Theta^{O}$ for the same $t$. Hence, $\Theta_{I}\subseteq\Theta^{O}$.

Second, because $\mu_{0\mid x}$ is Dirac at $g_{0}\left(x\right)$,
we can compute the OT cost in (\ref{Def::Theta_o Conditioning}) as
\begin{align*}
 & \int\left[\inf_{\mu_{10\mid x}\in\mathcal{M}\left(\mu_{1\mid x},\mu_{0\mid x}\right)}\int\int t^{\top}m\left(y_{1},y_{0},x;\theta\right)d\mu_{10\mid x}\right]d\mu_{X}\\
=\; & \int\left[\int t^{\top}m\left(y_{1},g_{0}\left(x\right),x;\theta\right)d\mu_{1\mid x}\left(y_{1}\right)\right]d\mu_{X}\left(x\right)\\
=\; & \int\int t^{\top}m\left(y_{1},g_{0}\left(x\right),x;\theta\right)d\mu_{1X}\left(y_{1},x\right)=\mathbb{E}\left[t^{\top}m\left(Y_{1},g_{0}\left(X\right),X;\theta\right)\right],
\end{align*}
where the second equality follows from the definition of $\mu_{1\mid x}$
and $\mu_{X}$. Denote the set $\left\{ \theta\in\Theta:\mathbb{E}\left[m\left(Y_{1},g_{0}\left(X\right),X;\theta\right)\right]=\boldsymbol{0}\right\}$ in the proposition as $\Theta^{P}$.
Since $\Theta_{I}=\Theta_{o}$ by Theorem \ref{Thm:: Equivalence of Theta I and Theta o},
in the following, we show that $\Theta_{o}\subseteq\Theta^{P}$ and
$\Theta^{P}\subseteq\Theta_{o}$. If $\theta_{o}\in\Theta_{o}$, then for any $t\in\mathbb{S}^{k}$
we have that $\mathbb{E}\left[t^{\top}m\left(Y_{1},g_{0}\left(X\right),X;\theta_{o}\right)\right]\leq0.$
If $t\in\mathbb{S}^{k}$, then $-t\in\mathbb{S}^{k}$ holds as well.
In consequence, we have that for any $t\in\mathbb{S}^{k}$, 
\[
0\geq\mathbb{E}\left[t^{\top}m\left(Y_{1},g_{0}\left(X\right),X;\theta_{o}\right)\right]=-\mathbb{E}\left[-t^{\top}m\left(Y_{1},g_{0}\left(X\right),X;\theta_{o}\right)\right]\geq0.
\]
This shows that $\mathbb{E}\left[t^{\top}m\left(Y_{1},g_{0}\left(X\right),X;\theta_{o}\right)\right]=0$
for any $t\in\mathbb{S}^{k}$, which implies that $\mathbb{E}\left[m\left(Y_{1},g_{0}\left(X\right),X;\theta_{o}\right)\right]=\boldsymbol{0}.$
Thus, we have that $\theta_{o}\in\Theta^{P}$. The other side $\Theta^{P}\subseteq\Theta_{o}$
is straightforward. Hence, the proposition holds. \end{proof}

\begin{proof}[Proof of Theorem \ref{Thm:: Theta_o equivalent expression}]

Theorem \ref{Thm:: Equivalence of Theta I and Theta o} implies that
$\Theta_{o}=\Theta_{I}$. The first decomposition of $m$
provided by Assumption \ref{assmp:: m linear in theta} provides that
that 
\begin{align*}
\int\int t^{\top}m\left(y_{1},y_{0},x;\theta\right)d\mu_{10\mid x} & =\int\int t^{\top}m_{a}\left(y_{0},x\right)\theta d\mu_{10\mid x}+\int\int t^{\top}m_{b}\left(y_{1},y_{0},x\right)d\mu_{10\mid x}\\
 & =t^{\top}\int\int m_{a}\left(y_{0},x\right)\theta d\mu_{0\mid x}+\int\int t^{\top}m_{b}\left(y_{1},y_{0},x\right)d\mu_{10\mid x}\\
 & =t^{\top}\mathbb{E}\left[m_{a}\left(Y_{0},X\right)\mid X\right]\theta+\int\int t^{\top}m_{b}\left(y_{1},y_{0},x\right)d\mu_{10\mid x},
\end{align*}
where the second equality holds by Assumption \ref{Assu:: DGP}. Thus,
we have for any $t\in\mathbb{S}^{k}$, 
\begin{align*}
 & \int\mathcal{KT}_{t^{\top}m}\left(\mu_{1\mid x},\mu_{0\mid x};x,\theta\right)d\mu_{X}\leq0\\
\iff & \int t^{\top}\mathbb{E}\left[m_{a}\left(Y_{0},X\right)\mid X\right]\theta d\mu_{X}+\int\mathcal{KT}_{t^{\top}m_{b}}\left(\mu_{1\mid x},\mu_{0\mid x};x\right)d\mu_{X}\leq0\\
\iff & t^{\top}\mathbb{E}\left[m_{a}\left(Y_{0},X\right)\right]\theta+\int\mathcal{KT}_{t^{\top}m_{b}}\left(\mu_{1\mid x},\mu_{0\mid x};x\right)d\mu_{X}\leq0\\
\iff & t^{\top}\mathbb{E}\left[m_{a}\left(Y_{0},X\right)\right]\theta\leq-\int\mathcal{KT}_{t^{\top}m_{b}}\left(\mu_{1\mid x},\mu_{0\mid x};x\right)d\mu_{X}.
\end{align*}
Thus, we show that each inequality in (\ref{Def::Theta_o Conditioning})
can be alternatively expressed as the inequality in part (i) of the
theorem.

To prove that $\Theta_{I}$ is closed, note that it is the intersection of $\Theta$ and the infinite number of closed half-spaces. Then, it is closed because the (uncountable) intersection of closed half-spaces is still closed. The convexity of $\Theta_{I}$ holds because the constraints
in the expression of $\Theta_{I}$ are affine in $\theta$ and $\Theta$
is convex by Assumption \ref{assump::Parameter space}. Thus, part
(ii) of the theorem holds. 
\end{proof}

\begin{lemma}\label{lem:: Theta_o equivalent expression-2} Under
Assumptions \ref{assump::Smoothness-conditioning}, \ref{assmp:: m linear in theta},
and \ref{assump::Full rank}, the identified set $\Theta_{I}$ in
Theorem \ref{Thm:: Equivalence of Theta I and Theta o} (i) can be
rewritten as $\left\{ \theta\in\Theta:q^{\top}\theta\leq s\left(q\right)\text{ for all }q\in\mathbb{S}^{d_{\theta}}\right\} $,
where $s\left(\cdot\right)$ is defined in (\ref{eq:support function - general}).\end{lemma}

\begin{proof}[Proof of Lemma \ref{lem:: Theta_o equivalent expression-2}]

Denote the set in the lemma as $\Theta_{I}^{\dagger}$. We aim to
show that $\Theta_{I}^{\dagger}\subseteq\Theta_{I}$ and $\Theta_{I}\subseteq\Theta_{I}^{\dagger}$.
Under Assumption \ref{assump::Full rank}, matrix $\mathbb{E}\left[m_{a}\left(Y_{0},X\right)\right]^{-1}$
exists. For any $t\in\mathbb{S}^{k}$, let $q=\mathbb{E}\left[m_{a}\left(Y_{0},X\right)\right]^{\top}t$.
Then for any $t\in\mathbb{S}^{k}$, we have 
\begin{align*}
\int\mathcal{KT}_{t^{\top}m_{b}}\left(\mu_{1\mid x},\mu_{0\mid x};x\right)d\mu_{X} & =\int\mathcal{KT}_{t^{\top}\mathbb{E}\left[m_{a}\left(Y_{0},X\right)\right]\mathbb{E}\left[m_{a}\left(Y_{0},X\right)\right]^{-1}m_{b}}\left(\mu_{1\mid x},\mu_{0\mid x};x\right)d\mu_{X}\\
 & =\int\mathcal{KT}_{q^{\top}\mathbb{E}\left[m_{a}\left(Y_{0},X\right)\right]^{-1}m_{b}}\left(\mu_{1\mid x},\mu_{0\mid x};x\right)d\mu_{X}.
\end{align*}
In consequence, for any $t\in\mathbb{S}^{k}$, if $\theta\in\Theta$
satisfies the inequality constraint 
\[
t^{\top}\mathbb{E}\left[m_{a}\left(Y_{0},X\right)\right]\theta\leq-\int\mathcal{KT}_{t^{\top}m_{b}}\left(\mu_{1\mid x},\mu_{0\mid x};x\right)d\mu_{X},
\]
then it would also satisfy 
$
q^{\top}\theta\leq-\int\mathcal{KT}_{q^{\top}\mathbb{E}\left[m_{a}\left(Y_{0},X\right)\right]^{-1}m_{b}}\left(\mu_{1\mid x},\mu_{0\mid x};x\right)d\mu_{X},
$
where $q$ does not necessarily belong to $\mathbb{S}^{d_{\theta}}$.
On the other hand, because the constraint is positive homogeneous
in $q$, it is equivalent to letting $\left\Vert q\right\Vert =1$.
Thus, we have shown that for any constraint in $\Theta_{I}$, there
is a corresponding constraint in $\Theta_{I}^{\dagger}$. It holds
that $\Theta_{I}^{\dagger}\subseteq\Theta_{I}$. The other direction
$\Theta_{I}\subseteq\Theta_{I}^{\dagger}$ follows from a similar
argument by letting $t=\left[\mathbb{E}\left[m_{a}\left(Y_{0},X\right)\right]^{-1}\right]^{\top}q$.
Hence, $\Theta_{I}=\Theta_{I}^{\dagger}$. 
\end{proof}

\begin{proof}[Proof of Proposition \ref{Prop:: Support function -  general form}]

Assumption \ref{assump:: Theta_o interior} and Lemma \ref{lem:: Theta_o equivalent expression-2}
imply that the identified set $\Theta_{I}$ in Theorem \ref{Thm:: Equivalence of Theta I and Theta o}
(i) can be written as $\left\{ \theta\in\mathbb{R}^{d_{\theta}}:q^{\top}\theta\leq s\left(q\right)\text{ for all }q\in\mathbb{S}^{d_{\theta}}\right\} $.
Note that the function $s(q)$ can be expressed as:
\[
\sup_{\mu\in\mathcal{M}\left(\mu_{1X},\mu_{0X}\right)} \mathbb{E}_{\mu} \left[-q^{\top} \mathbb{E}\left[m_{a}\left(Y_{0},X\right)\right]^{-1} m_{b}\left(Y_{1},Y_{0},X\right) \right].
\]
It can be seen that $s(q)$ is positively homogeneous. Assumptions
\ref{assump::Smoothness-conditioning} and \ref{assump::Full rank} imply that $\mathbb{E}_{\mu}\left[m_{b}\left(Y_{1},Y_{0},X\right)\right]$
is finite for every $\mu\in\mathcal{M}\left(\mu_{1X},\mu_{0X}\right)$.
Because $s(q)$ is the supremum of affine functions with $s\left(0\right)=0$,
it is a proper and closed convex function by Theorem 9.4 of \cite{rockafellar1997ConvexAnalysis}.
Corollary 13.2.1 of \cite{rockafellar1997ConvexAnalysis} implies
that $s\left(q\right)$ is the support function of $\Theta_{I}$.
This concludes the proof. 
\end{proof}

\begin{proof}[Proof of Proposition \ref{Prop:: Linear Projection Model2 identified set}]

Because both $Y_{1}$ and $Y_{0}$ have finite second moments, Assumption
\ref{assump::Smoothness-conditioning} holds and Theorem \ref{Thm:: Equivalence of Theta I and Theta o}
shows that $\Theta_{I}=\Theta_{o}$. Additionally, Assumption \ref{assmp:: m linear in theta}
is verified with $m_{a}\left(y_{0},x\right)$ being the identity matrix.
By Theorem \ref{assmp:: m linear in theta}, we obtain that the identified
set $\Theta_{I}$ for $\theta^{*}$ can be expressed as (\ref{eq: LP Theta I general}).
Equation (\ref{eq:LP Full OT}) can be obtained by writing down the
expression of the OT cost. The representation
of the OT map via the gradient of a convex function is ensured by the existence
of the conditional Monge mapping (\cite{carlier2016vector} and Proposition 3.8 in \cite{hosseini2025conditional}).
\end{proof}

\begin{proof}[Proof of Proposition \ref{Prop:: Linear Projection Model identified set}]

Assumptions \ref{assump::Smoothness-conditioning} (i) and \ref{assmp:: m linear in theta} hold by the nature of the moment function. Since $\left\Vert \ensuremath{y_{0}}y_{1}\right\Vert \leq\frac{1}{2}\left(\left|y_{1}\right|^{2}+\left\Vert y_{0}\right\Vert ^{2}\right)$,
Assumption \ref{assump::Smoothness-conditioning} (ii) holds by $\mathbb{E}\left[\left|Y_{1}\right|^{2}\right]=\int\int\left|y_{1}\right|^{2}d\mu_{1|x}\left(y_{1}\right)d\mu_{X}\left(x\right)$
and $\mathbb{E}\left[\left\Vert Y_{0}\right\Vert ^{2}\right]=\int\int\left\Vert y_{0}\right\Vert ^{2}d\mu_{0|x}\left(y_{0}\right)d\mu_{X}\left(x\right)$ both being finite. Thus, Theorem \ref{Thm:: Theta_o equivalent expression}
together with Proposition 2.17 in \cite{santambrogio2015optimal}
imply (\ref{eq: DHau Theta I}). Since we let $\Theta=\mathbb{R}^{d_{0}}$,
both Assumptions \ref{assump::Parameter space} and \ref{assump:: Theta_o interior}
are satisfied. We obtain the support function as (\ref{eq:DHau Theta I suppor}).
Because $\left(A_{00}^{\top},A_{p0}^{\top}\right)^{\top}\theta^{\ast}+\left(A_{0p}^{\top},A_{pp}^{\top}\right)^{\top}\mathbb{E}\left[X_{p}Y_{1}\right]$
is an affine map, for any $q\equiv\left(q_{0}^{\top},q_{X_{p}}^{\top}\right)^{\top}\in\mathbb{S}^{d_{0}+d_{x_{p}}}$
such that $q_{0}\in\mathbb{R}^{d_{0}}$ and $q_{X_{p}}\in\mathbb{R}^{d_{x_{p}}}$,
it holds that $h_{\Delta_{I}}\left(q\right)=h_{\Delta_{I}}\left(q_{0},q_{X_{p}}\right)=h_{\Theta_{I}}\left(A_{00}^{\top}q_{0}+A_{p0}^{\top}q_{X_{p}}\right)+\left(q_{0}^{\top}A_{0p}+q_{X_{p}}^{\top}A_{pp}\right)\mathbb{E}\left[Y_{1}X_{p}\right]$.
By plugging in the expression of $h_{\Theta_{I}}\left(\cdot\right)$,
we obtain the result in the proposition. 
\end{proof}

\begin{proof}[Proof of Proposition \ref{Prop:: DD identified set}]

Since $Y_{1}$ and $Y_{0}$ are both discrete, Assumption \ref{assump::Smoothness-conditioning}
is satisfied. Furthermore, Assumption \ref{assmp:: m linear in theta}
holds with $m_{a}\left(y_{0},x\right)$ and $m_{b}\left(y_{0},y_{1},x\right)$
discussed in Section \ref{sec: Identified set - DD}. Thus, Theorem
\ref{Thm:: Theta_o equivalent expression} (i) applies. We have
\[
\mathbb{E}\left[m_{a}\left(Y_{0},X\right)\right]=\mathrm{diag}\left(\Pr\left(Y_{0}=a_{1}\right),\ldots,\Pr\left(Y_{0}=a_{J}\right)\right).
\] For the first part of the proposition, it remains to compute $\mathcal{KT}_{t^{\top}m_{b}}\left(\mu_{1\mid x},\mu_{0\mid x};x\right)$. Plugging in the functional form of $m_{b}\left(y_{0},y_{1},x\right)$, we have the following. 
\begin{eqnarray*}
 \mathcal{KT}_{t^{\top}m_{b}}\left(\mu_{1\mid x},\mu_{0\mid x};x\right)
  = &\inf_{\mu_{10\mid x}\in\mathcal{M}\left(\mu_{1\mid x},\mu_{0\mid x}\right)}\int\int\sum_{j=1}^{J}-t_{j}\mathds{1}\left\{ y_{1}=1,y_{0}=a_{j}\right\} d\mu_{10\mid x}\\
  = &\inf_{\mu_{10\mid x}\in\mathcal{M}\left(\mu_{1\mid x},\mu_{0\mid x}\right)}\int\int-\mathds{1}\left\{ y_{1}=1\right\} \left[\sum_{j=1}^{J}t_{j}\mathds{1}\left\{ y_{0}=a_{j}\right\} \right]d\mu_{10\mid x}.
\end{eqnarray*}
Let $\mathtt{\mathtt{\boldsymbol{d}}}\left(y_{1}\right)=\mathds{1}\left\{ y_{1}=1\right\} $
and $\mathtt{\mathtt{\boldsymbol{d}}}_{t}\left(y_{0}\right)=\sum_{j=1}^{J}t_{j}\mathds{1}\left\{ y_{0}=a_{j}\right\} $.
Proposition 2.17 in \citet{santambrogio2015optimal} provides
\begin{align*}
 & \inf_{\mu_{10\mid x}\in\mathcal{M}\left(\mu_{1\mid x},\mu_{0\mid x}\right)}\int\int-\mathtt{\mathtt{\boldsymbol{d}}}\left(y_{1}\right)\times\mathtt{\mathtt{\boldsymbol{d}}}_{t}\left(y_{0}\right)d\mu_{10\mid x}\left(y_{1},y_{0}\right)\\
 & =-\int_{0}^{1}F_{D\mid x}^{-1}\left(u\right)F_{D_{t}\mid x}^{-1}\left(u\right)du=-\int_{\Pr\left(Y_{1}=0\mid x\right)}^{1}F_{D_{t}\mid x}^{-1}\left(u\right)du,
\end{align*}
where $D\equiv\mathds{1}\left\{ Y_{1}=1\right\} $. We have obtained
each term in Theorem \ref{Thm:: Theta_o equivalent expression} (i).
The first part of the proposition hence follows.

For the second part of the proposition, we apply Proposition \ref{Prop:: Support function -  general form}.
Because the elements of $\theta^{\ast}$ are probabilities, we have
$\Theta=\left[0,1\right]^{J}$. Both Assumptions \ref{assump::Parameter space}
and \ref{assump:: Theta_o interior} hold. Assumption \ref{assump::Full rank}
is satisfied by the assumption that $\Pr\left(Y_{0}=a_{j}\right)>0$
for $j=1,\ldots,J$. Proposition \ref{Prop:: Support function -  general form}
implies that for any $q\in\mathbb{S}^{J}$, the support function of
$\Theta_{I}$ can be expressed as 
$
h_{\Theta_{I}}\left(q\right)=\int\int_{\Pr\left(Y_{1}=0\mid x\right)}^{1}F_{D_{q}\mid x}^{-1}\left(u\right)dud\mu_{X},
$
where $D_{q}\equiv\sum_{j=1}^{J}q_{j}\mathds{1}\left\{ Y_{0}=a_{j}\right\} \Pr\left(Y_{0}=a_{j}\right)^{-1}$.
Since $\Theta_{I}$ is convex and $E$ is a linear map, $\Delta_{DD}$
is convex. We obtain the support function of $\Delta_{DD}$ as $h_{\Delta_{DD}}\left(p\right)=h_{\Theta_{I}}\left(E^{\top}p\right)=\int\int_{\Pr\left(Y_{1}=0\mid x\right)}^{1}F_{D_{E^{\top}p}\mid x}^{-1}\left(u\right)dud\mu_{X}$. This completes the second part of the proposition.
\end{proof}

\begin{proof}[Proof of Lemma \ref{lem:: Partial transport problem}]

Because $c\left(\cdot,\cdot\right)$ only takes finitely different values and all the constraints are equalities and weak inequalities, there exists a solution to (\ref{eq:Partial transport problem}). Let $\mu_{10\mid x}^{\bigtriangleup}\left(\cdot\right)$ be a solution. We
aim to show that there exists $\mu_{10\mid x}^{\ast}\left(\cdot\right)$
with monotone support such that 
\begin{equation}
\sum_{j=1}^{J}\sum_{i=0}^{1}c\left(i,j\right)\mu_{10\mid x}^{\ast}\left(i,1,j\right)=\sum_{j=1}^{J}\sum_{i=0}^{1}c\left(i,j\right)\mu_{10\mid x}^{\bigtriangleup}\left(i,1,j\right).\label{eq: mu ast equal to mu triangle}
\end{equation}
Set $\mu_{0|x}^{\bigtriangleup}\left(j\right)\equiv\sum_{i=0}^{1}\mu_{10\mid x}^{\bigtriangleup}\left(i,1,j\right)$.
Consider the following optimization problem: 
\begin{align}
 & \min_{\mu_{10\mid x}}\sum_{j=1}^{J}\sum_{i=0}^{1}c\left(i,j\right)\mu_{10\mid x}\left(i,1,j\right)\label{eq: full transport problem}\\
\textrm{s.t. (i) } & \mu_{10\mid x}\left(i,1,j\right)\geq0\textrm{ for }i=0,1\textrm{ and }j=1,\ldots,J\textrm{,}\nonumber \\
\textrm{(ii) } & \sum_{i=0}^{1}\mu_{10\mid x}\left(i,1,j\right)=\mu_{0|x}^{\bigtriangleup}\left(j\right)\textrm{ for }j=1,\ldots,J\textrm{, and}\nonumber \\
\textrm{(iii) } & \sum_{j=1}^{J}\mu_{10\mid x}\left(i,1,j\right)=\mu_{1\mid x}\left(i,1\right)\textrm{ for }i=0,1.\nonumber 
\end{align}
Because $\sum_{i=0}^{1}\mu_{1\mid x}\left(i,1\right)=\sum_{j=1}^{J}\mu_{0|x}^{\bigtriangleup}\left(j\right)$,
(\ref{eq: full transport problem}) is a full optimal transport problem
between $\mu_{0|x}^{\bigtriangleup}\left(\cdot\right)$ and $\mu_{1\mid x}\left(\cdot\right)$
by ignoring the normalizing constant $1/\sum_{i=0}^{1}\mu_{1\mid x}\left(i,1\right)$.
We claim that $\mu_{10\mid x}^{\bigtriangleup}\left(\cdot\right)$
then must solve (\ref{eq: full transport problem}). This holds because
if there is another $\mu_{10\mid x}^{\bigtriangledown}\left(\cdot\right)$
that $\sum_{j=1}^{J}\sum_{i=0}^{1}c\left(i,j\right)\mu_{10\mid x}^{\bigtriangledown}\left(i,1,j\right)<\sum_{j=1}^{J}\sum_{i=0}^{1}c\left(i,j\right)\mu_{10\mid x}^{\bigtriangleup}\left(i,1,j\right)$ and satisfies the above constraints, then it clearly satisfies the constraints in (\ref{eq:Partial transport problem}),
and therefore $\mu_{10\mid x}^{\bigtriangleup}\left(\cdot\right)$
cannot be optimal in (\ref{eq:Partial transport problem}). This establishes
the claim.

Second, it is well known that the full optimal transport problem (\ref{eq: full transport problem})
has a solution with monotone support. Therefore, if the support of
$\mu_{10\mid x}^{\bigtriangleup}\left(\cdot\right)$ is not monotone,
there exists another solution $\mu_{10\mid x}^{\ast}\left(\cdot\right)$
to (\ref{eq: full transport problem}) whose support is monotone.
Since $\mu_{10\mid x}^{\ast}\left(\cdot\right)$ satisfies the constraints
in (\ref{eq: full transport problem}), it must also satisfy the less
stringent constraints in (\ref{eq:Partial transport problem}). By
the optimality in (\ref{eq: full transport problem}), Equality (\ref{eq: mu ast equal to mu triangle})
holds. Hence, $\mu_{10\mid x}^{\ast}\left(\cdot\right)$ is optimal
in (\ref{eq:Partial transport problem}), which proves the desired
result. 
\end{proof}

\section{Additional Materials}\label{sec::additional}

\subsection{Additional Detail on Linear Projection Models }\label{SA::1}

\subsubsection{Alternative Formulation}\label{subsec:LP - Alternative-Formulation}

In this subsection, we provide an alternative formulation such that
the identified set $\Theta_{I}$ consists of a Wasserstein distance
between the conditional distribution of $\left(t_{1}^{\top}Y_{1},\cdots,t_{d_{0}}^{\top}Y_{1}\right)$
and that of $Y_{0}$ given $X=x$ with $t_{j}\in\mathbb{R}^{d_{1}}$
for $j=1,\ldots,d_{0}$. Specifically, we define $\theta^{*}$ as
\[
\theta^{\ast}\equiv\left(\begin{array}{c}
\theta_{1}^{\ast}\\
\vdots\\
\theta_{d_{0}}^{\ast}
\end{array}\right)\equiv\left(\begin{array}{c}
\mathbb{E}_{o}\left[Y_{1}Y_{0,1}\right]\\
\vdots\\
\mathbb{E}_{o}\left[Y_{1}Y_{0,d_{0}}\right]
\end{array}\right)\in\mathbb{R}^{d_{1}d_{0}},
\]
where $Y_{0,j}$ for $j=1,\ldots,d_{0}$ are the elements of $Y_{0}$.
It is easy to see that $\theta^{*}$ satisfies the moment condition:
$\mathbb{E}_{o}\left[m\left(Y_{1},Y_{0},X;\theta^{\ast}\right)\right]=\boldsymbol{0}$,
where 
\[
m\left(y_{1},y_{0},x;\theta\right)=\theta-\left(y_{1}^{\top}y_{0,1},y_{1}^{\top}y_{0,2},\ldots,y_{1}^{\top}y_{0,d_{0}}\right)^{\top}\text{.}
\]
Let $\theta_{i,j}^{\ast}$ denote the $j$-th element of $\theta_{i}^{\ast}$.
Define $\theta_{r,j}^{\ast}\equiv\left[\theta_{1,j+1}^{\ast},\ldots,\theta_{d_{0},j+1}^{\ast}\right]^{\top}$
for $j=1,\ldots,d_{1}-1$ and $\theta_{r}^{\ast}\equiv\left[\theta_{r,1}^{\ast},\ldots,\theta_{r,\left(d_{1}-1\right)}^{\ast}\right]\in\mathbb{R}^{\left(d_{1}-1\right)\times d_{0}}$.
Then, for $d_{1}>1$, we can express $\delta^{\ast}$ as 
\[
\delta^{\ast}=\left(\begin{array}{ccc}
\mathbb{E}\left[\ensuremath{Y_{0}}Y_{0}^{^{\top}}\right] & \mathbb{E}\left[\ensuremath{Y_{0}}X_{p}^{\top}\right] & \theta_{r}^{\ast}\\
\mathbb{E}\left[X_{p}Y_{0}^{^{\top}}\right] & \mathbb{E}\left[X_{p}X_{p}^{\top}\right] & \mathbb{E}\left[X_{p}Y_{1r}^{\top}\right]\\
\theta_{r}^{\ast\top} & \mathbb{E}\left[Y_{1r}X_{p}^{\top}\right] & \mathbb{E}\left[Y_{1r}Y_{1r}^{\top}\right]
\end{array}\right)^{-1}\left(\begin{array}{c}
\theta_{s}^{\ast}\\
\mathbb{E}\left[X_{p}Y_{1s}\right]\\
\mathbb{E}\left[Y_{1r}Y_{1s}\right]
\end{array}\right)\equiv G\left(\theta^{\ast}\right).
\]

Following the discussion in Section \ref{sec:Identified set: Example-2:-LP}, we have that
\[
m_{b}\left(y_{1},y_{0},x\right)\equiv-\left(y_{1}^{\top}y_{0,1},y_{1}^{\top}y_{0,2},\ldots,y_{1}^{\top}y_{0,d_{0}}\right)^{\top}.
\]
The identified set for $\theta^{\ast}$ is given by
\[
\Theta_{I}=\left\{ \theta\in\Theta:t^{\top}\theta\leq-\int\mathcal{KT}_{t^{\top}m_{b}}\left(\mu_{1\mid x},\mu_{0\mid x};x\right)d\mu_{X}\text{ for all }t\in\mathbb{S}^{d_{0}d_{1}}\right\} .
\]
Partition $t$ as $\left(t_{1}^{\top},\ldots,t_{d_{0}}^{\top}\right)^{\top}$
with $t_{j}\in\mathbb{R}^{d_{1}}$ for $j=1,\ldots,d_{0}$. We have
that 
\begin{align*}
\mathcal{KT}_{t^{\top}m_{b}}\left(\mu_{1\mid x},\mu_{0\mid x};x\right)= & \inf_{\mu_{10\mid x}\in\mathcal{M}\left(\mu_{1\mid x},\mu_{0\mid x}\right)}\int\int-\left(t_{1}^{\top}y_{1},\cdots,t_{d_{0}}^{\top}y_{1}\right)y_{0}d\mu_{10\mid x}\\
= & \frac{1}{2}\inf_{\mu_{10\mid x}\in\mathcal{M}\left(\mu_{1\mid x},\mu_{0\mid x}\right)}\int\int\left\Vert \left(t_{1}^{\top}y_{1},\cdots,t_{d_{0}}^{\top}y_{1}\right)-y_{0}\right\Vert ^{2}d\mu_{10\mid x}\\
 & -\frac{1}{2}\int\left\Vert \left(t_{1}^{\top}y_{1},\cdots,t_{d_{0}}^{\top}y_{1}\right)\right\Vert ^{2}d\mu_{0\mid x}-\frac{1}{2}\int\left\Vert y_{0}\right\Vert ^{2}d\mu_{1\mid x}.
\end{align*}
The first term on the right-hand side of the above equation is half
of the Wasserstein distance between the conditional distribution of
$\left(t_{1}^{\top}Y_{1},\cdots,t_{d_{0}}^{\top}Y_{1}\right)$ and
that of $Y_{0}$ given $X=x$. As a result, the `effective' dimension
of the marginal measures is equal to the dimension of $Y_{0}$ regardless
of the dimension of $Y_{1}$.

\subsubsection{Outer Set and Comparison with \citet{hwang2023bounding}}

\label{Appendix:: LP Hwang}

In this subsection, we construct a computationally efficient subset
of $\Theta_{I}$ based on our characterization in (\ref{eq: LP Theta I general}).
We show that our outer set is contained in the superset proposed in
\citet{hwang2023bounding}. Denote $a\equiv\left(a_{1},\ldots,a_{d_{1}}\right)^{\top}\in\mathbb{R}^{d_{1}}$
and define 
\[
\mathbb{V}^{d_{0}d_{1}}\equiv\left\{ \left(a_{1}t_{0}^{\top},a_{2}t_{0}^{\top},\ldots,a_{d_{1}}t_{0}^{\top}\right)^{\top}\in\mathbb{S}^{d_{0}d_{1}}:t_{0}\in\mathbb{S}^{d_{0}}\textrm{ and }a\in\mathbb{S}^{d_{1}}\right\} .
\]
Our outer set $\Theta_{I}^{F}$ is defined as 
\[
\Theta_{I}^{F}\equiv\left\{ \theta\in\Theta:t^{\top}\theta\leq-\int\mathcal{KT}_{t^{\top}m_{b}}\left(\mu_{1\mid x},\mu_{0\mid x};x\right)d\mu_{X}\text{ for all }t\in\mathbb{V}^{d_{0}d_{1}}\right\} .
\]
For any $t\in\mathbb{V}^{d_{0}d_{1}}$, the COT cost $\mathcal{KT}_{t^{\top}m_{b}}\left(\mu_{1\mid x},\mu_{0\mid x};x\right)$
has a closed-form expression: 
\begin{align*}
 & \inf_{\mu_{10\mid x}\in\mathcal{M}\left(\mu_{1\mid x},\mu_{0\mid x}\right)}\int\int-\left(a_{1}t_{0}^{\top}y_{0},a_{2}t_{0}^{\top}y_{0},\cdots,a_{d_{1}}t_{0}^{\top}y_{0}\right)y_{1}d\mu_{10\mid x}\\
= & \inf_{\mu_{10\mid x}\in\mathcal{M}\left(\mu_{1\mid x},\mu_{0\mid x}\right)}\int\int-\left(t_{0}^{\top}y_{0}\right)\left(a^{\top}y_{1}\right)d\mu_{10\mid x}=-\int_{0}^{1}F_{t_{0}^{\top}Y_{0}\mid x}^{-1}\left(u\right)F_{a^{\top}Y_{1}\mid x}^{-1}\left(u\right)du.
\end{align*}
In consequence, we can compute $\Theta_{I}^{F}$ via 
\[
\Theta_{I}^{F}=\left\{ \theta\in\Theta:t^{\top}\theta\leq\int\int_{0}^{1}F_{t_{0}^{\top}Y_{0}\mid x}^{-1}\left(u\right)F_{a^{\top}Y_{1}\mid x}^{-1}\left(u\right)dud\mu_{X}\text{ for all }t\in\mathbb{V}^{d_{0}d_{1}}\right\} ,
\]
where $t\equiv\left(a_{1}t_{0}^{\top},a_{2}t_{0}^{\top},\ldots,a_{d_{1}}t_{0}^{\top}\right)^{\top}$.
Define the following two sets: 
\begin{align*}
\mathbb{U}^{d} & \equiv\left\{ u\in\mathbb{S}^{d}:\textrm{only one element in }u\textrm{ is nonzero}\right\} \textrm{ and }\\
\mathbb{W}^{d_{0}d_{1}} & \equiv\left\{ \left(a_{1}t_{0}^{\top},a_{2}t_{0}^{\top},\ldots,a_{d_{1}}t_{0}^{\top}\right)^{\top}\in\mathbb{S}^{d_{0}d_{1}}:t_{0}\in\mathbb{U}^{d_{0}}\textrm{ and }a\in\mathbb{U}^{d_{1}}\right\} .
\end{align*}
The superset, denoted as $\Theta_{I}^{S}$, proposed
in \citet{hwang2023bounding} is equivalent to 
\[
\Theta_{I}^{S}=\left\{ \theta\in\Theta:t^{\top}\theta\leq-\int\mathcal{KT}_{t^{\top}m_{b}}\left(\mu_{1\mid x},\mu_{0\mid x};x\right)d\mu_{X}\text{ for all }t\in\mathbb{W}^{d_{0}d_{1}}\right\} .
\]
Because $\mathbb{W}^{d_{0}d_{1}}\subset\mathbb{V}^{d_{0}d_{1}}\subset\mathbb{S}^{d_{0}d_{1}}$,
it holds that $\Theta_{I}\subset\Theta_{I}^{F}\subset\Theta_{I}^{S}$.
While both $\Theta_{I}^{S}$ and $\Theta_{I}^{F}$ are computationally
less costly than the identified set $\Theta_{I}$, our outer set $\Theta_{I}^{F}$
is contained in $\Theta_{I}^{S}$.

\subsubsection{Comparison with \citet{pacini2019two} }

\label{appendix:: LP Pacini}

We perform two numerical exercises to compare our results with \citet{pacini2019two}.
The first one is to show the difference between our identified set
$\Delta_{I}$ and the set provided in \citet{pacini2019two} when
$d_{0}>1$. In the second exercise, we illustrate the effect of the
dependence between $Y_{0}$ and $X$ on the size of the identified
set.

\paragraph*{Simulation 1. }

Define $\overline{X}\equiv\left(1,X\right)$. The linear projection model is defined as 
\begin{equation}
Y_{1}=\left(Y_{0}^{^{\top}},\overline{X}^{\top}\right)\delta^{\ast}+\epsilon\text{ and }\mathbb{E}\left[\epsilon\left(Y_{0}^{^{\top}},\overline{X}^{\top}\right)\right]=0.\label{eq:Linear projection simulation}
\end{equation}
Let $\left(Y_{0}^{^{\top}},X\right)\equiv\left(Y_{0a},Y_{0b},X\right)\sim\mathcal{N}\left(\mathbf{0},\Omega\right)$,
where $\Omega=\left(\begin{array}{ccc}
1 & \rho & 0\\
\rho & 1 & 0\\
0 & 0 & 1
\end{array}\right)$ is the covariance matrix. For $\delta^{\ast}\equiv\left(\alpha_{a}^{\ast},\alpha_{b}^{\ast},\beta_{1}^{\ast},\beta_{2}^{\ast}\right)=\left(1,1,0,1\right)$,
we let 
\begin{align*}
Y_{1} & =\alpha_{a}^{\ast}Y_{0a}+\alpha_{b}^{\ast}Y_{0b}+\beta_{1}^{\ast}+\beta_{2}^{\ast}X+\epsilon=Y_{0a}+Y_{0b}+X+\epsilon,
\end{align*}
where $\epsilon$ follows a standard normal distribution and is independent
of $\left(Y_{0a},Y_{0b},X\right)$.

\begin{figure}[h]
\includegraphics[width=5cm,height=5cm]{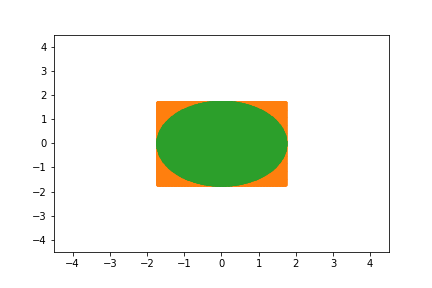}\includegraphics[width=5cm,height=5cm]{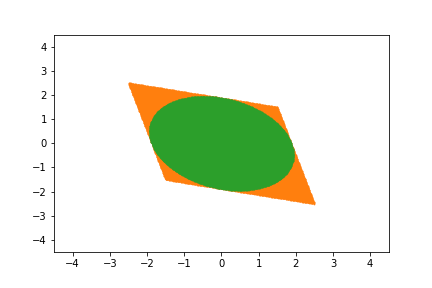}\includegraphics[width=5cm,height=5cm]{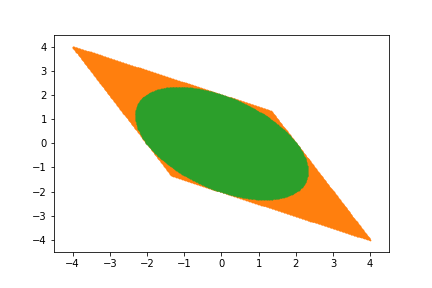}

\includegraphics[width=5cm,height=5cm]{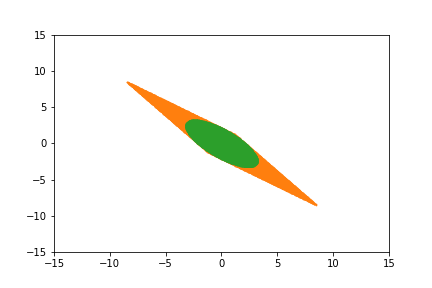}\includegraphics[width=5cm,height=5cm]{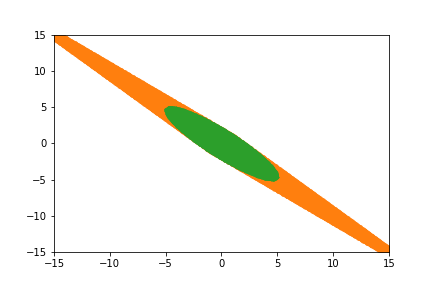}\includegraphics[width=5cm,height=5cm]{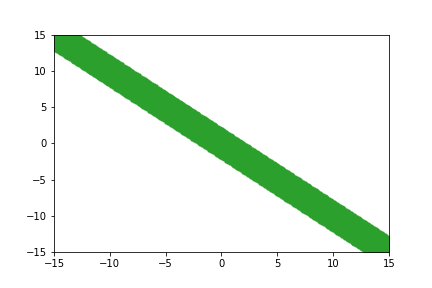}\caption{Top-left: $\rho=0$; top-middle: $\rho=0.25$; top-right: $\rho=0.5$;
bottom-left: $\rho=0.75$; bottom-middle: $\rho=0.9$; bottom-right:
$\rho=1$}
\label{Figure: Simulation - regression - comparison} 
\end{figure}

Figure \ref{Figure: Simulation - regression - comparison} compares
the identified sets for $\left(\alpha_{a}^{\ast},\alpha_{b}^{\ast}\right)$
obtained from our method (green area) with the one from \citet{pacini2019two}
(orange area) for different values of $\rho$. When $\rho=0$, random
variables $Y_{0a}$ and $Y_{0b}$ are independent. However, even in
this case, the approach in \citet{pacini2019two} provides a larger
identified set. This is because the method in \citet{pacini2019two}
ignores the possible connection between the dependence within $\left(Y_{0a},Y_{1}\right)$
and the dependence within $\left(Y_{0b},Y_{1}\right)$. For example,
if the dependence within $\left(Y_{0a},Y_{1}\right)$ reaches the
Fr\'{e}chet--Hoeffding bound, then dependence within $\left(Y_{0b},Y_{1}\right)$
cannot reach the bound anymore, because that would imply a perfect
correlation between $Y_{0a}$ and $Y_{0b}$.

The bottom-right of Figure \ref{Figure: Simulation - regression - comparison}
illustrates the identified set when $\rho=1$. When $\left|\rho\right|=1$,
the model is not identified even if we have the joint distribution
of $\left(Y_{1},Y_{0},X\right)$. In the case where $\rho=1$, we
can only identify $\alpha_{a}^{\ast}+\alpha_{b}^{\ast}$ but not individually.
The approach in \citet{pacini2019two} is not applicable, because
it requires $\mathbb{E}\left[\ensuremath{Y_{0}}Y_{0}^{^{\top}}\right]$
be invertible, which no longer holds when $\rho=1$. On the other hand,
our construction of $\Theta_I$ in (\ref{eq: DHau Theta I})
still applies and provides the upper and lower bounds for $\alpha_{a}^{\ast}+\alpha_{b}^{\ast}$.
Because we do not observe the distribution of $\left(Y_{1},Y_{0}\right)=\left(Y_{1},Y_{0a},Y_{0b}\right)$,
we do not point identify $\alpha_{a}^{\ast}+\alpha_{b}^{\ast}$.

\paragraph*{Simulation 2.}

Consider the same linear projection model as defined in (\ref{eq:Linear projection simulation}).
We now let $X\sim\mathcal{N}\left(0,4\right)$ and $Y_{0}\equiv\left(Y_{0a},Y_{0b}\right)$,
where $Y_{0a}=X^{2}+\eta_{a}$ and $Y_{0b}=Y_{0a}^{2}+\eta_{b}$ 
with $\eta_{a}\sim\mathcal{N}\left(0,\sigma_{a}^{2}\right)$ and $\eta_{b}\sim\mathcal{N}\left(0,\sigma_{b}^{2}\right)$.
Random variables $X$, $\eta_{a}$, and $\eta_{b}$ are mutually independent.
The value of $\sigma_{a}$ controls the dependence between $Y_{0a}$
and $X$; and the value $\sigma_{b}$ controls the dependence between
$Y_{0a}$ and $Y_{0b}$. We let $Y_{1}=\alpha_{a}^{\ast}Y_{0a}+\alpha_{b}^{\ast}Y_{0b}+\beta_{1}^{\ast}+\beta_{2}^{\ast}X+\epsilon$,
where $\left(\alpha_{a}^{\ast},\alpha_{b}^{\ast},\beta_{1}^{\ast},\beta_{2}^{\ast}\right)=\left(1,0.2,1,1\right)$
and $\epsilon$ follows a standard normal distribution and is independent
of $\left(Y_{0a},Y_{0b},X\right)$.

\begin{figure}[!ht]
\includegraphics[width=5cm,height=5cm]{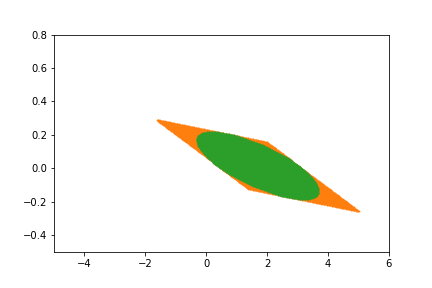}\includegraphics[width=5cm,height=5cm]{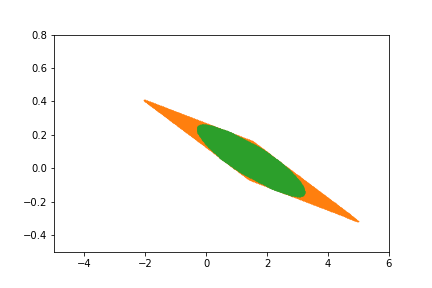}\includegraphics[width=5cm,height=5cm]{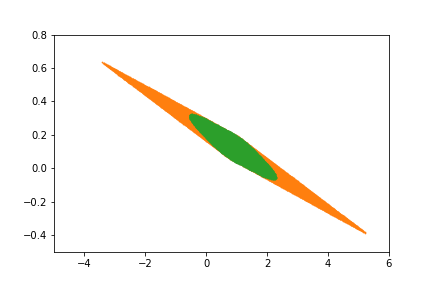}

\includegraphics[width=5cm,height=5cm]{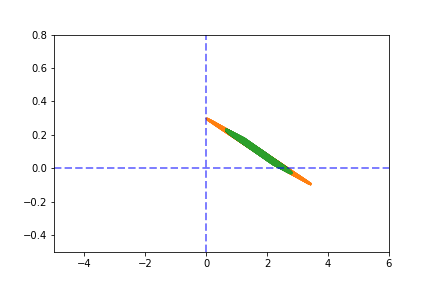}\includegraphics[width=5cm,height=5cm]{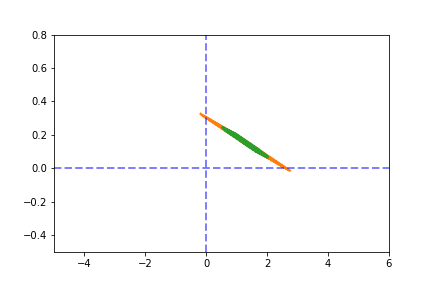}\includegraphics[width=5cm,height=5cm]{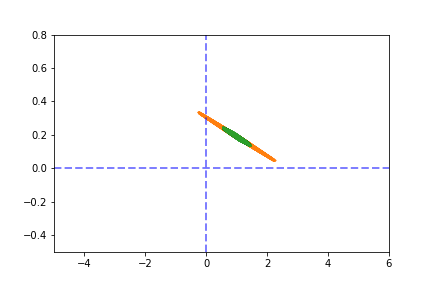}\caption{Top-left: $\sigma_{a}=2,\sigma_{b}=40$; top-middle: $\sigma_{a}=2,\sigma_{b}=20$
top-right: $\sigma_{a}=2,\sigma_{b}=2$; bottom-left: $\sigma_{a}=0.5,\sigma_{b}=20$;
bottom-middle: $\sigma_{a}=0.5,\sigma_{b}=4$ bottom-right: $\sigma_{a}=0.5,\sigma_{b}=0.1$}
\label{Figure: Set vs dependence} 
\end{figure}

Figure \ref{Figure: Set vs dependence} shows the identified sets
for $\left(\alpha_{a}^{\ast},\alpha_{b}^{\ast}\right)$ based on our
method (green area) and the ones based on \citet{pacini2019two} (orange
area) for different values of $\sigma_{a}$ and $\sigma_{b}$. First,
we see that as $\sigma_{a}$ and $\sigma_{b}$ decrease, the green
identified sets become smaller. This agrees with our discussion in
Section \ref{subsec: Identification - general} on the role of $X$
in reducing the size of the identified set because the dependence
between $X$ and $Y_{0a}$ and the dependence between $X$ and $Y_{0b}$
become stronger when $\sigma_{a}$ and $\sigma_{b}$ decrease. Second,
the identified sets obtained from our method are always smaller than
the ones from \citet{pacini2019two}. Such a difference can be crucial
in cases such as those described in the last two graphs. When $\sigma_{a}=0.5$
and $\sigma_{b}=4$, our identified set becomes small enough to lie
in the first quadrant. On the other hand, the identified set from
\citet{pacini2019two} includes both $\alpha_{a}^{\ast}=0$ and $\alpha_{b}^{\ast}=0$.
When $\sigma_{a}=0.5$ and $\sigma_{b}=0.1$, both $Y_{0a}$ and $Y_{0b}$
are strongly dependent on $X$. However, because the method in \citet{pacini2019two}
ignores the dependence between $Y_{0a}$ and $Y_{0b}$, the orange
identified set still cannot exclude $\alpha_{a}^{\ast}=0$ even in
this case.

\subsection{Additional Detail on Demographic Disparity Measures}\label{SA::2}

\subsubsection{Single DD Measure}

\label{appendix:: Single DD}

Our closed-form expression of the support function of $\Delta_{DD}$
in (\ref{eq:DD support function}) provides an easy way to obtain
the identified set for any single DD measure $\delta_{DD}\left(j,j^{\dagger}\right)$
for $J\geq2$. For example, if we are interested in $\delta_{DD}\left(1,2\right)$,
then we can let $K=1$ and $E=\left(1,-1,0,\ldots,0\right)$. Then
$p$ can be $1$ or $-1$. We first let $p=1$. Plugging $p$ in the
support function $h_{\Delta_{DD}}\left(\cdot\right)$, we obtain the
following equalities: 
\[
h_{\Delta_{DD}}\left(1\right)=h_{\Theta_{I}}\left(E^{\top}\right)=\sup_{\theta\in\Theta_{I}}E^{\top}\theta=\sup_{\theta\in\Theta_{I}}\left(\theta_{1}-\theta_{2}\right)=\sup_{\theta\in\Theta_{I}}\delta_{DD}\left(1,2\right).
\]
As a result, it suffices to compute $h_{\Theta_{I}}\left(E^{\top}\right)$
to obtain the tight upper bound of $\delta_{DD}\left(1,2\right)$.
The tight lower bound can be obtained similarly with $p=-1$. Since any one-dimensional convex set is either an interval or a degenerate interval, i.e., a point, the tight upper and lower bounds characterize
the identified set for $\delta_{DD}\left(1,2\right)$. The following
corollary provides the expression for the identified set for $\delta_{DD}\left(j,j^{\dagger}\right)$
for $J\geq2$.

\begin{corollary} \label{Cor:: DD single DD} For any $J\geq2$,
the identified set for any single DD measure: $\delta_{DD}\left(j,j^{\dagger}\right)$
is given by $\left[\delta_{DD}^{L}\left(j,j^{\dagger}\right),\delta_{DD}^{U}\left(j,j^{\dagger}\right)\right]$,
where 
\begin{align*}
\delta_{DD}^{U}\left(j,j^{\dagger}\right)\equiv & \frac{\mathbb{E}\left[\min\left\{ P_{A}\left(X\right),P_{B}\left(X\right)\right\} \right]}{\Pr\left(Y_{0}=a_{j}\right)}-\frac{\mathbb{E}\left[\max\left\{ P_{A}\left(X\right)+P_{C}\left(X\right)-1,0\right\} \right]}{\Pr\left(Y_{0}=a_{j^{\dagger}}\right)}\\
\delta_{DD}^{L}\left(j,j^{\dagger}\right)\equiv & \frac{\mathbb{E}\left[\max\left\{ P_{A}\left(X\right)+P_{B}\left(X\right)-1,0\right\} \right]}{\Pr\left(Y_{0}=a_{j}\right)}-\frac{\mathbb{E}\left[\min\left\{ P_{A}\left(X\right),P_{C}\left(X\right)\right\} \right]}{\Pr\left(Y_{0}=a_{j^{\dagger}}\right)},
\end{align*}
in which $P_{A}\left(x\right)\equiv\Pr\left(Y_{1}=1\mid X=x\right)$,
$P_{B}\left(x\right)\equiv\Pr\left(Y_{0}=a_{j}\mid X=x\right)$, and
$P_{C}\left(x\right)\equiv\Pr\left(Y_{0}=a_{j^{\dagger}}\mid X=x\right)$.
\end{corollary}

Because $Y_{1}$ and $Y_{0}$ are both one-dimensional, the Fr\'{e}chet-Hoeffding
inequalities provide the identified set for $\theta_{j}^{\ast}$ as
a closed interval $\left[\theta_{j}^{L},\theta_{j}^{U}\right]$, where
\begin{align*}
\theta_{j}^{L} & =\frac{\mathbb{E}\left[\max\left\{ \Pr\left(Y_{1}=1\mid X\right)+\Pr\left(Y_{0}=a_{j}\mid X\right)-1,0\right\} \right]}{\Pr\left(Y_{0}=a_{j}\right)}\textrm{ and }\\
\theta_{j}^{U} & =\frac{\mathbb{E}\left[\min\left\{ \Pr\left(Y_{1}=1\mid X\right),\Pr\left(Y_{0}=a_{j}\mid X\right)\right\} \right]}{\Pr\left(Y_{0}=a_{j}\right)}.
\end{align*}
Since $\delta_{DD}\left(j,j^{\dagger}\right)=\theta_{j}^{\ast}-\theta_{j^{\dagger}}^{\ast}$,
we have $\theta_{j}^{L}-\theta_{j^{\dagger}}^{U}\leq\delta_{DD}\left(j,j^{\dagger}\right)\leq\theta_{j}^{U}-\theta_{j^{\dagger}}^{L}$.
When $J=2$, KMZ prove that there is a joint probability on $\left(Y_{1},Y_{0},X\right)$
that allows $\theta_{j}^{\ast}$ to achieve its lower (upper) bound
and $\theta_{j^{\dagger}}^{\ast}$ to achieve its upper (lower) bound
simultaneously. Consequently, interval $\left[\theta_{j}^{L}-\theta_{j^{\dagger}}^{U},\theta_{j}^{U}-\theta_{j^{\dagger}}^{L}\right]$
is in fact the identified set for $\delta_{DD}\left(j,j^{\dagger}\right)$
when $J=2$. Corollary \ref{Cor:: DD single DD} shows that the interval
$\left[\theta_{j}^{L}-\theta_{j^{\dagger}}^{U},\theta_{j}^{U}-\theta_{j^{\dagger}}^{L}\right]$
is the identified set for $\delta_{DD}\left(j,j^{\dagger}\right)$
even for $J>2$.

\begin{proof}[Proof of Corollary \ref{Cor:: DD single DD}]
Following the discussion in Section \ref{sec: Identified set - DD},
we know that the identified set for $\delta_{DD}\left(j,j^{\dagger}\right)=\theta_{j}-\theta_{j^{\dagger}}$
is an interval or a degenerated interval. We aim to find the two endpoints
of the interval. Let $E=e_{+}\left(j\right)+e_{-}\left(j^{\dagger}\right)$.

We first derive the upper bound. Let $p=1$. The discussion in Section
\ref{appendix:: Single DD} shows that $h_{\Delta_{DD}}\left(1\right)=\sup_{\theta\in\Theta_{I}}\delta_{DD}\left(j,j^{\dagger}\right)$.
Thus, it suffices to compute 
\begin{align*}
 & h_{\Theta_{I}}\left(E^{\top}\right)=\int\int_{\Pr\left(Y_{1}=0\mid x\right)}^{1}F_{D_{E^{\top}p}\mid x}^{-1}\left(u\right)dud\mu_{X}\textrm{, where }\\
 & D_{E^{\top}}=\mathds{1}\left\{ Y_{0}=a_{j}\right\} \Pr\left(Y_{0}=a_{j}\right)^{-1}-\mathds{1}\left\{ Y_{0}=a_{j^{\dagger}}\right\} \Pr\left(Y_{0}=a_{j^{\dagger}}\right)^{-1}.
\end{align*}

Define $\delta_{DD}^{U}\left(j,j^{\dagger}\mid x\right)\equiv\int_{\Pr\left(Y_{1}=0\mid x\right)}^{1}F_{D_{E^{\top}p}\mid x}^{-1}\left(u\right)du$.
We now compute $\delta_{DD}^{U}\left(j,j^{\dagger}\mid x\right)$
for each $x$. To simply the notation, we use $\Pr\left(\cdot\mid x\right)$
to denote $\Pr\left(\cdot\mid X=x\right)$. We also let $p_{j}\equiv\Pr\left(Y_{0}=a_{j}\right)^{-1}$
and $p_{j^{\dagger}}\equiv\Pr\left(Y_{0}=a_{j^{\dagger}}\right)^{-1}$.
A simple calculation would show that 
\begin{align*}
F_{D_{E^{\top}}\mid x}\left(d\right) & =\begin{cases}
0 & d<-p_{j^{\dagger}}\\
\Pr\left(Y_{0}=a_{j^{\dagger}}\mid x\right) & -p_{j^{\dagger}}\leq d<0\\
\Pr\left(Y_{0}\neq a_{j}\mid x\right) & 0\leq d<p_{j}\\
1 & p_{j}\leq d
\end{cases}.
\end{align*}
This implies that 
\begin{align*}
F_{D_{E^{\top}}\mid x}^{-1}\left(u\right) & =\begin{cases}
-p_{j^{\dagger}} & 0 < u\leq\Pr\left(Y_{0}=a_{j^{\dagger}}\mid x\right)\\
0 & \Pr\left(Y_{0}=a_{j^{\dagger}}\mid x\right)<u\leq\Pr\left(Y_{0}\neq a_{j}\mid x\right)\\
p_{j} & \Pr\left(Y_{0}\neq a_{j}\mid x\right)<u\leq1
\end{cases}.
\end{align*}
Therefore, for any given $x$, it holds that $\delta_{DD}^{U}\left(j,j^{\dagger}\mid x\right)$
\begin{align*}
= & \begin{cases}
-p_{j^{\dagger}}\left[P_{A}\left(x\right)+P_{C}\left(x\right)-1\right]+p_{j}P_{B}\left(x\right) & \textrm{if }1-P_{A}\left(x\right)<P_{C}\left(x\right)\\
p_{j}P_{B}\left(x\right) & \textrm{if }P_{C}\left(x\right)\leq1-P_{A}\left(x\right)<1-P_{B}\left(x\right),\\
p_{j}P_{A}\left(x\right) & \textrm{if }1-P_{B}\left(x\right)\leq1-P_{A}\left(x\right)
\end{cases}
\end{align*}
where the conditional probabilities $P_{A}\left(x\right)$, $P_{B}\left(x\right)$,
and $P_{C}\left(x\right)$ are defined in the corollary. By plugging
the values of $p_{j}$ and $p_{j^{\dagger}}$ into the above expression
and combining terms, we obtain the expression of $\delta_{DD}^{U}\left(j,j^{\dagger}\mid x\right)$:
\[
\delta_{DD}^{U}\left(j,j^{\dagger}\mid x\right)=\frac{\min\left\{ P_{A}\left(x\right),P_{B}\left(x\right)\right\} }{\Pr\left(Y_{0}=a_{j}\right)}-\frac{\max\left\{ P_{A}\left(x\right)+P_{C}\left(x\right)-1,0\right\} }{\Pr\left(Y_{0}=a_{j^{\dagger}}\right)}.
\]
Therefore, we obtain that the upper bound for $\delta_{DD}\left(j,j^{\dagger}\right)$
as 
\begin{align*}
\int\delta_{DD}^{U}\left(j,j^{\dagger}\mid x\right)d\mu_{X}= & \frac{\mathbb{E}\left[\min\left\{ P_{A}\left(X\right),P_{B}\left(X\right)\right\} \right]}{\Pr\left(Y_{0}=a_{j}\right)} 
-\frac{\mathbb{E}\left[\max\left\{ P_{A}\left(X\right)+P_{C}\left(X\right)-1,0\right\} \right]}{\Pr\left(Y_{0}=a_{j^{\dagger}}\right)}.
\end{align*}

The lower bound can be obtained in the same way by letting $p=-1$.
\end{proof}

\subsubsection{Time Complexity of Constructing the Identified Set for DD Measures}

\label{subsec:Time-Complexity-of-DD}

As discussed in Section \ref{subsec:Identified-Set-Theta-DD}, the
identified set $\Delta_{DD}$ can be constructed via its support function
or vertex representation. For both methods, the fundamental step is
to compute $\int_{\Pr\left(Y_{1}=0\mid x\right)}^{1}F_{D_{E^{\top}p}\mid x}^{-1}\left(u\right)du$.
The following lemma provides the time complexity of computing such
an integration.

\begin{lemma} \label{lem:: time complexity - DD - FPS} The time
complexity for computing $\int_{\Pr\left(Y_{1}=0\mid x\right)}^{1}F_{D_{E^{\top}p}\mid x}^{-1}\left(u\right)du$
is $\frac{1}{2}J^{2}+\frac{1}{2}J+2K$ approximately.\footnote{We ignore the effect of the bit-length, which refers to the number
of bits required to represent the input data of the problem.} \end{lemma}

\begin{proof}[Proof of Lemma \ref{lem:: time complexity - DD - FPS}]
We count the number of elementary operations when computing $\int_{\Pr\left(Y_{1}=0\mid x\right)}^{1}F_{D_{E^{\top}p}\mid x}^{-1}\left(u\right)du$
for given $p$ and $x$. First, we compute $E^{\top}p$. Because $E^{\star}$
has only $2K$ non-zero elements and $p$ is one dimensional, it takes
$2K$ operations to compute. The discrete random variable $D_{E^{\top}p}$
has $J$ values, where each value requires $1$ multiplication to
compute. The function $F_{D_{E^{\top}p}\mid x}^{-1}\left(u\right)$
is a step function with $J$ jumps. Sorting its $J$ jumps needs $\frac{1}{2}J\left(J-1\right)$
number of operations. The integration in $\int_{\Pr\left(Y_{1}=0\mid x\right)}^{1}F_{D_{E^{\top}p}\mid x}^{-1}\left(u\right)du$
is essentially a summation of at most $J+1$ terms, where each term
requires $1$ multiplication. In total, we need $2K+J+\frac{1}{2}J\left(J-1\right)+J+1\simeq\frac{1}{2}\left(J^{2}+J\right)+2K$
operations. 
\end{proof}

To construct $\Delta_{DD}$ via its support function, we first sample
$N_{p}$ vectors $p_{1},\ldots,p_{N_{p}}$ uniformly from the $K$-dimensional
unit sphere and $N_{\delta}$ vectors $\delta_{1},\ldots,\delta_{N_{\delta}}$
uniformly from $\left[-1,1\right]^{K}$. Then, we construct the set:
\[
\widehat{\Delta}_{DD}\equiv\left\{ \delta\in\left\{ \delta_{1},\ldots,\delta_{N_{\delta}}\right\} :p^{\top}\delta\leq h_{\Delta_{DD}}\left(p\right)\textrm{ for all }p=p_{1},\ldots,p_{N_{p}}\right\} 
\]
as an approximate to $\Delta_{DD}$. Given any $p$, we compute $h_{\Delta_{DD}}\left(p\right)$
based on (\ref{eq:DD support function}). Suppose a grid-based method
is employed for the numerical integration with respect to measure
$\mu_{X}$. Let $N_{x}$ denote the number of grid points used in
the numerical integration. Then for any given $p$, computing $h_{\Delta_{DD}}\left(p\right)$
requires $\left(\frac{1}{2}J^{2}+\frac{1}{2}J+2K\right)N_{x}$ number
of operations. We obtain the overall time complexity as $\left(\frac{1}{2}J^{2}+\frac{1}{2}J+2K\right)N_{x}N_{p}+N_{\delta}N_{p}$.

Constructing $\Delta_{DD}$ based on its vertex representation: $\textrm{conv}\left\{ Ev_{1},\ldots,Ev_{\left|\mathfrak{S}_{M}\right|}\right\} $
requires an even fewer number of operations. Instead of sampling $N_{p}$
direction vectors in the support function approach, we only need to
compute the value of the support function $J!$ number of times. Therefore,
based on Lemma \ref{lem:: time complexity - DD - FPS}, the time complexity
of the vertex representation approach is $J!\left(\frac{1}{2}J^{2}+\frac{1}{2}J+2K\right)N_{x}$.
Because we usually sample much more direction vectors in $\mathbb{S}^{J}$
than $J!$. The vertex representation approach is much faster to compute
than the support function approach.

The time complexity of the method in KMZ varies depending on the algorithm
used to solve the linear programming in their Proposition 10. For
relatively small $J$ (e.g. $J<5$), the simplex method is more efficient
and requires roughly $\left(5J\right)^{2}=25J^{2}$ number of operations
(\citet{dantzig2016linear}); while for moderate $J$, interior-point
methods are often computationally cheaper and require about $33\left(5J\right)^{3}\approx4000J^{3}$
number of operations (\citet{karmarkar1984new,wright1997primal}).\footnote{The algorithms used in modern solvers, such as Mosek and \cite{gurobi}, share the ideas of the simplex method and interior-point methods. The information on the exact time complexity of these solvers is often not public.} When applied to $E^{\star}\theta^{\ast}$, Lemma \ref{lem:: time complexity - DD - FPS}
shows that our procedure computes $\varPhi_{K}\left(p,x\right)$ for
any given $p\in\mathbb{S}^{J-1}$ and $x\in\mathcal{X}$ with $\frac{1}{2}\left(J^{2}+5J\right)$
number of operations. As a result, even our support function approach
is advantageous compared to the fastest theoretical running time for
solving the linear programming in KMZ, which is roughly $\widetilde{O}\left(\left(2J\right)^{2.37}\right)$,
where the notation $\widetilde{O}$ hides polylogarithmic factors
(\citet{van2020deterministic,cohen2021solving,williams2024new}).

\subsection{Additional Detail on the True-Positive Rate Disparity Measures}\label{SA::3}

\subsubsection{Single TPRD Measure }

\label{Appendix::TPRD Single}

Using the simple expression for the support function (\ref{eq:TPRD support function})
of $\Theta_{I}$, we can derive the closed-form expressions for the
lower and upper endpoints of the interval, extending the result in
KMZ established for $J=2$.

\begin{corollary} \label{Cor:: TPRD single TPRD} For any $J\geq2$,
the identified set for any single TPRD measure: $\delta_{TPRD}\left(j,j^{\dagger}\right)$
is given by $\left[\delta_{TPRD}^{L}\left(j,j^{\dagger}\right),\delta_{TPRD}^{U}\left(j,j^{\dagger}\right)\right]$,
where 
\begin{align*}
\delta_{TPRD}^{U}\left(j,j^{\dagger}\right) & =\frac{\theta_{j}^{U}}{\theta_{j}^{U}+\theta_{J+j}^{L}}-\frac{\theta_{j^{\dagger}}^{L}}{\theta_{j^{\dagger}}^{L}+\theta_{J+j^{\dagger}}^{U}}\textrm{ and }\\
\delta_{TPRD}^{L}\left(j,j^{\dagger}\right) & =\frac{\theta_{j}^{L}}{\theta_{j}^{L}+\theta_{J+j}^{U}}-\frac{\theta_{j^{\dagger}}^{U}}{\theta_{j^{\dagger}}^{U}+\theta_{J+j^{\dagger}}^{L}},
\end{align*}
where the expressions for each term in $\delta_{TPRD}^{U}\left(j,j^{\dagger}\right)$
and $\delta_{TPRD}^{L}\left(j,j^{\dagger}\right)$ are provided in the proof of the corollary. \end{corollary}

For any $j\in\left\{ 1,\ldots,J\right\} $, $\theta_{j}^{U}$ and
$\theta_{j}^{L}$ are the Fr\'{e}chet-Hoeffding upper and lower bounds
of $\Pr\left(Y_{1}=(1,1),Y_{0}=a_{j}\right)$; and $\theta_{J+j}^{U}$
and $\theta_{J+j}^{L}$ are the Fr\'{e}chet-Hoeffding upper and lower
bounds of $\Pr\left(Y_{1}=(0,1),Y_{0}=a_{j}\right)$. The proof reduces
to computing $h_{\Theta_{I}}\left(q\right)$ for two carefully chosen
values of $q$, which essentially solves two optimal transport problems.
Although any TPRD measure is a non-linear function of $\theta^{\ast}$:
$\delta_{TPRD}\left(j,j^{\dagger}\right)=g_{j,j^{\dagger}}\left(\theta^{\ast}\right)$,
Corollary \ref{Cor:: TPRD single TPRD} shows that the identified
set for $\delta_{TPRD}\left(j,j^{\dagger}\right)$ is a closed interval
and provides the closed-form expression of the two end-points of the
interval.

\begin{proof}[Proof of Corollary \ref{Cor:: TPRD single TPRD}]
We first provide the definitions of each term in the corollary. For
any $j\in\left\{ 1,\ldots,J\right\} $, define 
\begin{align*}
\theta_{j}^{U} & \equiv\mathbb{E}\left[\min\left\{ \Pr\left(Y_{1s}=1,Y_{1r}=1\mid X\right),\Pr\left(Y_{0}=a_{j}\mid X\right)\right\} \right],\\
\theta_{j}^{L} & \equiv\mathbb{E}\left[\max\left\{ \Pr\left(Y_{1s}=1,Y_{1r}=1\mid X\right)+\Pr\left(Y_{0}=a_{j}\mid X\right)-1,0\right\} \right],\\
\theta_{J+j}^{U} & \equiv\mathbb{E}\left[\min\left\{ \Pr\left(Y_{1s}=0,Y_{1r}=1\mid X\right),\Pr\left(Y_{0}=a_{j}\mid X\right)\right\} \right]\textrm{, and }\\
\theta_{J+j}^{L} & \equiv\mathbb{E}\left[\max\left\{ \Pr\left(Y_{1s}=0,Y_{1r}=1\mid X\right)+\Pr\left(Y_{0}=a_{j}\mid X\right)-1,0\right\} \right].
\end{align*}
Without loss of generality, we prove the result for $\delta_{TPRD}\left(1,2\right)$
with any $J\geq2$. Since any one-dimensional connected set is an
interval, it suffices to derive the tight upper and lower bounds of
the identified set. We focus on the tight upper bound. The tight lower
bound can be obtained in the same way.

By definition, we have that $\delta_{TPRD}\left(1,2\right)\equiv\frac{\theta_{1}^{\ast}}{\theta_{1}^{\ast}+\theta_{J+1}^{\ast}}-\frac{\theta_{2}^{\ast}}{\theta_{2}^{\ast}+\theta_{J+2}^{\ast}}$.
It is easy to see that $\delta_{TPRD}\left(1,2\right)$ is an increasing
function of $\theta_{1}^{\ast}$ and $\theta_{J+2}^{\ast}$ and a
decreasing function of $\theta_{J+1}^{\ast}$ and $\theta_{2}^{\ast}$.
Let $\theta_{1}^{U}$, $\theta_{J+2}^{U}$, $\theta_{J+1}^{L}$, and
$\theta_{2}^{L}$ denote the tight upper bounds of $\theta_{1}^{\ast}$
and $\theta_{J+2}^{\ast}$ and tight lower bounds of $\theta_{J+1}^{\ast}$
and $\theta_{2}^{\ast}$, respectively. Let $\Theta_{I}^{\triangle}$
be the set obtained from projecting $\Theta_{I}$ onto its first,
second, $\left(J+1\right)$-th, and $\left(J+2\right)$-th elements
$\left(\theta_{1},\theta_{2},\theta_{J+1},\theta_{J+2}\right)$. In
the following, we first provide the expressions of $\theta_{1}^{U}$,
$\theta_{J+2}^{U}$, $\theta_{J+1}^{L}$, and $\theta_{2}^{L}$. Then
we show that the bounds can be achieved simultaneously: $\left(\theta_{1}^{U},\theta_{2}^{L},\theta_{J+1}^{L},\theta_{J+2}^{U}\right)\in\Theta_{I}^{\triangle}$.

By the support function expressed in (\ref{eq:TPRD support function}),
we can let $q=\left[1,0,\ldots,0\right]^{\top}$ to obtain the tight
upper bound of $\theta_{1}^{\ast}$. For any $\theta\in\Theta_{I}$,
its first element $\theta_{1}$ satisfies that 
\[
\int\left[\sup_{\mu_{10\mid x}\in\mathcal{M}\left(\mu_{1\mid x},\mu_{0\mid x}\right)}\int\int\mathds{1}\left\{ y_{1s}=1,y_{1r}=1,y_{0}=a_{1}\right\} d\mu_{10\mid x}\right]d\mu_{X}\geq\theta_{1}.
\]
In addition, the bound is tight by Lemma \ref{lem:: Tight bounds}.
We obtain that 
\begin{align*}
 & \int\left[\sup_{\mu_{10\mid x}\in\mathcal{M}\left(\mu_{1\mid x},\mu_{0\mid x}\right)}\int\int\mathds{1}\left\{ y_{1s}=1,y_{1r}=1,y_{0}=a_{1}\right\} d\mu_{10\mid x}\right]d\mu_{X}\\
= & \int\left[\sup_{\mu_{10\mid x}\in\mathcal{M}\left(\mu_{1\mid x},\mu_{0\mid x}\right)}\int\int\mathds{1}\left\{ y_{0}=a_{1}\right\} \mathds{1}\left\{ y_{1s}=1,y_{1r}=1\right\} d\mu_{10\mid x}\right]d\mu_{X}\\
= & \int\left[\int_{0}^{1}F_{D_{0}\mid x}^{-1}\left(u\right)F_{D_{1}\mid x}^{-1}\left(u\right)du\right]d\mu_{X},
\end{align*}
where $D_{0}=\mathds{1}\left\{ Y_{0}=a_{1}\right\} $ and $D_{1}=\mathds{1}\left\{ Y_{1s}=1,Y_{1r}=1\right\} $.
The last equality follows from the monotone rearrangement inequality.
To simply the notation, we use $\Pr\left(\cdot\mid x\right)$ to denote
$\Pr\left(\cdot\mid X=x\right)$. We have that 
\begin{align*}
F_{D_{0}\mid x}^{-1}\left(u\right) & =\begin{cases}
0 & 0 < u \le 1 - \Pr(Y_0 = a_1 | x) \\
0 & 1 - \Pr\left(Y_{0}=a_{1}\mid x\right)<u\leq1
\end{cases}\textrm{, and }\\
F_{D_{1}\mid x}^{-1}\left(u\right) & =\begin{cases}
0 & 0 < u \le 1 - \Pr\left(Y_{1s}=1,Y_{1r}=1\mid x\right)\\
1 & 1-\Pr\left(Y_{1s}=1,Y_{1r}=1\mid x\right)<u\leq1
\end{cases}.
\end{align*}
We thus obtain the tight upper bound of $\theta_{1}^{\ast}$ as 
\[
\theta_{1}^{U}=\mathbb{E}\left[\min\left\{ \Pr\left(Y_{1s}=1,Y_{1r}=1\mid X\right),\Pr\left(Y_{0}=a_{1}\mid X\right)\right\} \right].
\]
Applying the same technique, we can obtain that 
\begin{align*}
\theta_{J+1}^{L} & =\mathbb{E}\left[\max\left\{ \Pr\left(Y_{1s}=0,Y_{1r}=1\mid X\right)+\Pr\left(Y_{0}=a_{1}\mid X\right)-1,0\right\} \right],\\
\theta_{2}^{L} & =\mathbb{E}\left[\max\left\{ \Pr\left(Y_{1s}=1,Y_{1r}=1\mid X\right)+\Pr\left(Y_{0}=a_{2}\mid X\right)-1,0\right\} \right]\textrm{, and }\\
\theta_{J+2}^{U} & =\mathbb{E}\left[\min\left\{ \Pr\left(Y_{1s}=0,Y_{1r}=1\mid X\right),\Pr\left(Y_{0}=a_{2}\mid X\right)\right\} \right].
\end{align*}

Next, we show that $\left(\theta_{1}^{U},\theta_{2}^{L},\theta_{J+1}^{L},\theta_{J+2}^{U}\right)\in\Theta_{I}^{\triangle}$.
Let $q^{\dagger}=\left[1,-1,0, \dotsc, 0,-1,1,0,\ldots,0\right]^{\top}$ be such
that its first and $\left(J+2\right)$-th elements are $1$ and its
second and $\left(J+1\right)$-th elements are $-1$. We ignore the
normalizing factor that makes $q\in\mathbb{S}^{2J}$ hold to simplify
the derivation. Then, by the definition of support function, we have
that 
\[
h_{\Theta_{I}}\left(q^{\dagger}\right)\geq\theta_{1}-\theta_{2}-\theta_{J+1}+\theta_{J+2}.
\]
By Lemma \ref{lem:: Tight bounds}, the left-hand side of the inequality
is the tight upper bound of $\theta_{1}-\theta_{2}-\theta_{J+1}+\theta_{J+2}$.
Because $\theta_{1}-\theta_{2}-\theta_{J+1}+\theta_{J+2}$ is an increasing
function of $\theta_{1}$ and $\theta_{J+2}$ and a decreasing function
of $\theta_{2}$ and $\theta_{J+1}$, the following equality 
\[
h_{\Theta_{I}}\left(q^{\dagger}\right)=\theta_{1}^{U}-\theta_{2}^{L}-\theta_{J+1}^{L}+\theta_{J+2}^{U}
\]
holds if and only if $\left(\theta_{1}^{U},\theta_{2}^{L},\theta_{J+1}^{L},\theta_{J+2}^{U}\right)\in\Theta_{I}^{\triangle}$.
In the following, we solve the optimal transport problem on the left-hand side and verify that the equality holds.

Plugging in the expression of $m_{b}$, we have that 
\begin{align*}
h_{\Theta_{I}}\left(q^{\dagger}\right)= & \int\left[\sup_{\mu_{10\mid x}\in\mathcal{M}\left(\mu_{1\mid x},\mu_{0\mid x}\right)}\left[\int\int\mathds{1}\left\{ y_{1s}=1,y_{1r}=1,y_{0}=a_{1}\right\} d\mu_{10\mid x}\right.\right.\\
 & \qquad\qquad\qquad\qquad\quad-\int\int\mathds{1}\left\{ y_{1s}=0,y_{1r}=1,y_{0}=a_{1}\right\} d\mu_{10\mid x}\\
 & \qquad\qquad\qquad\qquad\quad-\int\int\mathds{1}\left\{ y_{1s}=1,y_{1r}=1,y_{0}=a_{2}\right\} d\mu_{10\mid x}\\
 & \qquad\qquad\qquad\qquad\quad\left.\left.+\int\int\mathds{1}\left\{ y_{1s}=0,y_{1r}=1,y_{0}=a_{2}\right\} d\mu_{10\mid x}\right]\right\} d\mu_{X}.
\end{align*}
Let $\mu_{10\mid x}^{\star}$ denotes the optimal coupling conditional
on $X=x$. We have that $h_{\Theta_{I}}\left(q^{\dagger}\right)$
equals to 
\begin{align}
 & \int\left\{ \left[\int\int\mathds{1}\left\{ y_{1s}=1,y_{1r}=1,y_{0}=a_{1}\right\} d\mu_{10\mid x}^{\star}-\int\int\mathds{1}\left\{ y_{1s}=0,y_{1r}=1,y_{0}=a_{1}\right\} d\mu_{10\mid x}^{\star}\right.\right.\nonumber \\
 & \left.\left.-\int\int\mathds{1}\left\{ y_{1s}=1,y_{1r}=1,y_{0}=a_{2}\right\} d\mu_{10\mid x}^{\star}+\int\int\mathds{1}\left\{ y_{1s}=0,y_{1r}=1,y_{0}=a_{2}\right\} d\mu_{10\mid x}^{\star}\right]\right\} d\mu_{X}.\label{eq: Proof of lemma TRPD}
\end{align}
Next, we solve for $\mu_{10\mid x}^{\star}$.

Note that the cost function assigns only the values of $\pm1$. We
must couple all of the mass of the points $\left(y_{1s},y_{1r}\right)\in\left\{ \left(1,1\right),\left(0,1\right)\right\} $
with the points $y_{0}\in\left\{ a_{1},a_{2}\right\} $. Note that the mass satisfies 
\begin{equation}
\mu_{1\mid x}\left(\left(1,1\right)\right)+\mu_{1\mid x}\left(\left(0,1\right)\right)\leq1=\mu_{0\mid x}\left(a_{1}\right)+\mu_{0\mid x}\left(a_{2}\right).\label{eqn: mass imbalance}
\end{equation}
The extra mass on the $\mu_{1\mid x}$ side (at points $\left(1,0\right)$
and $\left(0,0\right)$) is assigned with cost $0$, regardless of
where it couples to.

The cost of coupling $\left(1,1\right)$ with $a_{1}$ and $\left(0,1\right)$
with $a_{2}$ is 1, while the cost of $\left(1,1\right)$ with
$a_{2}$ and $\left(0,1\right)$ with $a_{1}$ is -1. It is therefore
optimal to assign as much mass to the former configurations as possible.
As a result, the optimizer $\mu_{10\mid x}^{\star}$ will satisfy
\begin{align*}
\mu_{10\mid x}^{\star}\big(\left(1,1\right),a_{1}\big) & =\min\left\{ \mu_{1\mid x}\left(\left(1,1\right)\right),\mu_{0\mid x}\left(a_{1}\right)\right\} \textrm{ and }\\
\mu_{10\mid x}^{\star}\big(\left(0,1\right),a_{2}\big) & =\min\left\{ \mu_{1\mid x}\left(\left(0,1\right)\right),\mu_{0\mid x}\left(a_{2}\right)\right\} .
\end{align*}

Now, if either $\mu_{1\mid x}\left(\left(1,1\right)\right)>\mu_{0\mid x}\left(a_{1}\right)$
or $\mu_{1\mid x}\left(\left(0,1\right)\right)>\mu_{0\mid x}\left(a_{2}\right)$
(note that by \eqref{eqn: mass imbalance}, at most one of these can
be true), then these configurations do not use up all of the mass from
$(1,1)$ and $(0,1)$ and we will have to couple either the remaining
mass from $(1,1)$ with $a_{2}$ or the remaining mass from $(0,1)$
with $a_{1}$, at cost -1. The optimal coupling therefore satisfies
\begin{align*}
\mu_{10\mid x}^{\star}\big(\left(1,1\right),a_{2}\big) & =\max\left\{ \mu_{1\mid x}\left(\left(1,1\right)\right)-\mu_{0\mid x}\left(a_{1}\right),0\right\} =\max\left\{ \mu_{1\mid x}\left(\left(1,1\right)\right)+\mu_{0\mid x}\left(a_{2}\right)-1,0\right\} \textrm{ and }\\
\mu_{10\mid x}^{\star}\big(\left(0,1\right),a_{1}\big) & =\max\left\{ \mu_{1\mid x}\left(\left(0,1\right)\right)-\mu_{0\mid x}\left(a_{2}\right),0\right\} =\max\left\{ \mu_{1\mid x}\left(\left(0,1\right)\right)+\mu_{0\mid x}\left(a_{1}\right)-1,0\right\} .
\end{align*}

Computing each term in (\ref{eq: Proof of lemma TRPD}) with the optimal
coupling $\mu_{10\mid x}^{\star}$, we have that the optimal cost
conditioning on $X=x$ is 
\[
\mu_{10\mid x}^{\star}\big(\left(1,1\right),a_{1}\big)+\mu_{10\mid x}^{\star}\big(\left(0,1\right),a_{2}\big)-\mu_{10\mid x}^{\star}\big(\left(1,1\right),a_{2}\big)-\mu_{10\mid x}^{\star}\big(\left(0,1\right),a_{1}\big).
\]
Integrating over $x$ and rearranging terms, we obtain that 
\[
h_{\Theta_{I}}\left(q^{\dagger}\right)=\theta_{1}^{U}-\theta_{2}^{L}-\theta_{J+1}^{L}+\theta_{J+2}^{U}.
\]
This completes the proof of the corollary. 
\end{proof}

\begin{lemma} \label{lem:: Tight bounds} Under Assumptions \ref{assump::Smoothness-conditioning}-\ref{assump::Full rank},
for any $q\in\mathbb{S}^{d_{\theta}}$ there exists some $\theta^{\dagger}\in\Theta_{I}$
such that $s\left(q\right) = \sup_{\theta \in \Theta_I} q^{\top} \theta = q^{\top} \theta^{\dagger}$.
\end{lemma}

\begin{proof}[Proof of Lemma \ref{lem:: Tight bounds}]

Proposition \ref{Prop:: Support function -  general form} implies that $s\left(q\right)$ is the support function of $\Theta_I$. Since $\Theta_I$ is closed and $\Theta$ is compact, $\Theta_I$ is also compact. Therefore, the exist $\theta^{\dagger} \in \Theta_I$ such that $s\left(q\right) = \sup_{\theta \in \Theta_I} q^{\top} \theta = q^{\top} \theta^{\dagger}$. The lemma follows. 
\end{proof}

\subsubsection{Dual Rank Equilibration AlgorithM (DREAM) }

\label{subsec:DREAM}

In this section, we describe each of the three steps in DREAM and
the equivalent optimization problem that it solves.

During Initialization, DREAM relabels $j$ so that $c\left(i,j\right)$
is submodular. Lemma \ref{lem:: Partial transport problem} shows that there is
a solution with monotone support. For any $jj\in\left\{ 1,...,J\right\} $,
after imposing the structure that $\mu_{10\mid x}\left(1,1,j\right)=0$
for all $j<jj$ and $\mu_{10\mid x}\left(0,1,j\right)=0$ for all
$j>jj$, the optimization problem (\ref{eq:Partial transport problem})
reduces to (\ref{eq:partial OT simplified}) below: 
\begin{align}
 & \min_{jj\in\left\{ 1,...,J\right\} }\min_{\mu_{10\mid x}}\left[\sum_{j=1}^{jj}c\left(0,j\right)\mu_{10\mid x}\left(0,1,j\right)+\sum_{j=jj}^{J}c\left(1,j\right)\mu_{10\mid x}\left(1,1,j\right)\right]\label{eq:partial OT simplified}\\
\textrm{s.t. } & \textrm{\textrm{(i) }}0\leq\mu_{10\mid x}\left(0,1,j\right)\leq\mu_{0\mid x}\left(j\right)\textrm{ for }j=1,\ldots,jj\textrm{;}\nonumber \\
 & \textrm{(ii) }\sum_{j=1}^{jj}\mu_{10\mid x}\left(0,1,j\right)=\mu_{1\mid x}\left(0,1\right)\textrm{;}\nonumber \\
 & \textrm{(iii) }\mu_{10\mid x}\left(0,1,jj\right)+\mu_{10\mid x}\left(1,1,jj\right)\leq\mu_{0\mid x}\left(jj\right)\textrm{;}\nonumber \\
 & \textrm{(iv) }0\leq\mu_{10\mid x}\left(1,1,j\right)\leq\mu_{0\mid x}\left(j\right)\textrm{ for }j=jj,\ldots,J\textrm{;}\nonumber \\
 & \textrm{(v) }\sum_{j=jj}^{J}\mu_{10\mid x}\left(1,1,j\right)=\mu_{1\mid x}\left(1,1\right).\nonumber 
\end{align}

Steps 1-3 of DREAM solve the inner and outer minimization problems
in (\ref{eq:partial OT simplified}) sequentially. In Step 1, we narrow
down the range of $jj$ in the outer minimization problem using the
constraints only. Combining Constraints (i) and (ii), we have that
$\sum_{j=1}^{jj}\mu_{0\mid x}\left(j\right)\geq\mu_{1\mid x}\left(0,1\right)$.
Because the left-hand-side of the inequality is an increasing function
of $jj$, we obtain the lower bound of $jj$, denoted as $JL$, by
starting with $JL=1$ and gradually increasing $JL$ until inequality
$\sum_{j=1}^{JL}\mu_{0\mid x}\left(j\right)\geq\mu_{1\mid x}\left(0,1\right)$
holds. Similarly, Constraints (iv) and (v) imply that $\sum_{j=jj}^{J}\mu_{0\mid x}\left(j\right)\geq\mu_{1\mid x}\left(1,1\right)$.
We obtain the upper bound $JU$ of $jj$ by starting with $JU=J$
and gradually decreasing $JU$ until $\sum_{j=JU}^{J}\mu_{0\mid x}\left(j\right)\geq\mu_{1\mid x}\left(1,1\right)$
holds. This is achieved in Step 1 of DREAM.

Given $jj$, directly solving the inner minimization problem in (\ref{eq:partial OT simplified}) can still be complicated. Instead, DREAM solves it via three minimizations,
all of which share the same structure and can be solved by the same
procedure. We name the procedure Linear Programming Solver (LPS).
Given any vector $cost$, non-negative vector $ineq$, and non-negative
scalar $eq$, $\textrm{LPS}\left(cost,ineq,eq\right)$ solves a generic
minimization problem of the following structure: 
\begin{align*}
 & \min_{m}\sum_{w=1}^{W}cost\left(w\right)m\left(w\right)\\
\textrm{s.t. } & \textrm{\textrm{(i) }}0\leq m\left(w\right)\leq ineq\left(w\right)\textrm{ for }w=1,\ldots,W\textrm{;}\\
 & \textrm{\textrm{(ii) }}\sum_{w=1}^{W}m\left(w\right)=eq.
\end{align*}
More specifically, LPS ranks $cost\left(w\right)$ from the smallest
to the largest and assigns mass to $m\left(w\right)$ following the
rank. During each assignment, LPS allocate $ineq\left(w\right)$ to
$m\left(w\right)$ until the total mass $eq$ is exhausted. The detailed
steps of LPS is provided in Algorithm \ref{LSP}.

Based on LPS, we solve the inner minimization problem in (\ref{eq:partial OT simplified})
during Steps 2 and 3. We first ignore Constraint (iii) in (\ref{eq:partial OT simplified})
because it involves both $\mu_{10\mid x}\left(0,1,j\right)$ and $\mu_{10\mid x}\left(1,1,j\right)$
for the same $j$. Without it, the inner minimization problem in (\ref{eq:partial OT simplified})
becomes 
\begin{align}
 & \min_{\mu_{10\mid x}}\left[\sum_{j=1}^{jj}c\left(0,j\right)\mu_{10\mid x}\left(0,1,j\right)+\sum_{j=jj}^{J}c\left(1,j\right)\mu_{10\mid x}\left(1,1,j\right)\right]\label{eq:partial OT simplified step 2}\\
\textrm{s.t. } & \textrm{(i), (ii), (iv), and (v) in (\ref{eq:partial OT simplified}) hold. }\nonumber 
\end{align}
Since the objective function and the constraints in (\ref{eq:partial OT simplified step 2})
can be separated in two parts, (\ref{eq:partial OT simplified step 2})
is equivalent to the following two minimization problems: 
\begin{align*}
\textbf{Problem 1: } & \min_{\mu_{10\mid x}}\sum_{j=1}^{jj}c\left(0,j\right)\mu_{10\mid x}\left(0,1,j\right)\\
\textrm{s.t. } & \textrm{\textrm{(i) }}0\leq\mu_{10\mid x}\left(0,1,j\right)\leq\mu_{0\mid x}\left(j\right)\textrm{ for }j=1,\ldots,jj\textrm{; }\\
 & \textrm{(ii) }\sum_{j=1}^{jj}\mu_{10\mid x}\left(0,1,j\right)=\mu_{1\mid x}\left(0,1\right);\\
\textbf{Problem 2: } & \min_{\mu_{10\mid x}}\sum_{j=jj}^{J}c\left(1,j\right)\mu_{10\mid x}\left(1,1,j\right)\\
\textrm{s.t. } & \textrm{\textrm{(i) }}0\leq\mu_{10\mid x}\left(1,1,j\right)\leq\mu_{0\mid x}\left(j\right)\textrm{ for }j=jj,\ldots,J\textrm{;}\\
 & \textrm{(ii) }\sum_{j=jj}^{J}\mu_{10\mid x}\left(1,1,j\right)=\mu_{1\mid x}\left(1,1\right).
\end{align*}
Problem 1 is solved by $\textrm{LPS}\left(c\left(0,1:jj\right),\mu_{0\mid x}\left(1:jj\right),\mu_{1\mid x}\left(0,1\right)\right)$
where $c\left(0,1:jj\right)$ is the vector consisting of $c\left(0,j\right)$
for $j=1,\ldots,jj$ and $\mu_{0\mid x}\left(1:jj\right)$ is the
vector consisting of $\mu_{0\mid x}\left(j\right)$ for $j=1,\ldots,jj$.
Similarly, we solve Problem 2 through LPS. This is accomplished during
Step 2 of DREAM.

If Constraint (iii) in (\ref{eq:partial OT simplified}) is satisfied
automatically after solving Problems 1 and 2, then we obtain the solution
to the inner minimization problem in (\ref{eq:partial OT simplified}).
If it is violated, then we proceed to Step 3 of DREAM.

In Step 3, we move the extra mass defined as 
\[
slack\equiv\mu_{10\mid x}\left(0,1,jj\right)+\mu_{10\mid x}\left(1,1,jj\right)-\mu_{0\mid x}\left(jj\right)
\]
away from $\mu_{10\mid x}\left(0,1,jj\right)$ and $\mu_{10\mid x}\left(1,1,jj\right)$
to make Constraint (iii) satisfied. Let $\alpha\left(0,j\right)\geq0$
for $j=1,\ldots,jj-1$ and $\alpha\left(1,j\right)\geq0$ for $j=jj+1,\ldots,J$
be the increment of the mass at each corresponding $\mu_{10\mid x}\left(0,1,j\right)$
and $\mu_{10\mid x}\left(1,1,j\right)$, respectively. We aim to find
$\alpha\left(0,j\right)$ and $\alpha\left(1,j\right)$ that solve the following minimization problem: 
\begin{align}
 & \min_{\alpha}\left[\sum_{j=1}^{jj-1}\left[c\left(0,j\right)-c\left(0,jj\right)\right]\alpha\left(0,j\right)+\sum_{j=jj+1}^{J}\left[c\left(1,j\right)-c\left(1,jj\right)\right]\alpha\left(1,j\right)\right]\label{eq:Step 3 optimization}\\
\textrm{s.t. } & \textrm{\textrm{(i) }}0\leq\alpha\left(0,j\right)\leq\mu_{0\mid x}\left(j\right)-\mu_{10\mid x}\left(0,1,j\right)\textrm{ for }j=1,\ldots,jj-1\textrm{;}\nonumber \\
 & \textrm{(ii) }0\leq\alpha\left(1,j\right)\leq\mu_{0\mid x}\left(j\right)-\mu_{10\mid x}\left(1,1,j\right)\textrm{ for }j=jj+1,\ldots,J\textrm{;}\nonumber \\
 & \textrm{(iii) }\sum_{j=1}^{jj-1}\alpha\left(0,j\right)+\sum_{j=jj+1}^{J}\alpha\left(1,j\right)=slack.\nonumber 
\end{align}
The objective function in (\ref{eq:Step 3 optimization}) measures
the increase in cost due to moving mass away from $\mu_{10\mid x}\left(0,1,jj\right)$
to $\mu_{10\mid x}\left(0,1,j\right)$ for $j\neq jj$ and from $\mu_{10\mid x}\left(1,1,jj\right)$
to $\mu_{10\mid x}\left(1,1,j\right)$ for $j\neq jj$. The increase
shall be made as small as possible. Constraints (i) and (ii) in (\ref{eq:Step 3 optimization}) correspond to the range of possible increments, and Constraint (iii) requires the total increment in mass to be equal to the extra mass $slack$. The
structure of (\ref{eq:Step 3 optimization}) allows LPS to apply.

Step 3 of DREAM implements (\ref{eq:Step 3 optimization}) and updates
the $\mu_{10\mid x}\left(0,1,j\right)$ and $\mu_{10\mid x}\left(1,1,j\right)$
after obtaining $\alpha\left(0,j\right)$ and $\alpha\left(1,j\right)$.
After solving (\ref{eq:partial OT simplified step 2}) by separately
solving Problems 1 and 2, we obtain $\mu_{10\mid x}\left(i,1,j\right)$
for each $i=0,1$ and $j=1,\ldots,jj$. However, because Constraint
(iii) in (\ref{eq:partial OT simplified}) does not appear in (\ref{eq:partial OT simplified step 2}),
it can be violated, making $\mu_{10\mid x}\left(i,1,j\right)$ not
a solution for (\ref{eq:partial OT simplified}). As a result, we
need to adjust $\mu_{10\mid x}\left(i,1,j\right)$ so that Constraint
(iii) in (\ref{eq:partial OT simplified}) is satisfied while minimize
the objective function.

We define $slack$ as the total amount of mass at $\mu_{10\mid x}\left(0,1,jj\right)$
and $\mu_{10\mid x}\left(1,1,jj\right)$ that needs to be moved away.
Values of $\alpha\left(0,j\right)$ and $\alpha\left(1,j\right)$
are the adjustment that we aim to make to $\mu_{10\mid x}\left(0,1,j\right)$
and $\mu_{10\mid x}\left(1,1,j\right)$. First, we have $\alpha\left(0,j\right)\geq0$
and $\alpha\left(1,j\right)\geq0$ because $\mu_{10\mid x}\left(0,1,j\right)$
for $j=1,\ldots,jj-1$ and $\mu_{10\mid x}\left(1,1,j\right)$ for
$j=jj+1,\ldots,J$ will only take mass from $\mu_{10\mid x}\left(0,1,jj\right)$
and $\mu_{10\mid x}\left(1,1,jj\right)$. Second, given the mass at
$\mu_{10\mid x}\left(0,1,j\right)$ for $j=1,\ldots,jj-1$ assigned
after Step 2, the largest possible increment is bounded by $\mu_{0\mid x}\left(j\right)-\mu_{10\mid x}\left(0,1,j\right)$.
Therefore, we obtain Constraint (i) in (\ref{eq:Step 3 optimization}).
Following from the similar argument, we obtain Constraint (ii). At
last, the sum of the increment should equal the total amount of
mass $slack$ that we plan to relocate. This renders Constraint (iii).
Because $\sum_{j=1}^{jj}\mu_{0|x}\left(j\right)\geq\mu_{1\mid x}\left(0,1\right)$
and $\sum_{j=jj}^{J}\mu_{0|x}\left(j\right)\geq\mu_{1\mid x}\left(1,1\right)$ for any $jj\in\left\{ JL,\ldots,JU\right\} $, Constraints (i) and
(ii) are consistent with Constraint (iii).

The minimization problem (\ref{eq:Step 3 optimization}) can be solved
via LPS by letting 
\begin{align*}
cost & =\left[c\left(0,1:jj-1\right)-c\left(0,jj\right),c\left(1,jj+1:J\right)-c\left(1,jj\right)\right],\\
ineq & =\left[\mu_{0\mid x}\left(1:jj-1\right)-\mu_{10\mid x}\left(0,1,jj-1\right),\mu_{0\mid x}\left(jj+1:J\right)-\mu_{10\mid x}\left(1,1,jj+1:J\right)\right],
\end{align*}
and $eq=slack$, where $\left[v_{1},v_{2}\right]$ combines two vectors
$v_{1}$ and $v_{2}$. Once we obtain $\alpha\left(0,j\right)$ and
$\alpha\left(1,j\right)$, we adjust $\mu_{10\mid x}\left(0,1,j\right)$
for $j=1,\ldots,jj-1$ by adding extra mass $\alpha\left(0,j\right)$
and adjust $\mu_{10\mid x}\left(1,1,j\right)$ for $j=jj+1,\ldots,J$
by adding extra mass $\alpha\left(1,j\right)$. Because $\alpha\left(1,j\right)\geq0$
for $j=jj+1,\ldots,J$, we have $\sum_{j=1}^{jj-1}\alpha\left(0,j\right)\leq slack$.
Moreover, since $\mu_{10\mid x}\left(1,1,jj\right)$ from Step 2 is
less than or equal to $\mu_{0\mid x}\left(jj\right)$, we have that
$slack$ is less than or equal to $\mu_{10\mid x}\left(0,1,jj\right)$
obtained from Step 2. Therefore, $\sum_{j=1}^{jj-1}\alpha\left(0,j\right)$
is less than or equal to $\mu_{10\mid x}\left(0,1,jj\right)$ from
Step 2. Thus, we can remove $\sum_{j=1}^{jj-1}\alpha\left(0,j\right)$
amount of mass away from $\mu_{10\mid x}\left(0,1,jj\right)$ to obtain
the updated value of $\mu_{10\mid x}\left(0,1,jj\right)$. Similarly,
we remove $\sum_{j=jj+1}^{J}\alpha\left(1,j\right)$ amount of mass
away from $\mu_{10\mid x}\left(1,1,jj\right)$. Because of Constraint
(iii) in (\ref{eq:Step 3 optimization}), the updated $\mu_{10\mid x}\left(0,1,jj\right)$
and $\mu_{10\mid x}\left(1,1,jj\right)$ will satisfies Constraint
(iii) in (\ref{eq:partial OT simplified}). 

After Steps 2 and 3, we obtain the solution $\mu_{10\mid x}\left(i,1,j\right)$
to the inner minimization problem in (\ref{eq:partial OT simplified}).
We then compute the optimal value of the inner minimization for the
given $jj$ by $t\_cost\left(jj\right)=\sum_{j=1}^{J}\sum_{i=0}^{1}c\left(i,j\right)\mu_{10\mid x}\left(i,1,j\right)$. The optimal value of the outer minimization problem, or equivalently
of the original optimization problem (\ref{eq:Partial transport problem}), is then calculated as $\min\left\{ t\_cost\left(JL\right),\ldots,t\_cost\left(JU\right)\right\} $.

\subsubsection{Time Complexity of Constructing the Identified Set for TPRD Measures}

\label{subsec:Time-Complexity-TPRD}

We first prove the time complexity of DREAM.

\begin{lemma} \label{lem:: time complexity - TPRD - FPS} The time
complexity for computing $\mathcal{KT}_{q^{\top}m_{b}}\left(\mu_{1\mid x},\mu_{0\mid x};x\right)$
based on DREAM is approximately $\frac{9}{4}J^{3}+\frac{21}{2}J^{2}+\frac{3}{2}J$.
\end{lemma}

\begin{proof}[Proof of Lemma \ref{lem:: time complexity - TPRD - FPS}]

We compute the number of basic operations in DREAM. We often overestimate
the required operations to simplify the calculation.

For vectors $cost$ and $ineq$ of length $len$, LPS first ranks
$cost\left(\cdot\right)$, which takes $\frac{1}{2}len\left(len-1\right)$
operations. During the for-loop, finding the index requires $len$
operations; assigning values takes in total $4$ operations. Thus,
the for-loop needs $len\left(len+4\right)$ operations. In total,
LPS requires $\frac{3}{2}len^{2}+\frac{7}{2}len$ operations.

During initialization, we first compute $J$ differences and then
sort them. This requires in total $\frac{1}{2}J\left(J+1\right)$
operations. In the worst case where $JL=JU$, Step 1 requires $J\left(J+1\right)$
operations.

Step 2 of DREAM calls LPS two times with the lengths of the cost functions
being $jj$ and $J-jj+1$, respectively. Thus, the time complexity
of Step 2 is at most $\frac{3}{4}J^{2}+\frac{7}{2}J$.

In Step 3, we first construct two arrays: $d\_cost$ and $d\_mass$,
which requires $J$ operations. Calling LPS takes $\frac{3}{2}J^{2}+\frac{7}{2}J$
operations. The subsequence value assignment needs about $2J$ operations.
Thus, Step 3 requires $\frac{3}{2}J^{2}+\frac{11}{2}J$ operations
in total.

We repeat Steps 2 and 3 at most $J$ times. As a result, the total
number of operations is $\frac{9}{4}J^{3}+\frac{21}{2}J^{2}+\frac{3}{2}J$.
Note that this is a greatly amplified upper bound because we only
consider worst cases during the calculation and ignore results that
can be reused across steps. 
\end{proof}

The original linear programming (\ref{eq:Partial transport problem})
has $2J$ variables and about $3J$ inequality constraints. By the
discussion in Section \ref{subsec:Time-Complexity-of-DD}, solving
such a linear programming requires about $33\left(3J\right)^{3}$
number of basic operations when $J$ is large. Our DREAM can be much faster.

We approximate $\Delta_{TPRD}$ via $\widehat{\Delta}_{TPRD}=\left\{ \delta=G\left(\theta\right):\theta\in\widehat{\Theta}_{I}\right\}$ as discussed in Section \ref{subsec: DREAM and Kallus}. Assume that a grid-based method is used for the numerical integration
when computing $h_{\Theta_{I}}\left(q\right)$ for any given $q$.
Let $N_{x}$ denote the total number of grid points. Then we need
$\left(\frac{9}{4}J^{3}+\frac{21}{2}J^{2}+\frac{3}{2}J\right)N_{x}$
basic operations for computing $h_{\Theta_{I}}\left(q\right)$ for
one given $q$. As a result, constructing $\widehat{\Theta}_{I}$
requires $\left(\frac{9}{4}J^{3}+\frac{21}{2}J^{2}+\frac{3}{2}J\right)N_{x}N_{q}+N_{q}N_{\theta}$
number of operations. By definition, for any $\theta$, $G\left(\theta\right)$
takes $5K$ operations. Thus, in total, it requires at most $
\left(\frac{9}{4}J^{3}+\frac{21}{2}J^{2}+\frac{3}{2}J\right)N_{x}N_{q}+N_{q}N_{\theta}+5KN_{\theta}
$ basic operations to construct $\widehat{\Delta}_{TPRD}$.

\subsection{Additional Details on the Empirical Study}\label{SA::4}

\label{subsec:Additional-Details-Empirical}

This section provides further implementation details for the empirical
studies discussed in Sections \ref{subsec:Comparison-DD} and \ref{subsec: DREAM and Kallus}.
All the computations and numerical comparison are carried out using Python 3.13.3, with NumPy
2.2.5, SciPy 1.15.3, Pandas 2.2.3, Scikit-learn 1.6.1, Joblib 1.5.0, Gurobi
12.0.2 and Matplotlib 3.10.3. Experiments are run on a laptop equipped with an Apple M3
Max processor (14 cores) and 36 GB of RAM. The code in KMZ is adapted
from Python 2 to Python 3. We use the pre-computed
conditional probabilities $\Pr\left(Y_{1}\mid X\right)$ and $\Pr\left(Y_{0}\mid X\right)$
for each observation on $X$ in KMZ, instead of computing them from
the raw data. 

In the comparison reported in Section \ref{subsec:Comparison-DD},
we implemented the method from KMZ without smoothing. The 100 direction
vectors are provided in their replication code. While KMZ mentions
using a 0.1\% subsample (14,903 observations), their code indicates
the use of a 1\% subsample (149,032 observations). We follow the latter
setting in our replication. In addition, we corrected obvious typos in their code. Our methods and KMZ are executed in a
single core for time comparison.

For the comparison in Section \ref{subsec: DREAM and Kallus}, we
set $N_{q}=10^{3}$ and $N_{\theta}=10^{8}$ for both our approach
and \cite{gurobi}. Both our method and linear programming via \cite{gurobi}
are executed using parallel computing across 8 cores. $\theta$ is
drawn sequentially, such that we draw 5000 numbers of $\theta$'s
for each loop and check if these $\theta$'s are in $\Theta_{I}$.
The running time for obtaining $\widehat{\Delta}_{TPRD}$ when solving
the linear programming in (\ref{eq:Partial transport problem}) via
\cite{gurobi} for genetic proxy is about 2 minutes 34 seconds. Our
method takes about 1 minute 13 seconds, which is about half the time.  

\end{appendices}

\bibliographystyle{ecta}
\bibliography{FPPS}

\end{document}